\ifdefined\pdfoutput\pdfoutput=1\fi
\PassOptionsToPackage{sort&compress}{natbib}
\documentclass[prx,aps,10pt,twocolumn,notitlepage,superscriptaddress,nofootinbib,longbibliography]{revtex4-2}

\usepackage{amsmath,amsthm}
\usepackage[charter,cal=cmcal,sfscaled=false]{mathdesign}
\usepackage{etoolbox}
\usepackage{booktabs}
\usepackage{graphicx}
\usepackage{caption}
\usepackage[colorlinks=true,urlcolor=blue,citecolor=blue,linkcolor=blue]{hyperref}

\makeatletter
\patchcmd{\@bibdataout@aps}{pages="0"}{pages="1"}{}{
  \PackageError{bibliography-layout}{Unexpected APS bibliography control}{}}
\makeatother

\DeclareMathOperator{\Tr}{Tr}
\newcommand{\id}{\mathbb{I}}
\newtheorem{theorem}{Theorem}
\newtheorem{lemma}[theorem]{Lemma}
\newtheorem{proposition}[theorem]{Proposition}
\newtheorem{corollary}[theorem]{Corollary}
\newcommand{\Nmax}{N_{\max}}
\newcommand{\Nexp}{N_{\mathrm{exp}}}
\newcommand{\Prob}{\mathbb{P}}
\newcommand{\condsep}{\,\mathord{\vert}\,}
\newcommand{\Eprefix}[1]{\mathbb E\ifstrempty{#1}{}{_{#1}}\,}
\newcommand{\Ebr}[2][]{\Eprefix{#1}[#2]}
\newcommand{\budgetopsep}{\;}

\newif\ifIncludePaperAbstract
\newif\ifIncludePaperSupplement
\IncludePaperAbstracttrue
\IncludePaperSupplementtrue
\hypersetup{pdftitle={Sharp Transitions of Copy Complexity in Local Certification of Typical Pure States},
  pdfauthor={Yimeng Cao, Ge Bai}}

\begin{document}
\title{Sharp Transitions of Copy Complexity in Local Certification of Typical Pure States}
\author{Yimeng Cao}
\author{Ge Bai}
\email{gebai@hkust-gz.edu.cn}
\affiliation{Thrust of Artificial Intelligence, Information Hub,\\
The Hong Kong University of Science and Technology (Guangzhou), Guangzhou 511453, China}

\ifIncludePaperAbstract
\begin{abstract}
We determine the asymptotic copy complexity of $n$-qudit state certification
with tensor-product measurements. For given acceptance and rejection fidelities
$a$ and $b$ ($b<a$), the optimal copy complexity is $\Theta(n^r)$.
For each qudit dimension $d$, $r$
is a piecewise constant function of \mbox{$g=(a-b)/(1-b)$}.
This scaling holds for a family of sets of pure target states whose Haar
measures tend to one as $n\to\infty$. We characterize the range of $g$
corresponding to each optimal exponent and provide an algorithm for
computing its endpoints exactly. The optimal exponent grows logarithmically with
$1/g$ as $g$ tends to zero.
The upper bound is achieved by a nonadaptive protocol using only single-qudit
projective measurements and classical postprocessing. The matching lower
bound is established by constructing pairs of states that are difficult to
distinguish under arbitrary tensor-product measurements.
Our result can thus guide the design of certification tests when the number of available copies is limited.
\end{abstract}
\fi

\maketitle

\begingroup
\setlength{\parskip}{2pt}
\par\noindent\textbf{Introduction.}---In experiments designed to prepare
a pure target state, measurements are needed to determine whether the
actual output meets the quality requirements. Noise and control errors
generally cause the prepared state to differ from the target, so we need
to set a standard on how much error can be tolerated. Given such a standard,
quantum state certification uses measurements on copies of the prepared
state to assess whether it is sufficiently close to the target, as
quantified by fidelity~\cite{EisertEtAl2020Certification,MontanaroDeWolf2016Survey}.
A useful test should accept imperfect preparations whose fidelity meets
an acceptance threshold and reject states below a rejection threshold.
This task is known as tolerant certification~\cite{FawziEtAl2024}.

Certifying many-body quantum states also requires attention to the
experimental feasibility of the measurements. Measuring each qudit
separately and processing the complete measurement records classically
avoids entangling operations between qudits during readout, reducing the
demands on quantum control~\cite{ZhangEtAl2020Verification,HuangPreskillSoleimanifar2025,ColadangeloLiSlote2026}.
Systematic methods have been developed for constructing verification
tests using such measurements~\cite{LiuEtAl2023Homogeneous}.
However, restricting the measurements may increase the number of copies
required for certification, so this experimental advantage must be weighed
against the copy cost.

Previous work has used the structure of the target state to design
certification tests based on single-qubit measurements~\cite{HayashiMorimae2015}. Pallister, Linden,
and Montanaro constructed single-qubit Pauli tests for stabilizer states,
including graph states~\cite{PallisterLindenMontanaro2018}. These tests
accept the ideal target and reliably reject states whose fidelity with it
is at most a prescribed threshold. With this rejection threshold fixed,
the required number of copies can be bounded independently of the number
of qubits in each copy.
For graph states, Li, Zhu, and Hayashi also obtained constant-copy
local-Pauli tests that accept sufficiently accurate imperfect
preparations~\cite{LiZhuHayashi2023RobustGraph}.

For more general targets, Huang, Preskill, and Soleimanifar developed a
nonadaptive single-qubit Pauli protocol for a Haar-typical family of
classically known pure multiqubit targets~\cite{HuangPreskillSoleimanifar2025}.
Here, Haar-typical refers to families of target sets whose Haar measures
tend to one as the number of qubits grows. Gupta, He, and O'Donnell
developed a protocol for any known pure multiqubit target using adaptive
single-qubit measurements within each copy~\cite{GuptaHeODonnell2026}.
For a fixed rejection threshold, each of these two results specifies an
acceptance threshold that approaches one as the number of qubits increases.
With these respective threshold choices, the nonadaptive Pauli protocol
uses polynomially many copies, while the adaptive protocol requires at most
a linear number.

For Haar-typical pure multiqubit targets, Coladangelo et al.\ proposed a
protocol using only single-qubit measurements. They conjectured that this
protocol would enable constant-copy certification for suitable fixed fidelity
thresholds~\cite{ColadangeloLiSloteWu2026}. Coladangelo, Li, and Slote
subsequently proved this conjecture and showed that nonadaptive single-qubit
Pauli measurements suffice for this task~\cite{ColadangeloLiSlote2026}.
Let $a$ and $b$ denote the acceptance and rejection fidelity thresholds,
respectively. There is a constant $C>1$ such that, for fixed thresholds
satisfying $(1-a)/(1-b)\le1/C$, the required number of copies can be
bounded independently of the number of qubits~\cite{ColadangeloLiSlote2026}.
Du et al.\ also obtained
constant-copy tests for Haar-typical pure targets and almost all graph
states using nonadaptive single-qubit Pauli measurements, provided that
the fixed ratio $(1-a)/(1-b)$ is sufficiently small~\cite{DuEtAl2026}.

Building on these results, we ask: What is the optimal copy complexity for an
arbitrary fixed pair of acceptance and rejection fidelity thresholds?
We consider Haar-typical $n$-qudit
targets under local product measurements, allowing adaptation across copies.
We show that the optimal copy complexity scales as $n^r\log(1/\delta)$,
where $\delta$ bounds both error probabilities and $r$ is a nonnegative
integer. The exponent depends only on
the local dimension $d$ and the normalized gap \mbox{$g=(a-b)/(1-b)$}. We characterize
the ranges of $g$ corresponding to each exponent and show that $r$
grows logarithmically with $1/g$ as $g\to0$.
The optimal scaling is attained by a nonadaptive protocol using only
single-qudit projective measurements. Together, this hierarchy and the
protocol provide a quantitative basis for designing certification tests
with limited copy resources.\par
\endgroup
\allowbreak

\begin{figure*}[t]
\centering
\includegraphics[width=0.90\textwidth]{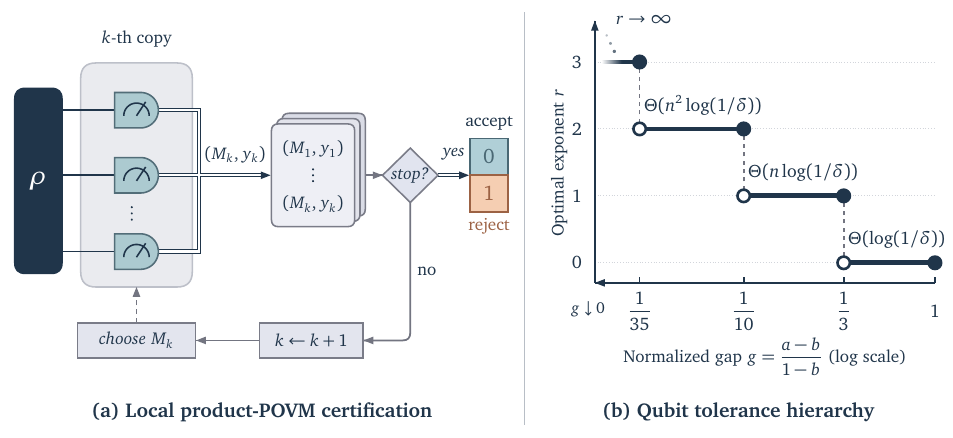}
\caption{Certification protocol and optimal copy complexity.
(a) An allowed protocol measures one copy at a time, recording the product
measurement $M_k$ and its outcome $y_k$ for copy $k$. It may use these
records to choose the next measurement and decide when to stop.
(b) Optimal exponent $r$ versus normalized gap $g$ for Haar-typical qubit
targets. Open and filled circles mark excluded and included endpoints,
respectively.}
\label{fig:local-certification-model}
\end{figure*}

\par\noindent\textbf{Tolerant certification from local readout.}---Consider
an $n$-qudit system with fixed local dimension $d\ge2$. For a known pure target
$|\psi\rangle$, we have access to independent, identically distributed
copies of an unknown state $\rho$. Let $P=|\psi\rangle\langle\psi|$, so that
the fidelity of $\rho$ with the target is $F(\rho)=\Tr(P\rho)$.

\begingroup
\setlength{\parskip}{2.4pt}
\begin{samepage}
For acceptance and rejection thresholds $a$ and $b$, with $0<b<a\le1$,
the task is to distinguish the following hypotheses:
\begin{equation}
H_0:F(\rho)\ge a,\qquad H_1:F(\rho)\le b.
\label{eq:promise}
\end{equation}
\end{samepage}%
For $0<\delta<1/2$, we require both error probabilities to be at most
$\delta$. We impose no requirements on the acceptance or rejection
probabilities for states with $b<F(\rho)<a$. Define the normalized gap
\begin{equation}
g=\frac{a-b}{1-b}=1-\frac{1-a}{1-b}.
\label{eq:normalized-gap}
\end{equation}
For fixed local dimension $d$, this gap determines the optimal copy-complexity exponent.

A certification protocol $\Pi$ measures the copies one at a time. On the
$k$th copy, it measures each qudit separately using an arbitrary positive
operator-valued measure (POVM). We allow randomization and adaptation across copies,
but no measurement feedback within a copy.

Each measurement record
includes the measurement setting and the outcomes on all qudits. The
protocol processes these records classically to decide whether to
continue measuring and, upon stopping, whether to accept or reject; see
\hyperref[fig:local-certification-model]{Fig.~\ref*{fig:local-certification-model}(a)}.

Let $\mathcal S$ denote the union of the two hypothesis classes. A protocol
$\Pi$ is valid if, for every $\rho\in\mathcal S$, it terminates almost
surely and returns an incorrect decision with probability at most $\delta$.
Classical computation and storage are unrestricted. We count only the copies measured.

Let $N$ denote the total number of copies measured by the protocol, and
let $N_\Pi(h)$ be its value for a complete execution record $h$ of $\Pi$.
The record $h$ includes all internal random choices and measurement
outcomes. For a fixed protocol $\Pi$, its worst-case copy cost is
$\sup_h N_\Pi(h)$, where $h$ ranges over all possible execution records.
We define $\Nmax$ as the infimum of this cost over all valid protocols.
\begin{equation}
\Nmax
=\mathop{\mathrm{inf}\vphantom{\mathrm{sup}}}\limits_{\Pi\,\mathrm{valid}}
\budgetopsep
\mathop{\mathrm{sup}\vphantom{\mathrm{inf}}}\limits_{h}
\,N_\Pi(h).
\label{eq:max-budget}
\end{equation}
The optimal worst-case expected copy cost is
\begin{equation}
\Nexp
=\mathop{\mathrm{inf}\vphantom{\mathrm{sup}}}
  \limits_{\Pi\,\mathrm{valid}\vphantom{\rho\in\mathcal S}}
\budgetopsep
\mathop{\mathrm{sup}\vphantom{\mathrm{inf}}}
  \limits_{\rho\in\mathcal S\vphantom{\Pi\,\mathrm{valid}}}
\,\Ebr{N\condsep\Pi,\rho},
\label{eq:expected-budget}
\end{equation}
where the expectation is over the measurement outcomes and the protocol's
internal randomness. Thus $\Nexp\le\Nmax$.

\par\noindent\textbf{The exact tolerance hierarchy.}---We now determine
the optimal copy complexity for Haar-typical pure targets. For fixed
local dimension $d$, the optimal exponent of $n$ depends on the
normalized gap $g$. We write $\gamma_j(d)$ for the step point
separating optimal exponents $j$ and $j+1$, where $j=0,1,\ldots$.
These values are positive, strictly decreasing, and converge to zero as $j\to\infty$. Their
mathematical definition is given in Sec.~\ref{sec:supp-duality} of the
Supplemental Material.

The first three step points are listed in
Table~\ref{tab:critical-values-dimension}.
Section~\ref{sec:supp-exact-certificates} of the Supplemental Material
gives a general procedure for computing the step points exactly.
Proposition~\ref{prop:supp-low-degree-values} derives the values in
Table~\ref{tab:critical-values-dimension}.
For qubit targets,
\hyperref[fig:local-certification-model]{Fig.~\ref*{fig:local-certification-model}(b)}
shows the first few step points and the \mbox{corresponding} \mbox{optimal exponents}.
\par\endgroup

\begin{center}
\begin{minipage}{\columnwidth}
\centering
\captionsetup{type=table,position=top,labelsep=period,justification=centering}
\setlength{\tabcolsep}{10pt}
\vspace*{3pt}
\caption{First three step points for different~local~dimensions.
The formula for $\gamma_2(d)$ with $d\ge3$ does not give $\gamma_2(2)=1/35$.}
\label{tab:critical-values-dimension}
\begin{tabular}{ccc}
\toprule
$j$ & $\gamma_j(2)$ & $\gamma_j(d)$, $d\ge3$ \\
\midrule
$0$ & $\dfrac{1}{3}$ & $\dfrac{1}{d+1}$ \\
\addlinespace[4pt]
$1$ & $\dfrac{1}{10}$ & $\dfrac{1}{(d+1)^2+1}$ \\
\addlinespace[4pt]
$2$ & $\dfrac{1}{35}$ & $\dfrac{1}{(d+1)\bigl((d+1)^2+3\bigr)}$ \\
\bottomrule
\end{tabular}
\end{minipage}
\end{center}

\begingroup
\setlength{\parskip}{2.5pt}
\setlength{\topsep}{4pt}
\addtolength{\abovedisplayskip}{-2.5pt}
\addtolength{\belowdisplayskip}{-2.5pt}
\makeatletter
\appto\thm@space@setup{\advance\thm@postskip by -\parskip}
\makeatother
\setlength{\lineskiplimit}{-1.25pt}
For a given normalized gap $g$, define
\begin{equation}
r_d(g)=\min\bigl\{j\in\mathbb N_0:\ g>\gamma_j(d)\bigr\}.
\label{eq:critical-degree}
\end{equation}

For each $n$, let $\Prob_{\mathrm{Haar}}$ denote the unitarily invariant
probability measure on pure $n$-qudit states, and let $\mathcal G_n$ be a set
of such states. We call the family $(\mathcal G_n)_{n\ge1}$ Haar-typical if
$\Prob_{\mathrm{Haar}}(\mathcal G_n)\to1$ as $n\to\infty$.

\begin{theorem}[Optimal copy complexity]
\label{thm:hierarchy}
Fix an integer $d\ge2$, thresholds $0<b<a\le1$, and set $r=r_d(g)$.
There exists a Haar-typical family $(\mathcal G_n)_{n\ge1}$ such that,
for every fixed $0<\delta_\star<1/2$, both $\Nexp$ and $\Nmax$ are
\begin{equation}
\Theta\!\left(n^r\log\frac{1}{\delta}\right),
\label{eq:hierarchy}
\end{equation}
uniformly over $0<\delta\le\delta_\star$ and $\psi\in\mathcal G_n$ as
$n\to\infty$.
\end{theorem}

\begingroup
\clubpenalty=-25\relax
For fixed $\delta$, a constant number of copies suffices for certification if and only if
$g>\gamma_0(d)=1/(d+1)$. Otherwise, the optimal copy complexity grows as a
positive integer power of $n$. At each step point $g=\gamma_j(d)$,
the optimal exponent is $j+1$.
\par\endgroup

The upper bound is attained by a nonadaptive measurement protocol using
a predetermined number of copies. The matching lower bound holds for
the worst-case expected copy cost of every valid protocol, including
those with adaptation across copies and random stopping. The Haar-typical
family must be chosen appropriately, since, when $r\ge1$, exceptional
targets such as product states admit constant-copy certification~\cite{DangniamHanZhu2020}. The proof is
given in Sec.~\hyperref[proof:supp-hierarchy]{\ref*{subsec:supp-matching}}
of the Supplemental Material.

In our notation, the sufficient condition for constant-copy certification
of Haar-typical qubit targets in Ref.~\cite{ColadangeloLiSlote2026}
becomes $g\ge1-1/C$, where $C$ is not given explicitly in the proof but
numerical experiments suggest $C<5.5$.
Theorem~\ref{thm:hierarchy} gives the exact condition $g>1/3$ under
arbitrary product measurements, allowing any $C>1.5$ in the earlier
sufficient condition. This range is optimal and substantially sharpens
the previous guarantee. More generally, it determines the
asymptotically optimal copy complexity for every fixed local dimension
and every fixed threshold pair $0<b<a\le1$. At a given dimension, the
normalized gap $g$ determines the optimal exponent. The absolute
gap $a-b$ alone does not.

Theorem~\ref{thm:hierarchy} treats fixed fidelity thresholds. We next allow
the thresholds $a$ and $b$ to vary with $n$, with their normalized gap near a step point.

\begin{corollary}
\label{cor:near-critical}
Fix integers $d\ge2$ and $j\ge0$. Let $(a_0,b_0)$
be any threshold pair with normalized gap $\gamma_j(d)$. There exist an
open neighborhood $\mathcal U$ of $(a_0,b_0)$ and a Haar-typical family
$(\mathcal G_n)_{n\ge1}$ with the following property.
For $(a,b)\in\mathcal U$, write
\begin{equation}
\varepsilon=\frac{a-b}{1-b}-\gamma_j(d).
\label{eq:critical-offset}
\end{equation}
Then, for every fixed $0<\delta_\star<1/2$, both $\Nexp$ and $\Nmax$ are
\addtolength{\abovedisplayskip}{-1.5pt}
\addtolength{\belowdisplayskip}{-1.5pt}
\begin{equation}
\label{eq:near-critical-cost}
\begin{cases}
\displaystyle
\Theta\!\left(\min\!\left\{\frac{n^j}{\varepsilon^2},\,n^{j+1}\right\}\log\frac{1}{\delta}\right),
  & \varepsilon>0,\\[6pt]
\displaystyle
\Theta\!\left(n^{j+1}\log\frac{1}{\delta}\right),
  & \varepsilon\le0,
\end{cases}
\end{equation}
uniformly over $(a,b)\in\mathcal U$, $0<\delta\le\delta_\star$,
and $\psi\in\mathcal G_n$ as~\mbox{$n\to\infty$}.
\end{corollary}

For fixed $\delta$, the optimal copy complexity is
$\Theta(n^j/\varepsilon^2)$ when $\varepsilon\ge n^{-1/2}$, and
$\Theta(n^{j+1})$ when $\varepsilon\le n^{-1/2}$. Thus, even when
$g>\gamma_j(d)$, the copy complexity scales as $n^{j+1}$ once
$\varepsilon$ is of order $n^{-1/2}$ or smaller. This identifies
$n^{-1/2}$ as the crossover scale near each step point.
The proof is given in Sec.~\ref{subsec:supp-near-critical} of the Supplemental Material.

As the normalized gap $g$ tends to zero, the optimal exponent in
Theorem~\ref{thm:hierarchy} increases in integer steps. We now determine
its asymptotic growth.

\begin{samepage}
\begin{theorem}[Growth of the optimal exponent]
\label{thm:sharp-hierarchy}
\setlength{\abovedisplayskip}{5pt plus 1pt minus 1pt}
\setlength{\belowdisplayskip}{5pt plus 1pt minus 1pt}
\setlength{\belowdisplayshortskip}{4pt plus 1pt minus 1pt}
For every fixed integer $d\ge2$, the step points satisfy
\begin{equation}
\gamma_j(d)=\Theta\!\left(\frac{(d+1)^{-j}}{\sqrt{j}}\right),
\qquad j\to\infty.
\label{eq:sharp-order}
\end{equation}
The optimal exponent satisfies
\postdisplaypenalty=10000
\begin{equation}
r_d(g)=L-\frac12\log_{d+1}L+O(1),
\qquad g\downarrow0,
\label{eq:degree-inversion}
\end{equation}
where $L=\log_{d+1}(1/g)$.
\end{theorem}
\end{samepage}

\par
Theorem~\ref{thm:sharp-hierarchy} gives the leading term
$\log_{d+1}(1/g)$ and the next correction to the optimal exponent, with a
bounded remainder. The interpolation bounds in Ref.~\cite{SzehrZarouf2021}
can also be used to derive the stated decay of the step points.
Section~\ref{sec:supp-asymptotics} of the Supplemental Material presents
this derivation and an alternative proof that does not use the
interpolation bounds.

\noindent\textbf{Discussion.}---We fully solve the problem of local
certification of Haar-typical multi-qudit states with any given acceptance
and rejection fidelity thresholds, by giving matching upper and lower
bounds on the worst-case and expected copy complexity. We discover
transitions in the copy complexity exponent determined by the normalized
gap Eq.~\eqref{eq:normalized-gap}, and that the exponent grows
logarithmically as the gap shrinks to zero.
Theorem~\ref{thm:hierarchy} provides
quantitative guidance for designing quantum state certification experiments.
For Haar-typical qubit targets, the threshold pairs $(a,b)=(0.99,0.98)$
and $(0.96,0.95)$ have the same $a-b=0.01$, but different normalized gaps
$g=1/2$ and $g=1/5$, yielding optimal copy complexities of $\Theta(1)$
and $\Theta(n)$, respectively, under tensor-product measurements at fixed
error probability. For $(a,b)=(0.96,0.95)$, this linear scaling cannot
be improved by arbitrary tensor-product POVMs or adaptation across copies.

\par\endgroup
\begingroup
\advance\baselineskip by -.25pt\relax
\setlength{\lineskiplimit}{-1.25pt}
\setlength{\parskip}{0pt}
\par\addvspace{.125\baselineskip}
\begingroup\clubpenalties 5 10000 10000 10000 10000 0\relax
This limitation can be overcome by allowing joint measurements.
If all qudits in each copy can be measured jointly,
the projective measurement $\{P,\id-P\}$ produces outcome $P$ with
probability $F(\rho)$, yielding a certification test using
$O\!\left((a-b)^{-2}\log(1/\delta)\right)$ copies~\cite{Fawzi2026Binary,Hoeffding1963}. For Haar-typical qubit
targets, constant-copy certification is possible even when joint
measurements are restricted to fixed-size blocks: partition the system
into $k$-qubit blocks, allowing arbitrary joint measurements within each
block and product measurements across blocks. Treating each block as a
$2^k$-dimensional subsystem shows that this holds whenever $g>1/(2^k+1)$.
For any fixed thresholds, $k$ can be chosen large enough and kept fixed as
the system grows (see~Sec.~\ref{subsec:supp-fixed-blocks} of the Supplemental
Material).\par\endgroup

\par\addvspace{.125\baselineskip}
Local state certification has been used to assess the quality of
experimentally prepared quantum states.
In particular, studies on photonic qubit and qutrit states have examined
how many copies are needed to certify that the fidelity exceeds a given
threshold~\cite{JiangEtAl2020Verification,XiaEtAl2022Verification}.
When such experiments are extended to Haar-typical multiqudit targets,
our results can help experimenters assess what certification requirements
are achievable with the \mbox{available copies}. The protocol constructed
in our upper-bound proof can also guide the design of such experiments.

\noindent\textbf{Formal verification and code availability.}--- We have formally
verified Theorem~\ref{thm:hierarchy}, Corollary~\ref{cor:near-critical}, and
Theorem~\ref{thm:sharp-hierarchy} in Lean~4~\cite{DeMouraUllrich2021Lean4},
using Mathlib~\cite{MathlibCommunity2020} and Lean-QIT~\cite{ZhuEtAl2026LeanQIT}.
The source code is available at
\url{https://github.com/cymhit/local-state-certification-lean}.

\begingroup
\makeatletter
\@booleanfalse\acknowledgments@sw
\makeatother
\par\addvspace{.125\baselineskip}
\begin{acknowledgments}
\noindent\textbf{\acknowledgmentsname.}---This work was supported by the
Guangdong Basic and Applied Basic Research Foundation (Grant Nos.\
2026A1515030035 and 2025A1515110223).
\end{acknowledgments}
\endgroup
\par\nobreak
\vskip-\prevdepth\hrule width 0pt height 1.5pt\relax
\endgroup

\makeatletter
\patchcmd{\bibsection}{\onecolumngrid@push}{\onecolumngrid@push\vskip 11pt}{}{%
  \PackageError{bibliography-layout}{Could not insert space above the reference rule}{}}
\patchcmd{\bibsection}{\addvspace{19\p@}}{\addvspace{8pt}}{}{%
  \PackageError{bibliography-layout}{Could not adjust space below the reference rule}{}}
\let\savedRaggedColumnSwitch\raggedcolumn@sw
\let\savedBalanceTwoColumns\balance@two
\@booleanfalse\raggedcolumn@sw
\patchcmd{\balance@two}{\dimen@ii<.5\p@}{\dimen@ii<1.01\baselineskip}{}{%
  \PackageError{bibliography-layout}{Unexpected column-balance upper test}{}}
\patchcmd{\balance@two}{\dimen@ii>-.5\p@}{\dimen@ii>-1.01\baselineskip}{}{%
  \PackageError{bibliography-layout}{Unexpected column-balance lower test}{}}
\makeatother
\apptocmd{\thebibliography}{\interlinepenalty=10000\relax}{}{%
  \PackageError{bibliography-layout}{Could not set the reference break penalty}{}}
\bibliography{references}
\onecolumngrid
\clearpage
\makeatletter
\let\balance@two\savedBalanceTwoColumns
\let\raggedcolumn@sw\savedRaggedColumnSwitch
\makeatother

\ifIncludePaperSupplement
\onecolumngrid
\clearpage
\interlinepenalty=0

\begin{center}
\textbf{\large Supplemental Material}
\end{center}

\setcounter{section}{0}
\renewcommand{\thesection}{\Roman{section}}
\renewcommand{\thesubsection}{\Alph{subsection}}
\renewcommand{\theHsection}{supp.\arabic{section}}
\renewcommand{\theHsubsection}{supp.\arabic{section}.\arabic{subsection}}
\setcounter{equation}{0}
\renewcommand{\theequation}{S\arabic{equation}}
\renewcommand{\theHequation}{supp.\arabic{equation}}

\setcounter{theorem}{0}
\renewcommand{\thetheorem}{S\arabic{theorem}}
\renewcommand{\theHtheorem}{supp.\arabic{theorem}}

\setcounter{table}{0}
\renewcommand{\thetable}{S\arabic{table}}
\renewcommand{\theHtable}{supp.\arabic{table}}

\begingroup
\newcommand{\smipar}{\par\addvspace{2pt}}

\section{Model, protocols, and copy complexity}
\label{sec:supp-model}

Fix integers $d\ge2$ and $n\ge1$, and let
$\mathcal H=(\mathbb C^d)^{\otimes n}$ and $D=d^n$.
For a known pure target $|\psi\rangle\in\mathcal H$, a certification
protocol $\Pi$ receives iid copies of an unknown state $\rho$ on
$\mathcal H$ as its quantum input. The state $\rho$ may depend on the target.
Let $P=|\psi\rangle\langle\psi|$. The fidelity of $\rho$ with the
target is $F(\rho)=\Tr(P\rho)$.
Classical computation and storage are unrestricted. We count only the copies
measured. Adversarial verification permits correlated registers
and relates the test's passing probability to the conditional
fidelity of an unmeasured register~\cite{ZhuHayashi2019Adversarial}.

\smipar
Fix acceptance and rejection thresholds $a$ and $b$, with $0<b<a\le1$,
and an error bound $0<\delta<1/2$. Define the acceptance class
$\mathcal S_0=\{\rho:F(\rho)\ge a\}$ and the rejection class
$\mathcal S_1=\{\rho:F(\rho)\le b\}$, and set
$\mathcal S=\mathcal S_0\cup\mathcal S_1$.
We impose no requirements on the acceptance or rejection
probabilities for states with $b<F(\rho)<a$.
The normalized gap is $g=(a-b)/(1-b)$.

\smipar
For each copy it measures, the protocol chooses all local POVMs $M_i$
before observing any outcomes from that copy. It then performs the product
measurement $M=\bigotimes_{i=1}^nM_i$. Let $\mathcal M_n$ denote the class
of all such product measurements, allowing arbitrary measurable local
outcome spaces, including discrete and continuous ones. For fixed $M$ and
$\rho$, the outcome distribution is the probability measure given on
measurable outcome sets $A$ by
\begin{equation}
q_M^\rho(A)=\Tr[\rho M(A)],\qquad M\in\mathcal M_n.
\label{eq:supp-povm-born}
\end{equation}
Here $M(A)$ denotes the positive operator associated with the outcome event $A$.
For a discrete POVM, $M(A)=\sum_{y\in A}M(\{y\})$.
The protocol records the setting and all local outcomes. To choose the
local POVMs, it may use shared classical randomness across sites and
records from earlier copies.
Within-copy feedback and joint quantum operations across sites or copies
are excluded.

Product-basis readout is the special case used by the upper-bound test.
Let $\mathcal U_n$ consist of $U=\bigotimes_{i=1}^nU_i$, with
$U_i\in\mathrm U(d)$. Measuring the projectors
$U|x\rangle\langle x|U^\dagger$, $x\in\{0,\ldots,d-1\}^n$, gives
\begin{equation}
q_U^\rho(x)=\langle x|U^\dagger\rho U|x\rangle,
\qquad
p_U(x)=q_U^P(x)=|\langle x|U^\dagger\psi\rangle|^2.
\label{eq:supp-born}
\end{equation}
The record for this copy is $(U,x)$. For Pauli readout, complete basis
outcomes retain compatible parity information used in fidelity
estimation~\cite{AcharyaEtAl2026}, whereas the Pauli-parity model supplies
only the selected string's binary outcome~\cite{Fawzi2026Binary}.

\smipar
Collect all auxiliary classical randomness, including that of the final
decision, in a seed $\zeta$ taking values in an arbitrary measurable
space. For a given protocol, the seed distribution may depend on the
known target and task parameters. With these data fixed, the distribution
is the same for all $\rho$, and the protocol draws $\zeta$ independently
of the experimental copies. The protocol rules are then measurable
functions of the seed and the available records.

\smipar
A slot for a measured copy records the setting and all local outcomes
in a measurable space.
After stopping, all subsequent slots hold a common empty record.
Let $\mathcal F_0=\sigma(\zeta)$ and let $\mathcal F_k$ be generated by
the seed and the first $k$ slots. Write $h_{<k}$ for the history consisting
of the seed and the first $k-1$ slots. This history determines whether
the protocol takes copy $k$ and, if it does, which product POVM $M_k$ it uses.
For each measurable outcome set $A$, the effect $M_k(h_{<k},A)$ is
measurable in $h_{<k}$. The seed distribution, the Born kernels
$\Tr[\rho M_k(h_{<k},\mathrm dy)]$ on continuation, and the point mass
at the empty record after stopping define the distribution of the
complete execution record by iteration.
Write $\Prob(\cdot\condsep\Pi,\rho)$ and
$\Ebr{\cdot\condsep\Pi,\rho}$ for probabilities and expectations
under this distribution, with $\Pi$ and $\rho$ fixed.

\smipar
A complete execution record $h$ of $\Pi$ consists of the seed and all
record slots. Let $N$ denote the number of copies measured and $N_\Pi(h)$
its value on $h$. Stopping before the first measurement gives $N_\Pi(h)=0$,
and continuing indefinitely gives \mbox{$N_\Pi(h)=\infty$}. The continuation
rule makes $N$ a stopping time for $(\mathcal F_k)_{k\ge0}$. When the
protocol stops, it returns an $\mathcal F_N$-measurable decision $J$,
with $J=0$ for acceptance and $J=1$ for rejection.
Call an allowed protocol $\Pi$ valid if it terminates almost surely for
every $\rho\in\mathcal S$ and satisfies
\begin{equation}
\sup_{\rho\in\mathcal S_0}\Prob(J=1\condsep\Pi,\rho)\le\delta,
\qquad
\sup_{\rho\in\mathcal S_1}\Prob(J=0\condsep\Pi,\rho)\le\delta.
\label{eq:supp-errors}
\end{equation}

\smipar
The smallest copy budget that a valid protocol can guarantee on every
run is
\begin{equation}
\Nmax
=\mathop{\mathrm{inf}\vphantom{\mathrm{sup}}}\limits_{\Pi\,\mathrm{valid}}
\budgetopsep
\mathop{\mathrm{sup}\vphantom{\mathrm{inf}}}\limits_{h}
\,N_\Pi(h).
\label{eq:supp-max-budget}
\end{equation}
Here $h$ ranges over all complete records allowed by the rules of $\Pi$,
including branches of probability zero. If no valid protocol has
a finite bound common to all runs, $\Nmax=\infty$.
For fixed $\Pi$ and promised input $\rho$, we average $N$ over the seed
and the measurement outcomes. The optimal
worst-case expected cost is
\begin{equation}
\Nexp
=\mathop{\mathrm{inf}\vphantom{\mathrm{sup}}}
  \limits_{\Pi\,\mathrm{valid}\vphantom{\rho\in\mathcal S}}
\budgetopsep
\mathop{\mathrm{sup}\vphantom{\mathrm{inf}}}
  \limits_{\rho\in\mathcal S\vphantom{\Pi\,\mathrm{valid}}}
\,\Ebr{N\condsep\Pi,\rho}.
\label{eq:supp-expected-budget}
\end{equation}
Expectations may be infinite. We use $\inf\varnothing=\infty$ in both
definitions. A bound on every run bounds each input's expectation, so
$\Nexp\le\Nmax$.

\smipar
In our typicality statements, the Haar probability is taken over the
target. For each fixed target, the error requirements apply to every
promised experimental state.
For fixed thresholds, Sec.~\ref{subsec:supp-matching} gives the target
family and size cutoff for Theorem~\ref{thm:hierarchy} and bounds
$\Nmax$ and $\Nexp$ as functions of $\delta$.
Section~\ref{subsec:supp-near-critical} gives a common family and cutoff
for thresholds varying near a pair whose normalized gap is a step point.

\smipar
In our copy-complexity bounds, we hold $d$ and any stated polynomial degree
or finite coefficient vector fixed as $n\to\infty$. We also hold the fidelity
thresholds fixed unless a result explicitly allows them to vary with $n$.
Constants and size cutoffs
may depend on these parameters. Limits in the degree or tolerance are
stated separately.
Unsubscripted logarithms are natural, and $\id$ denotes the identity operator on
the relevant space.

\par
\endgroup
\section{Polynomial envelopes and finite extremal duality}
\label{sec:supp-duality}

We establish the optimal polynomials and finite dual weights used in the
upper and lower bounds. Fix a real parameter $d>0$ and an integer $r\ge0$.
In the certification problem, $d$ is the integer local dimension, $d\ge2$.
Let $\mathcal P_r=\{p\in\mathbb R[w]:\deg p\le r,\ p(0)=1\}$.
For $p\in\mathcal P_r$, define its lower and upper envelopes by
\begin{equation}
\ell_d(p)=\inf_{w\in\mathbb N_{\ge1}}(d+1)^{-w}p(w),
\qquad
u_d(p)=\sup_{w\in\mathbb N_{\ge1}}(d+1)^{-w}p(w).
\label{eq:envelope}
\end{equation}
Since $(d+1)^{-w}p(w)\to0$ as $w\to\infty$, we have
$\ell_d(p)\le0\le u_d(p)$. The degree-$r$ step point is
\begin{equation}
\gamma_r(d)=\inf_{p\in\mathcal P_r}
\frac{u_d(p)-\ell_d(p)}{1-\ell_d(p)}.
\label{eq:critical-values}
\end{equation}
For $g>0$, the strict crossing index is
\begin{equation}
r_d(g)=\min\bigl\{j\in\mathbb N_0:\ g>\gamma_j(d)\bigr\}.
\label{eq:supp-critical-degree}
\end{equation}
For integer $d\ge2$, this is the index in Eq.~\eqref{eq:critical-degree}.
Proposition~\ref{prop:supp-critical-sequence} gives strict decrease.
For $d>1$, the elementary bound
in Proposition~\ref{prop:supp-elementary-decay} shows that the step
points tend to zero. For $0<d\le1$, convergence follows from
Proposition~\ref{prop:supp-sharp-decay}. Thus the minimum exists for every $g>0$.

\subsection{Envelopes and extrapolation}

Rescaling the two envelope endpoints turns a fractional optimization into
a bounded linear extrapolation problem. For the lower bound, we use its
signed evaluation weights to construct two states. When we expand the
difference between their local readout distributions, the low-order
terms cancel.

Let $\mathcal X=\mathbb N_{\ge1}\cup\{\infty\}$ have the one-point
compactification topology: finite points are isolated, and a neighborhood
of $\infty$ contains all but finitely many positive integers. Define
\[
V_r(d)=\{G(x)=h+(d+1)^{-x}R(x):h\in\mathbb R,\ R\in\mathbb R[x],\
\deg R\le r\},\qquad G(\infty)=h.
\]
This is a real $(r+2)$-dimensional subspace of $C(\mathcal X)$.
The same expression extends $G$ to finite real $x\ge0$, in particular
to the extrapolation point $0$ outside $\mathcal X$.
The norm $\|G\|_{\mathcal X}
=\sup_{x\in\mathcal X}|G(x)|$ constrains only the positive integers
and $\infty$. Write $V_r=V_r(d)$ and define
\begin{equation}
B_r=B_r(d)=\sup\{G(0):G\in V_r,\ \|G\|_{\mathcal X}\le1\}.
\label{eq:supp-extrapolation-norm}
\end{equation}

\begin{lemma}
\label{lem:supp-interpolation}
Every nonzero $G\in V_r$ has at most $r+1$ distinct zeros in
$[0,\infty]$, counting a zero at $\infty$ once. Evaluation at any
$r+2$ distinct nodes in $[0,\infty]$ is invertible. Evaluation vectors
at any set of at most $r+2$ distinct nodes in this interval are linearly
independent.
\end{lemma}

\begin{proof}
Write $G(x)=h+(d+1)^{-x}R(x)$. If $h=0$, the nonzero polynomial $R$
has at most $r$ finite zeros, and $\infty$ contributes one more.
If $h\ne0$, then $G(\infty)\ne0$ and
\[
\frac{\mathrm d^{\,r+1}}{\mathrm d x^{r+1}}\bigl((d+1)^xG(x)\bigr)
=h[\log(d+1)]^{r+1}(d+1)^x\ne0.
\]
Were there $r+2$ distinct finite zeros, repeated Rolle's theorem would
give a zero of this derivative. Thus the evaluation map at $r+2$
nodes has trivial kernel and, by equality of dimensions, is invertible.
A smaller collection of evaluation vectors extends to such a matrix,
so it is linearly independent.
\end{proof}

\begin{proposition}
\label{prop:supp-envelope-equivalence}
The supremum $B_r(d)$ is finite and attained. The infimum over
$p\in\mathcal P_r$ in Eq.~\eqref{eq:critical-values} is also attained, and
\begin{equation}
\gamma_r(d)=\frac{2}{1+B_r(d)}>0,
\qquad B_0(d)=2d+1,\qquad \gamma_0(d)=\frac1{d+1}.
\label{eq:supp-envelope-equivalence}
\end{equation}
\end{proposition}

\begin{proof}
Choose $r+2$ distinct nodes $x_i$ in $\mathcal X$.
By Lemma~\ref{lem:supp-interpolation}, the inequalities $|G(x_i)|\le1$
bound the coefficient vector of $G$ in any fixed basis. The remaining
pointwise constraints are closed. The feasible coefficient set is
therefore compact, so the supremum is finite and attained. In degree zero,
$G(0)=(d+1)G(1)-dG(\infty)$ gives $B_0\le2d+1$.
The function $G(x)=2(d+1)^{1-x}-1$ is feasible and attains this bound.
By inclusion of the spaces, $B_r\ge B_0=2d+1>1$.

For $p\in\mathcal P_r$, write $\ell=\ell_d(p)$ and $u=u_d(p)$.
The width $u-\ell$ is positive: otherwise the limit at infinity would
force $p$ to vanish at every positive integer. The function
\[
G(x)=\frac{2(d+1)^{-x}p(x)-\ell-u}{u-\ell}
\]
belongs to $V_r$, is feasible, and satisfies
\[
G(0)=\frac{2(1-\ell)}{u-\ell}-1\le B_r.
\]
Hence every envelope objective is at least $2/(1+B_r)$.

Conversely, let $G=h+(d+1)^{-x}R(x)$ attain $B_r$.
Since $|h|\le1<B_r$, we have $R(0)=B_r-h>0$, and
$p=R/(B_r-h)$ belongs to $\mathcal P_r$. Its true envelope satisfies
\[
L:=\frac{-1-h}{B_r-h}\le\ell_d(p)\le u_d(p)
\le U:=\frac{1-h}{B_r-h}<1.
\]
For $t<1$, the function $(U-t)/(1-t)$ is decreasing. Consequently,
\[
\frac{u_d(p)-\ell_d(p)}{1-\ell_d(p)}
\le\frac{U-\ell_d(p)}{1-\ell_d(p)}
\le\frac{U-L}{1-L}=\frac{2}{1+B_r}.
\]
Thus this $p$ attains the infimum in Eq.~\eqref{eq:critical-values}.
Substituting $B_0=2d+1$ gives $\gamma_0(d)=1/(d+1)$.
\end{proof}

\subsection{Alternating interpolation and a finite dual}

\begin{lemma}
\label{lem:supp-interpolation-signs}
For ordered nodes $0<x_1<\cdots<x_{r+2}\le\infty$, there are unique
coefficients $\mu_i$ such that, for every $G\in V_r$,
\begin{equation}
G(0)=\sum_{i=1}^{r+2}\mu_iG(x_i),
\qquad (-1)^{i-1}\mu_i>0.
\label{eq:supp-interpolation-signs}
\end{equation}
\end{lemma}

\begin{proof}
Fix a real basis of $V_r$ and write evaluation vectors as rows.
The determinant at increasingly ordered finite nonnegative nodes is continuous
and nonzero on a connected domain, by
Lemma~\ref{lem:supp-interpolation}, so its sign is constant.
This sign persists when the last node tends to $\infty$: the evaluation
matrix converges and its limit is still invertible.
Let $M$ be the evaluation matrix at $x_1,\ldots,x_{r+2}$, and let
$\widetilde M_i$ be obtained by replacing row $i$ by evaluation at
zero and moving that row to the first position. Cramer's rule gives
\[
\mu_i=(-1)^{i-1}\frac{\det\widetilde M_i}{\det M}.
\]
Both determinants have the same nonzero sign, since their nodes are
in increasing order. This proves the assertion.
\end{proof}

\begin{lemma}
\label{lem:supp-alternating-degree}
Let $N\ge2$, $x_1<\cdots<x_N$ be real, $a_i>0$, and $|h|\le1$.
If a real polynomial $R$ satisfies
$R(x_i)=a_i[(-1)^{i-1}-h]$, then $\deg R\ge N-1$.
\end{lemma}

\begin{proof}
Set $D_i=\prod_{k\ne i}(x_i-x_k)$ and $\varepsilon_i=(-1)^{i-1}$.
If $\deg R\le N-2$, the coefficient of $x^{N-1}$ in its Lagrange
interpolation formula gives $\sum_iR(x_i)/D_i=0$.
Since $\operatorname{sgn}D_i=(-1)^{N-i}$,
\begin{equation}
0=(-1)^{N-1}\sum_{i=1}^N\frac{R(x_i)}{D_i}
=\sum_{i=1}^N\frac{a_i(1-\varepsilon_i h)}{|D_i|}>0.
\label{eq:supp-alternating-degree}
\end{equation}
Indeed, every summand is nonnegative, and the two signs of
$\varepsilon_i$ both occur, so at least one summand is positive.
The contradiction proves the degree bound.
\end{proof}

\begin{theorem}
\label{thm:supp-extremal-duality}
For every maximizer $G_r$ of the extrapolation problem, there are
positive integers $j_1<\cdots<j_{r+1}$ and nonzero real weights
$\mu_1,\ldots,\mu_{r+1},\mu_\infty$ such that, for every $G\in V_r$,
\begin{gather}
G(0)=\sum_{i=1}^{r+1}\mu_iG(j_i)+\mu_\infty G(\infty),
\label{eq:supp-dual-extrapolation}\\
\sum_{i=1}^{r+1}|\mu_i|+|\mu_\infty|=B_r.
\label{eq:supp-dual-variation}
\end{gather}
The signs and contact values are
\begin{equation}
\begin{aligned}
\operatorname{sgn}\mu_i&=G_r(j_i)=(-1)^{i-1}
&& (1\le i\le r+1),\\
\operatorname{sgn}\mu_\infty&=G_r(\infty)=(-1)^{r+1}.
\end{aligned}
\label{eq:supp-dual-alternation}
\end{equation}
\end{theorem}

\begin{proof}
Put $m=r+2$. In a fixed basis of $V_r$, let $v(x)\in\mathbb R^m$
be the evaluation vector, so that $G_c(x)=\langle c,v(x)\rangle$
for coefficient vector $c$. Write $e=v(0)$ and
\[
K=\operatorname{conv}\{v(x),-v(x):x\in\mathcal X\}.
\]
The signed evaluation vectors form a compact set, and Carath\'eodory's
theorem expresses every point of $K$ using at most $m+1$ of them.
Together with the simplex of weights, these vector choices form a
compact parameter space whose continuous summation map has image $K$,
so $K$ is compact~\cite{Rockafellar1970Convex}.

We claim that $e/B_r\in K$. Otherwise finite-dimensional strict
separation gives a linear functional $L$ and a number $t$ with
$L(z)<t<L(e/B_r)$ for all $z\in K$.
Since $0\in K$, we have $t>0$; since both $v(x)$ and $-v(x)$ belong
to $K$, the function $H(x)=L(v(x))/t$ lies in $V_r$ and has norm at
most one on $\mathcal X$. But $H(0)>B_r$, a contradiction.

Let $c_r$ represent $G_r$, and choose a finite convex representation
$e/B_r=\sum_i\alpha_i z_i$, where $\alpha_i>0$,
$\sum_i\alpha_i=1$, and $z_i=\varepsilon_i v(x_i)$ with
$\varepsilon_i\in\{-1,1\}$. Feasibility gives
$\langle c_r,z_i\rangle\le1$, while
$\langle c_r,e/B_r\rangle=1$. Thus every term is a contact:
\[
\langle c_r,z_i\rangle=1,\qquad G_r(x_i)=\varepsilon_i.
\]
Combine repeated nodes; a node cannot carry opposite contact signs.
If more than $m$ vectors remain, choose a nonzero linear dependence
$\sum_i\beta_i z_i=0$. Pairing with $c_r$ gives $\sum_i\beta_i=0$.
Thus some $\beta_i$ are positive. Replace $\alpha_i$ by
$\alpha_i-s\beta_i$, where
$s=\min_{\beta_i>0}\alpha_i/\beta_i$.
The new coefficients are nonnegative, still sum to one, preserve the
represented vector, and at least one vanishes. Discard zeros and repeat.
We obtain a representation on $k\le m$ distinct nodes.

Set $\mu_i=B_r\alpha_i\varepsilon_i$ for the surviving coefficients.
These weights are nonzero and give
\[
e=\sum_{i=1}^k\mu_i v(x_i),\qquad
\sum_{i=1}^k|\mu_i|=B_r,\qquad
\mu_iG_r(x_i)=|\mu_i|.
\]
If $k<m$, the first identity is a nontrivial dependence among the
$k+1\le r+2$ evaluation vectors at $0,x_1,\ldots,x_k$, contradicting
Lemma~\ref{lem:supp-interpolation}. Hence $k=r+2$.
Order the nodes. Lemma~\ref{lem:supp-interpolation-signs} now forces
$\operatorname{sgn}\mu_i=G_r(x_i)=(-1)^{i-1}$.

Write $G_r(x)=h_r+(d+1)^{-x}R_r(x)$. If all $r+2$ nodes were finite,
the contact equations would give
$R_r(x_i)=(d+1)^{x_i}[(-1)^{i-1}-h_r]$ with $|h_r|\le1$.
Lemma~\ref{lem:supp-alternating-degree} would force
$\deg R_r\ge r+1$, a contradiction. The last node is therefore
$\infty$, and its already nonzero weight has sign $(-1)^{r+1}$.
Relabeling the finite nodes as $j_1,\ldots,j_{r+1}$ proves the theorem.
\end{proof}

Let $\mu$ denote the finite signed atomic measure with these weights
at $j_1,\ldots,j_{r+1},\infty$. The binomial coefficient
$\binom{x}{k}$ is understood as a polynomial, with $\binom{x}{0}=1$.
Applying Eq.~\eqref{eq:supp-dual-extrapolation} to the constant function
and to $(d+1)^{-x}\binom{x}{k}$, whose value at $\infty$ is zero, gives
\begin{gather}
\sum_{i=1}^{r+1}\mu_i+\mu_\infty=1,
\label{eq:supp-dual-mass}\\
\sum_{i=1}^{r+1}\mu_i(d+1)^{-j_i}\binom{j_i}{k}
=\mathbf1_{\{k=0\}},\qquad 0\le k\le r.
\label{eq:supp-dual-moments}
\end{gather}
The total mass is one and the total variation is $B_r$. Solving for
the positive and negative masses, including the atom at $\infty$, yields
\begin{equation}
\mu^+(\mathcal X)=\frac{B_r+1}{2}=\frac1{\gamma_r(d)},
\qquad
\mu^-(\mathcal X)=\frac{B_r-1}{2}=\frac1{\gamma_r(d)}-1.
\label{eq:supp-dual-part-masses}
\end{equation}

\subsection{Optimizers and step points}

\begin{proposition}
\label{prop:supp-critical-sequence}
The extrapolation and envelope problems have unique optimizers $G_r$
and $p_r$, respectively. The polynomial $p_r$ has degree exactly $r$,
and
\[
0<\gamma_{r+1}(d)<\gamma_r(d)\le\gamma_0(d)=1/(d+1)<1.
\]
The envelope endpoints $\ell^{(r)}=\ell_d(p_r)$ and $u^{(r)}=u_d(p_r)$ satisfy
\begin{equation}
(\ell^{(r)},u^{(r)})=
\begin{cases}
(0,\gamma_r(d)),&r\ \text{even},\\[2pt]
\left(-\dfrac{\gamma_r(d)}{1-\gamma_r(d)},\,0\right),&r\ \text{odd}.
\end{cases}
\label{eq:supp-parity-envelopes}
\end{equation}
If $d>0$ is rational, $p_r$ has rational coefficients, and $B_r(d)$,
$\gamma_r(d)$, and the weights in the selected finite dual are rational.
\end{proposition}

\begin{proof}
Fix a maximizer $G_r$ and choose nodes and weights as in
Theorem~\ref{thm:supp-extremal-duality}. Any other maximizer $H$ satisfies
\[
B_r=H(0)=\sum_i\mu_iH(j_i)+\mu_\infty H(\infty)
\le\sum_i|\mu_i|+|\mu_\infty|=B_r.
\]
Every summand must saturate its bound. Thus $H$ and $G_r$ agree at
all $r+2$ support nodes, and Lemma~\ref{lem:supp-interpolation}
makes them identical.
Write $G_r(x)=h_r+(d+1)^{-x}R_r(x)$. The inverse construction in
Proposition~\ref{prop:supp-envelope-equivalence} gives an optimal
polynomial $R_r/(B_r-h_r)$. Conversely, any optimal normalized
polynomial $p$, with true envelope $\ell,u$, maps to $G_r$ by that
proposition. Comparing polynomial parts and evaluating at zero gives
\[
R_r(x)=\frac{2p(x)}{u-\ell},\qquad
B_r-h_r=R_r(0)=\frac{2}{u-\ell}.
\]
Consequently the normalized optimizer is unique and equals
\begin{equation}
p_r(x)=\frac{R_r(x)}{B_r-h_r}.
\label{eq:supp-optimal-polynomial}
\end{equation}

For $r\ge1$, apply Lemma~\ref{lem:supp-alternating-degree} to the
$r+1$ finite contacts, using
$R_r(j_i)=(d+1)^{j_i}[(-1)^{i-1}-h_r]$ and $|h_r|\le1$.
It gives $\deg R_r\ge r$, hence equality. For $r=0$, the same
conclusion follows from $R_r(0)=B_r-h_r>0$. Normalization preserves
the degree. Inclusion $V_r\subset V_{r+1}$ gives $B_r\le B_{r+1}$.
If equality held, $G_r$ would also maximize over $V_{r+1}$. Applying
the same degree argument at order $r+1$ would give $\deg R_r=r+1$,
contradicting $\deg R_r=r$. Thus $B_r<B_{r+1}$, proving strict decrease of
$\gamma_r(d)$ through Eq.~\eqref{eq:supp-envelope-equivalence}.

The final contact gives $h_r=(-1)^{r+1}$. Both endpoints $-1$ and $1$
occur among the selected contacts, including $\infty$. Since
$(d+1)^{-x}p_r(x)=(G_r(x)-h_r)/(B_r-h_r)$, the true envelope is
\[
\ell^{(r)}=\frac{-1-h_r}{B_r-h_r},\qquad
u^{(r)}=\frac{1-h_r}{B_r-h_r}.
\]
An endpoint occurring at $\infty$ is the limit along the positive
integers, so it still gives the required infimum or supremum.
Substitution yields Eq.~\eqref{eq:supp-parity-envelopes}.

For rational $d>0$, use the basis
$1,(d+1)^{-x}\binom{x}{0},\ldots,(d+1)^{-x}\binom{x}{r}$.
Evaluation at $j_1,\ldots,j_{r+1},\infty$ gives an invertible rational
matrix. With the alternating values in Eq.~\eqref{eq:supp-dual-alternation}
as the right-hand side, this system gives rational coefficients for
$G_r$ in the chosen basis. Thus $B_r=G_r(0)$ and
$\gamma_r(d)$ are rational, and $p_r$ has rational coefficients
by Eq.~\eqref{eq:supp-optimal-polynomial}.
The transposed system for the extrapolation weights has a rational
right-hand side, so the selected dual weights are rational as well.
\end{proof}

\begin{proposition}
\label{prop:supp-elementary-decay}
For $d>1$,
\begin{equation}
0<\gamma_r(d)\le d^{-r-1}.
\label{eq:supp-critical-limit}
\end{equation}
In particular, $\gamma_r(d)\to0$ as $r\to\infty$ in this range, and
$r_d(g)$ is finite for every fixed integer $d\ge2$ and $g>0$.
\end{proposition}

\begin{proof}
Consider
\begin{equation}
\widetilde p_r(w)=\sum_{k=0}^{r}(-1)^k\binom{w}{k}
=(-1)^r\binom{w-1}{r}.
\label{eq:supp-decay-polynomial}
\end{equation}
Pascal's rule makes the sum telescope. The first expression gives
$\widetilde p_r(0)=1$, so this polynomial belongs to $\mathcal P_r$.
On positive integers it vanishes for $1\le w\le r$ and has sign
$(-1)^r$ thereafter. Put
$M_r=\sup\{(d+1)^{-w}\binom{w-1}{r}: w\in\mathbb Z,\ w\ge r+1\}$.
Differentiating the geometric series $r$ times and multiplying by
$z^{r+1}/r!$ gives, for $0<z<1$,
\[
\sum_{w=r+1}^{\infty}\binom{w-1}{r}z^w
=\frac{z^{r+1}}{(1-z)^{r+1}}.
\]
At $z=1/(d+1)$, the nonnegative sum bounds each term, so
$M_r\le d^{-r-1}$. The true envelope of $\widetilde p_r$ has
endpoints $0,M_r$ for even $r$, and $-M_r,0$ for odd $r$.
The objective in Eq.~\eqref{eq:critical-values} for $\widetilde p_r$
is $M_r$ for even $r$ and $M_r/(1+M_r)$ for odd $r$. This proves the
bound. Since $d>1$, the bound tends to zero, and hence
$r_d(g)<\infty$ for every $g>0$.
\end{proof}

\section{Uniform score bounds and achievable budgets}
\label{sec:supp-scores}

We assign a score to the product-basis readout of each copy. For typical
targets, the polynomial envelopes from Sec.~\ref{sec:supp-duality} bound
the mean score in each of the two promise classes up to a finite-size error. When
these bounds separate the classes, we use the second moment to bound the
number of copies needed to distinguish them.
Throughout this section, the integers $d\ge2$, $r\ge0$, and the real
weights $\lambda_0=1,\lambda_1,\ldots,\lambda_r$ are fixed. Put
$D=d^n$. Constants may depend on these fixed parameters.

\subsection{Local readout and complement scores}

At each site, independently, choose a uniformly random Pauli basis when $d=2$
and a Haar-random orthonormal basis when $d\ge3$. These choices are repeated
independently on different copies. To define the reference distribution
$\mu_d$ of the local basis and outcome, choose the basis as above and
then choose one of its $d$ outcomes uniformly. The record $s$ of one copy
contains all the basis labels and outcomes. Its reference distribution is
$\mu=\mu_d^{\otimes n}$. Write $P_s=\bigotimes_i P_{s_i}$, where $P_{s_i}$
projects onto the recorded vector at site $i$. For an input $\rho$, the
record distribution and the POVM $E$ that generates it are given by
\begin{equation}
\Prob(ds\condsep\rho)=D\Tr(\rho P_s)\,\mu(ds),
\qquad E(ds)=DP_s\,\mu(ds),\qquad \int E(ds)=\id.
\label{eq:supp-record-law}
\end{equation}
In each local basis, the uniform average of the outcome projectors is
$\id/d$. For fixed bases, multiplying the uniform outcome probabilities
by $D\Tr(\rho P_s)$ gives the Born probabilities of Sec.~\ref{sec:supp-model}.

The recorded bases and outcomes specify the subspaces used to compute
the score. For $J\subseteq[n]$, let
\begin{equation}
Q_s(J)=\bigotimes_{i=1}^n
 \begin{cases}\id-P_{s_i},&i\in J,\\ P_{s_i},&i\notin J,\end{cases}
\qquad
H_s=\sum_{j=0}^r\lambda_j\sum_{|J|=j}Q_s(J).
\label{eq:supp-complement-kernel}
\end{equation}
The mutually orthogonal projectors $Q_s(J)$ have rank $(d-1)^{|J|}$.
The complement $\id-P_{s_i}$ is the sum of the projectors for the other
outcomes in the recorded basis. The score is computed classically from
the record and the known target. Orthogonality gives
\begin{equation}
\begin{gathered}
H_s^2=\sum_{j=0}^r\lambda_j^2\sum_{|J|=j}Q_s(J),\\
h_n:=\Tr H_s=\sum_{j=0}^r\lambda_j\binom nj(d-1)^j,\qquad
\nu_n:=\Tr H_s^2=\sum_{j=0}^r\lambda_j^2\binom nj(d-1)^j,
\end{gathered}
\label{eq:supp-kernel-traces}
\end{equation}
with $\binom nj=0$ for $j>n$. Under the reference distribution, averaging the
local factors gives $\int H_s\,\mu(ds)=(h_n/D)\id$. Thus $h_n$ is the
reference mean of $D\Tr(PH_s)$ for every target. Subtracting this mean
defines the score $f_\psi$. The operators $B_\psi$ and $M_\psi$ below
give its mean and full second moment for the actual readout.
\begin{equation}
f_\psi(s)=D\Tr(PH_s)-h_n,\qquad
B_\psi=\int E(ds)f_\psi(s),\qquad
M_\psi=\int E(ds)f_\psi(s)^2.
\label{eq:supp-score-operators}
\end{equation}
Thus $\Ebr{f_\psi\condsep\rho}=\Tr(\rho B_\psi)$ and
$\Ebr{f_\psi^2\condsep\rho}=\Tr(\rho M_\psi)$, with the expectation over
basis choices and Born outcomes. For the second-moment estimate, define the
auxiliary operator $S_\psi$ and the coefficient bound $\Lambda_n$ by
\begin{equation}
S_\psi=2\int E(ds)\bigl[D\Tr(PH_s^2)-\nu_n\bigr],\qquad
\Lambda_n=\sum_{j=0}^r|\lambda_j|\binom nj(d-1)^j=O(n^r).
\label{eq:supp-score-auxiliary}
\end{equation}
We bound $S_\psi$ by the same method used for $B_\psi$. To bound
$M_\psi$, we also need an estimate for the remainder in the expansion
of the second moment.

\subsection{Weyl expansions}

Let $\omega=e^{2\pi i/d}$, $X|t\rangle=|t+1\bmod d\rangle$, and
$Z|t\rangle=\omega^t|t\rangle$. The local Weyl operators
$W_{a,b}=X^aZ^b$, with labels in $\mathbb Z_d^2$, obey
\[
W_{a,b}W_{c,e}=\omega^{bc}W_{a+c,b+e},\qquad
W_{a,b}^\dagger=\omega^{ab}W_{-a,-b}.
\]
For tensor products $W$, let $|W|$ count the nonidentity factors. Summing
phases and shifts yields
\begin{equation}
\Tr(W^\dagger V)=D\,\mathbf 1_{W=V},\qquad
\sum_W WAW^\dagger=D\Tr(A)\id.
\label{eq:supp-weyl-identities}
\end{equation}
For the second identity, averaging over local powers of $Z$ removes the
off-diagonal matrix entries, and averaging over powers of $X$ makes all
diagonal entries equal. Tensoring the local averages gives the formula.

For a one-site operator $A$, take the expectation over a local record
$s$ drawn from $\mu_d$, with $P_s$ the projector onto its recorded
vector. The two local twirls are
\begin{equation}
\mathcal T(A):=d\Ebr[s]{P_s\Tr(P_sA)}
 =\frac{A+\Tr(A)\id}{d+1},\qquad
\mathcal T_f(A):=d\Ebr[s]{P_s\Tr((\id-P_s)A)}
 =\frac{d\Tr(A)\id-A}{d+1}.
\label{eq:supp-local-twirls}
\end{equation}
For Haar readout, unitary invariance gives
$\mathbb E_s P_s^{\otimes2}=(\id+F)/[d(d+1)]$, where $F$ swaps the two
factors. The invariant average is proportional to the symmetric projector
and has trace one. Contracting with $A$ proves the first identity. For Pauli readout,
write $A=\alpha\id+\sum_{k=1}^3a_k\sigma_k$; averaging the three basis
dephasings gives $\alpha\id+\sum_k a_k\sigma_k/3$. The complement identity
follows by subtracting the first from $\Tr(A)\id$.
Equation~\eqref{eq:supp-weyl-identities} rewrites these maps as
\[
d\mathcal T(A)=A+\frac1{d+1}\sum_{W\ne\id}WAW^\dagger,\qquad
d\mathcal T_f(A)=\frac{d}{d+1}\sum_{W\ne\id}WAW^\dagger.
\]

Define
\begin{equation}
\begin{gathered}
p(w)=\sum_{j=0}^r d^j\lambda_j\binom wj,\qquad
c_w=(d+1)^{-w}p(w),\\
e_w=2(d+1)^{-w}\sum_{j=0}^r d^j\lambda_j^2\binom wj,\qquad
e_*:=\sup_{w\in\mathbb Z_{\ge0}}e_w<\infty.
\end{gathered}
\label{eq:supp-radial-coefficients}
\end{equation}
The binomial polynomials form a basis, so every real $p\in\mathcal P_r$
determines unique real weights with $\lambda_0=1$. Put
$\ell=\ell_d(p)$ and $u=u_d(p)$. The identity has coefficient $c_0=1$.
At every positive integer weight $w$, $\ell\le c_w\le u$ by the envelope
definition. For all integer weights $w\ge0$, $0\le e_w\le e_*$.

Tensoring the two twirls gives the exact expansions
\begin{equation}
B_\psi=\sum_W c_{|W|}WPW^\dagger-h_n\id,\qquad
S_\psi=\sum_W e_{|W|}WPW^\dagger-2\nu_n\id.
\label{eq:supp-weyl-score-expansions}
\end{equation}
For a string $W$ of weight $w$, the term with no complement sites has
coefficient $(d+1)^{-w}$ in the expansion for $B_\psi$. Complement sites
must lie among the $w$ nonidentity sites. At each selected site the
coefficient changes from $1/(d+1)$ to $d/(d+1)$, giving the additional factor
$d^j\binom{w}{j}$ after summing over choices of $j$ sites.
The $D$ in the score supplies the $n$ factors $d\mathcal T$ or
$d\mathcal T_f$. There are $\binom nw(d^2-1)^w$ operators of weight $w$.
Using $(d^2-1)/(d+1)=d-1$ and the binomial identity
$\binom nw\binom wj=\binom nj\binom{n-j}{w-j}$ therefore gives
\begin{equation}
\sum_Wc_{|W|}=Dh_n,\qquad
\sum_We_{|W|}=2D\nu_n,\qquad
D^{-1}\sum_W|c_{|W|}|\le\Lambda_n.
\label{eq:supp-weight-sums}
\end{equation}
For random single-qubit Haar readout, Chen and Zhang relate the
Pauli-weight distribution of an observable to lower bounds on the
fluctuations of its unbiased estimators~\cite{ChenZhang2025Fluctuation}.
Here the weight dependence of $c_w$ and $e_w$ allows us to bound the
score moments by treating low and high weights separately.

\subsection{Uniform operator estimates}

For matrices, $\|\cdot\|$ and $\|\cdot\|_{\mathrm F}$ denote the operator
and Frobenius norms, respectively. We split the Weyl sums for $B_\psi$
and $S_\psi$ into low and high weights. The low-weight vectors $W\psi$
are nearly orthogonal for typical targets, and the high-weight sums
are close to scalar multiples of the identity. To bound $M_\psi$, we also
estimate the remainder in a Gaussian expansion. Normalizing the Gaussian
vector gives the following bounds on a common set of Haar targets.

\begin{theorem}
\label{thm:supp-uniform-score}
For fixed $d,r,\lambda$, there are sequences $\eta_n,\varepsilon_n=O(n^{-3})$,
an integer $n_0$, and a family $(\mathcal G_n)_n$ of measurable target sets,
all depending only on these fixed parameters. For every $n\ge n_0$,
$\Prob_{\mathrm{Haar}}(\mathcal G_n)\ge1-n^{-2}$ and every
$\psi\in\mathcal G_n$ satisfies
\begin{gather}
P+\ell(\id-P)-\eta_n\id\preceq B_\psi
 \preceq P+u(\id-P)+\eta_n\id,
\label{eq:supp-uniform-mean}\\
(\nu_n-\varepsilon_n)\id\preceq M_\psi
 \preceq(\nu_n+e_*+\varepsilon_n)\id.
\label{eq:supp-uniform-second}
\end{gather}
\end{theorem}

\begin{lemma}
\label{lem:supp-weyl-chaos}
Let $z\sim\mathcal N_{\mathbb C}(0,\id_D)$, with independent real and
imaginary parts of variance $1/2$, and let $a_W$ be real. Set
\[
\begin{gathered}
K_a(z)=D^{-1}\sum_W a_WWzz^\dagger W^\dagger,\qquad s_a=\sum_Wa_W,\\
|\Phi_W\rangle=(\id\otimes W)D^{-1/2}\sum_i|i,i\rangle,\qquad
C_a=\sum_Wa_W|\Phi_W\rangle\langle\Phi_W|.
\end{gathered}
\]
If $\Gamma$ transposes the first tensor factor, then a universal $C$ satisfies
\begin{equation}
\mathbb E\bigl\|K_a(z)-(s_a/D)\id\bigr\|
\le C\log(2D)\max\left\{
 \left(D^{-1}\sum_Wa_W^2\right)^{1/2},\ \max_W|a_W|,
 \ \|C_a^\Gamma\|\right\}.
\label{eq:supp-weyl-chaos}
\end{equation}
\end{lemma}

\begin{proof}
We first establish the matrix Gaussian estimate used twice in the
argument; a standard form is given in
Ref.~\cite[Theorem~4.1.1]{Tropp2015MatrixConcentration}.
For Hermitian $H_i$ of order $N$, $X=\sum_i\xi_iH_i$, and independent
standard real Gaussians $\xi_i$, integration by parts gives
\[
\mathbb E\Tr X^{2p}
 =\sum_i\sum_{l=0}^{2p-2}\mathbb E\Tr(H_iX^lH_iX^{2p-2-l}).
\]
Diagonalizing $X$ and applying weighted arithmetic--geometric means to its
absolute eigenvalues shows, for even $k$ and $0\le l\le k$, that
$|\Tr(HX^lHX^{k-l})|\le\Tr(H^2X^k)$. The two endpoint sums agree after
interchanging the eigenvalue indices. Hence
\[
\mathbb E\Tr X^{2p}\le
 (2p-1)\Bigl\|\sum_iH_i^2\Bigr\|\mathbb E\Tr X^{2p-2}
 \le N(2p-1)!!\Bigl\|\sum_iH_i^2\Bigr\|^p.
\]
Taking $p=\lceil\log(2N)\rceil$ and applying this to the Hermitian dilation
of rectangular matrices $A_i\in\mathbb C^{R\times C}$ yields
\begin{equation}
\mathbb E\Bigl\|\sum_i\xi_iA_i\Bigr\|
 \le C_0\sqrt{\log(2(R+C))}
 \max\left\{\Bigl\|\sum_i A_iA_i^\dagger\Bigr\|^{1/2},
             \Bigl\|\sum_i A_i^\dagger A_i\Bigr\|^{1/2}\right\}.
\label{eq:supp-rectangular-gaussian}
\end{equation}

Let $x,y\in\mathbb R^D$ be independent standard Gaussian vectors and
$A_{ij}\in\mathbb C^{D\times D}$. Conditional on $x$, put
$B_j=\sum_i x_iA_{ij}$. Applying Eq.~\eqref{eq:supp-rectangular-gaussian}
to $\sum_j y_jB_j$ bounds its conditional expected norm by
$C_0\sqrt{\log(4D)}$ times the larger of the norms of the block row
$[B_1\ \cdots\ B_D]$ and the block column with entries $B_1,\ldots,B_D$.
Bound the expectation of this maximum by the sum of the two expectations.
Both blocks are linear in $x$, so the same Gaussian estimate applies to each.
For $T_{ij\alpha\beta}=(A_{ij})_{\alpha\beta}$, the block row yields the
two flattenings $\alpha\mid(i,j,\beta)$ and $(i,\alpha)\mid(j,\beta)$,
where the bar separates row from column indices. The block column yields
$(j,\alpha)\mid(i,\beta)$ and $(i,j,\alpha)\mid\beta$.
The blocks have dimensions $D\times D^2$ and $D^2\times D$.
Thus $\mathbb E\|\sum_{ij}x_i y_jA_{ij}\|$ is bounded by
$C_1\log(2D)$ times the largest of the four flattening norms.

Now put $L_{ij}=D^{-1}\sum_Wa_WW|i\rangle\langle j|W^\dagger$.
Then $L_{ji}=L_{ij}^\dagger$ and $K_a(z)=\sum_{ij}z_i\bar z_jL_{ij}$.
For an independent copy $z'$, Jensen's inequality in $z'$ gives
$\mathbb E\|K_a(z)-\mathbb E K_a\|\le\mathbb E\|K_a(z)-K_a(z')\|$.
The independent complex Gaussians
$v=(z+z')/\sqrt2$ and $w=(z-z')/\sqrt2$ turn this difference into
$Y+Y^\dagger$, where $Y=\sum_{ij}v_i\bar w_jL_{ij}$.
Writing each complex Gaussian as $(x+iy)/\sqrt2$ splits $Y$ into four
real bilinear Gaussian sums, with only a universal constant loss.
The preceding estimate thus applies, including the diagonal terms $i=j$.

Weyl orthogonality gives
\[
\sum_{ij}L_{ij}L_{ij}^\dagger
 =\sum_{ij}L_{ij}^\dagger L_{ij}
 =\left(D^{-1}\sum_Wa_W^2\right)\id.
\]
For example, summing over $j$ in the first product produces
$\Tr(W^\dagger V)=D\mathbf 1_{W=V}$. The flattening
$(i,\alpha)\mid(j,\beta)$ is
$\sum_{ij}|i\rangle\langle j|\otimes L_{ij}=C_a$.
Exchanging $i$ and $j$ gives $(j,\alpha)\mid(i,\beta)$ and hence $C_a^\Gamma$.
The Bell vectors are orthonormal, so $\|C_a\|=\max_W|a_W|$.
Finally $\mathbb E K_a=(s_a/D)\id$, proving the claim.
\end{proof}

\begin{lemma}
\label{lem:supp-low-orbit}
Let $\mathcal W_L=\{W:|W|\le L\}$, $m=|\mathcal W_L|$, and let $V$ have
columns $W\psi$, with the identity column first. For Haar $\psi$ and $t>0$,
\begin{equation}
\Prob_{\mathrm{Haar}}\{\|V^\dagger V-\id_m\|>t\}
 \le\frac{m(m-1)}{(D+1)t^2}.
\label{eq:supp-orbit-gram}
\end{equation}
On $\|V^\dagger V-\id_m\|\le t<1$, the matrix
$Q=V(V^\dagger V)^{-1/2}$ is an isometry and $\|V-Q\|\le t$.
\end{lemma}
\begin{proof}
The Haar second moment gives
$\mathbb E|\langle\psi,A\psi\rangle|^2=
[\Tr(AA^\dagger)+|\Tr A|^2]/[D(D+1)]$.
For distinct Weyl operators $W,W'$, the unitary $W^\dagger W'$ has trace zero.
Thus every off-diagonal Gram entry has second moment $1/(D+1)$ and
$\mathbb E\|V^\dagger V-\id_m\|_{\mathrm F}^2=m(m-1)/(D+1)$.
Markov's inequality and $\|\cdot\|\le\|\cdot\|_{\mathrm F}$ give the probability
bound. For the last assertion, use $V=Q(V^\dagger V)^{1/2}$ and
$|\sqrt{x}-1|\le t$ on $[1-t,1+t]$.
\end{proof}

To control the full second moment, the next two lemmas bound its
fourth-order Gaussian remainder. At sites untouched by the complement
factors, the fourth trace moment has four record vertices:
normal-ordered Gaussian contractions give degree four at each vertex,
and the output trace adds the four-cycle. This yields the degree-six
graphs below.

\begin{lemma}
\label{lem:supp-local-contraction}
Let $G$ be a loopless multigraph on four vertices, with degree six at every
vertex, containing the cycle $12,23,34,41$. For each readout considered
below, draw four local records independently from its reference
distribution and let $s_1,\ldots,s_4$ be their recorded unit vectors. Set
\[
I_d(G)=\mathbb E\prod_{\{i,j\}\in E(G)}|\langle s_i,s_j\rangle|.
\]
Then $d^4I_d(G)\le\kappa_d<1$ for Haar readout when $d\ge3$, and for uniform
readout from any supplied complete set of $d+1$ mutually unbiased bases (MUBs)
when $d\ge2$. In particular, the latter includes the three Pauli bases.
\end{lemma}
\begin{proof}
For Haar $s$ and any fixed unit vector $b$,
\[
m_6:=\mathbb E|\langle s,b\rangle|^6
 =\frac{6}{d(d+1)(d+2)}.
\]
Indeed, the squared overlap $|\langle s,b\rangle|^2$ has the
$\mathrm{Beta}(1,d-1)$ distribution.
At a vertex with $k\le6$ remaining edges, H\"older's inequality, padded
with constant factors, bounds its conditional integral by $m_6^{k/6}$.
Eliminating vertices counts each of the twelve edges once. Consequently
$I_d(G)\le m_6^2$, and we may take
\begin{equation}
\kappa_d=\frac{36d^2}{(d+1)^2(d+2)^2}<1\qquad(d\ge3).
\label{eq:supp-haar-contraction}
\end{equation}

For the finite ensemble, fix a vector $b$ among the $N=d(d+1)$ labeled
basis vectors. It has unit overlap with itself and squared overlap $1/d$ with the
$d^2$ vectors in the other bases.
Thus $\mathbb E|\langle s,b\rangle|^6=(1+d^2d^{-3})/N=d^{-2}$.
The vertex bound alone therefore gives $d^4I_d(G)\le1$. To obtain a
strict inequality, fix $e,f$ from distinct bases and condition on
$s_2=s_3=e$, $s_4=f$, an event of probability $N^{-3}$.
Let $k$ be the multiplicity of the edge $\{1,4\}$.
The prescribed four-cycle and the degree-six condition give $1\le k\le5$.
On this event, let $J_k$ be the mean over $s_1$ of the product on the
six edges incident to vertex 1. Only $s_1=e$ and $s_1=f$ contribute
from their two bases. Each of the $d(d-1)$ vectors in the other bases
contributes $d^{-3}$, giving
\[
J_k=\frac{d^{-k/2}+d^{-(6-k)/2}+(d-1)/d^2}{d(d+1)}.
\]
Convexity and symmetry in $k$ give $J_k\le J_1=J_5=d^{-2}-\beta_d$, where
\begin{equation}
\beta_d=\frac{(d^2+1)(1-d^{-1/2})}{d^3(d+1)}>0.
\label{eq:supp-mub-strict-loss}
\end{equation}
Let $R$ be the overlap product on the six edges remaining after removing
vertex 1. The same vertex elimination gives $\mathbb E R\le d^{-2}$,
whereas $R\ge d^{-3}$ on the conditioned event. On all other configurations,
the first vertex integral is at most $d^{-2}$. Hence
\[
I_d(G)\le d^{-2}\mathbb E R-\beta_dd^{-3}N^{-3}
 \le d^{-4}-\beta_dd^{-3}N^{-3}.
\]
The required constant is therefore
\begin{equation}
\kappa_d=1-\frac{\beta_d}{d^2(d+1)^3}\in(0,1).
\label{eq:supp-mub-contraction}
\end{equation}
For $d=2$, the Pauli ensemble has exactly these unbiased overlaps.
\end{proof}

\begin{lemma}
\label{lem:supp-normal-remainder}
For $z\sim\mathcal N_{\mathbb C}(0,\id_D)$, define
\begin{gather}
\mathcal W_z(A,B)=(z^\dagger Az-\Tr A)(z^\dagger Bz-\Tr B)
 -z^\dagger(AB+BA)z+\Tr(AB),
\label{eq:supp-normal-order}\\
Z_z=\int E(ds)\mathcal W_z(H_s,H_s).
\label{eq:supp-normal-remainder}
\end{gather}
For the Pauli and Haar readouts specified above, $Z_z$ is Hermitian and, when $n\ge8r$,
\begin{equation}
\mathbb E\Tr Z_z^4\le8!\,C_d^{8r}\Lambda_n^8\kappa_d^{n-8r},
\qquad C_d=d^4(d-1)^8,
\label{eq:supp-remainder-fourth}
\end{equation}
with $\kappa_d$ from Lemma~\ref{lem:supp-local-contraction}. The same bound
holds for a supplied complete-MUB ensemble.
\end{lemma}
\begin{proof}
We use a finite normal-order identity. In coordinates, set
\[
\begin{split}
N_{ab;cd}={}&\bar z_a\bar z_bz_cz_d
 -\delta_{ac}\bar z_bz_d-\delta_{ad}\bar z_bz_c
 -\delta_{bc}\bar z_az_d-\delta_{bd}\bar z_az_c\\
 &+\delta_{ac}\delta_{bd}+\delta_{ad}\delta_{bc}.
\end{split}
\]
Contracting with $A_{ac}B_{bd}$ gives $\mathcal W_z(A,B)$: the four single
contractions produce the two trace terms and $z^\dagger(AB+BA)z$, while
the double contractions produce $\Tr A\Tr B+\Tr(AB)$.
For any polynomial $R(z,\bar z)$, Gaussian integration by parts gives
$\Ebr{\bar z_i R(z,\bar z)}=\Ebr{\partial_{z_i}R(z,\bar z)}$.
Iterating this identity yields
\[
\mathbb E\prod_{j=1}^m\bar z_{a_j}\prod_{j=1}^m z_{b_j}
 =\sum_{\pi\in S_m}\prod_j\delta_{a_j,b_{\pi(j)}}.
\]
In the product of four $N$'s, fix a complete matching of the eight conjugate
and eight ordinary positions. If $E_{\mathrm{int}}$ is its set of within-block edges, its
coefficient is $\sum_{S\subseteq E_{\mathrm{int}}}(-1)^{|S|}$.
It vanishes unless there are no internal edges. The matchings are between
labeled positions, so repeated coordinate indices are counted with their
multiplicities. Let
$\mathfrak D_4\subset S_8$ be the set of permutations that send each block
$\{2j-1,2j\}$, $j=1,\ldots,4$, entirely outside itself. With $V_\pi$ the
permutation of tensor factors, the finite identity is
\begin{equation}
\mathbb E\prod_{j=1}^4\mathcal W_z(A_{j,1},A_{j,2})
 =\sum_{\pi\in\mathfrak D_4}
  \Tr\!\left[\left(\bigotimes_{j=1}^4\bigotimes_{a=1}^2A_{j,a}\right)V_\pi\right].
\label{eq:supp-finite-wick}
\end{equation}
There are at most $8!$ terms.

For fixed $n$, the record projectors and $H_s$ are uniformly bounded,
and Gaussian polynomials have finite moments. We may therefore interchange
expectation and the record integrals, obtaining
\[
\mathbb E\Tr Z_z^4
 =D^4\int\Tr(P_{s_1}P_{s_2}P_{s_3}P_{s_4})
    \mathbb E\prod_{j=1}^4\mathcal W_z(H_{s_j},H_{s_j})
    \prod_{j=1}^4\mu(ds_j).
\]
Insert Eq.~\eqref{eq:supp-finite-wick} and expand the eight $H$ factors
using Eq.~\eqref{eq:supp-complement-kernel}.
The eight chosen subsets have union of size $t\le8r$. The sum of the
absolute coefficients is $(\sum_{j=0}^r|\lambda_j|\binom nj)^8\le\Lambda_n^8$.
For each choice of subsets and matching $\pi$, the integral factors over
sites. At a site outside the union, input slot $a$ carries the projector
from record $v(a)=\lceil a/2\rceil$, for $1\le a\le8$. With the site
index suppressed and $s_5=s_1$, the absolute local integrand is
\[
d^4\prod_{a=1}^8|\langle s_{v(a)},s_{v(\pi(a))}\rangle|
    \prod_{j=1}^4|\langle s_j,s_{j+1}\rangle|.
\]
Each record has two incoming and two outgoing matching edges, all to
other records. The output trace supplies the four-cycle in the second
product. Lemma~\ref{lem:supp-local-contraction} therefore bounds the local
integral in absolute value by $\kappa_d$.

At a changed site, write $Q_s^{(0)}=P_s$ and $Q_s^{(1)}=\id-P_s$.
The local average for the two input factors and the output factor in one
block is
\[
K_{ab}=d\Ebr[s]{Q_s^{(a)}\otimes Q_s^{(b)}\otimes P_s},
\qquad a,b\in\{0,1\}.
\]
It is positive and has trace $d(d-1)^{a+b}\le d(d-1)^2$.
The four-block tensor therefore has trace at most $C_d$; contraction
against the permutation unitary implementing the Wick and output edges
has absolute trace at most $C_d$. Each expanded term is bounded by
$C_d^t\kappa_d^{n-t}\le C_d^{8r}\kappa_d^{n-8r}$.
Summing the coefficient masses and the permutations proves the bound.
Finally $\mathcal W_z(H_s,H_s)$ is real, so $Z_z$ is Hermitian.
\end{proof}

\begin{lemma}
\label{lem:supp-common-radius}
Let $t=\|z\|^2/D$ for $z\sim\mathcal N_{\mathbb C}(0,\id_D)$ and $D=d^n$.
For fixed $d,r$ and sufficiently large $n$,
\begin{equation}
\Prob\{|t-1|>n^{-r-4}\}
 \le\exp[-(\log d)n^3/4].
\label{eq:supp-radial-event}
\end{equation}
\end{lemma}
\begin{proof}
Write $z=(x+iy)/\sqrt2$ with $2D$ independent standard real Gaussian
coordinates. Since the $|z_i|^2$ are independent with mean and variance one,
$\|t-1\|_{L^2}=D^{-1/2}$. As a polynomial of degree two in the real
coordinates, $t-1$ satisfies $\|t-1\|_{L^p}\le(p-1)D^{-1/2}$ for every
$p\ge2$ by Gaussian hypercontractivity~\cite[Proposition~3.2]{GamarnikEtAl2024Hypercontractivity}.
Markov's inequality with $p=n^2$ gives
\[
\Prob\{|t-1|>n^{-r-4}\}
 \le\left[(n^2-1)n^{r+4}d^{-n/2}\right]^{n^2}.
\]
The logarithm of the right-hand side is at most
$-\tfrac12(\log d)n^3+(r+6)n^2\log n$,
which implies the claim for sufficiently large $n$.
\end{proof}

\begin{proof}[Proof of Theorem~\ref{thm:supp-uniform-score}]
We first bound the Bell matrices for the radial sequences $c$ and $e$.
On a single pair of $d$-dimensional spaces, let $\Phi$ be the normalized
maximally entangled projector and put
$A_0=\Phi+(d+1)^{-1}(\id-\Phi)$ and $R_0=(d+1)^{-1}(\id-\Phi)$.
If $K_i(J)=R_0$ for $i\in J$ and $A_0$ otherwise, then
\[
C_c=\sum_{j=0}^r d^j\lambda_j\sum_{|J|=j}\bigotimes_iK_i(J),\qquad
C_e=2\sum_{j=0}^r d^j\lambda_j^2\sum_{|J|=j}\bigotimes_iK_i(J).
\]
On a Bell basis vector of weight $w$, only sets $J$ of nonidentity
sites contribute. Each such tensor product has eigenvalue $(d+1)^{-w}$,
giving the coefficients in Eq.~\eqref{eq:supp-radial-coefficients}
after summing over $J$.
Since $\Phi^\Gamma=F/d$,
$A_0^\Gamma=(\id+F)/(d+1)$ and $R_0^\Gamma=(\id-F/d)/(d+1)$ have norms $2/(d+1)$
and $1/d$, respectively. Therefore
\begin{equation}
\|C_c^\Gamma\|\le\sum_{j=0}^r|\lambda_j|\binom nj\left(\frac{2}{d+1}\right)^{n-j},
\qquad
\|C_e^\Gamma\|\le2\sum_{j=0}^r\lambda_j^2\binom nj\left(\frac{2}{d+1}\right)^{n-j}.
\label{eq:supp-choi-tails}
\end{equation}

Choose
\begin{equation}
L=\lceil(2r+24)\log_{d+1} n\rceil,\qquad
m_n=\sum_{w=0}^{\min\{L,n\}}\binom nw(d^2-1)^w
 =\exp[O((\log n)^2)].
\label{eq:supp-weight-cutoff}
\end{equation}
For either sequence $a=c,e$, let $a_*:=\sup_{w\in\mathbb Z_{\ge0}}|a_w|<\infty$ and
$\sigma_a=D^{-1}\sum_{|W|>L}a_{|W|}$.
Equation~\eqref{eq:supp-weight-sums} gives
$D^{-1}\sum_W|a_{|W|}|=O(n^r)$. On the weights present at size $n$ the
polynomial factors are $O(n^r)$, so
\[
\max_{L<w\le n}|a_w|=O(n^r(d+1)^{-L})=O(n^{-r-24}),\qquad
\left(D^{-1}\sum_{|W|>L}a_{|W|}^2\right)^{1/2}=O(n^{-12}).
\]
Removing the $m_n$ low-weight Bell terms changes the partial-transpose norm
by at most $a_*m_n/D$, since a normalized Bell projector has
partial-transpose norm $1/D$. Equations~\eqref{eq:supp-weyl-chaos} and
\eqref{eq:supp-choi-tails} now imply
\begin{equation}
\mathbb E\left\|D^{-1}\sum_{|W|>L}a_{|W|}Wzz^\dagger W^\dagger
             -\sigma_a\id\right\|=O(n^{-11}).
\label{eq:supp-high-weight-expectation}
\end{equation}
Let $E_{a,z}$ denote the centered high-weight operator inside the norm in
Eq.~\eqref{eq:supp-high-weight-expectation}. By Markov's inequality,
both centered operators have norm at most $n^{-3}$ except on an event
of probability $O(n^{-8})$.

Put $\psi=z/\|z\|$ and $t=\|z\|^2/D$. The direction $\psi$ is Haar by
unitary invariance. When these two norm bounds and the estimate
$|t-1|\le n^{-r-4}$ of Lemma~\ref{lem:supp-common-radius} hold, we have
\[
\sum_{|W|>L}a_{|W|}WPW^\dagger-\sigma_a\id
 =t^{-1}E_{a,z}+(t^{-1}-1)\sigma_a\id=O(n^{-3})
\]
in operator norm, since $|\sigma_a|=O(n^r)$.

Apply Lemma~\ref{lem:supp-low-orbit} with tolerance $n^{-3}$ to the
low-weight columns. By Eq.~\eqref{eq:supp-orbit-gram}, the probability
that $\|V^\dagger V-\id_{m_n}\|>n^{-3}$ is at most
$n^6m_n(m_n-1)/(D+1)$. When the Gram bound holds, any real diagonal
matrix $D_a$ with entries bounded in absolute value by $a_*$ satisfies
\[
\|VD_aV^\dagger-QD_aQ^\dagger\|
 \le a_*\|V-Q\|(\|V\|+\|Q\|)\le3a_*n^{-3}.
\]
Now take $D_a=\operatorname{diag}(a_{|W|}:W\in\mathcal W_L)$ in the
order of the columns of $V$, so that
$VD_aV^\dagger=\sum_{|W|\le L}a_{|W|}WPW^\dagger$.
Let $\widehat P$ project onto the first column of $Q$.
Then $\|\widehat P-P\|\le2n^{-3}$.
For $a=c$, the first diagonal entry is one and all others lie in
$[\ell,u]$. On the orthogonal complement of the range of $Q$,
$QD_cQ^\dagger$ vanishes, and $0$ also lies in this interval. Thus
\[
\widehat P+\ell(\id-\widehat P)
 \preceq QD_cQ^\dagger
 \preceq\widehat P+u(\id-\widehat P).
\]
Combining the low- and high-weight parts of
Eq.~\eqref{eq:supp-weyl-score-expansions} leaves the scalar correction
$-D^{-1}\sum_{|W|\le L}c_{|W|}\id$, whose norm is at most $c_*m_n/D$.
This proves Eq.~\eqref{eq:supp-uniform-mean} with
$\eta_n=O(n^{-3}+m_n/D)$.
For $a=e$, all diagonal entries lie in $[0,e_*]$, so the same argument gives
\begin{equation}
-\zeta_n\id\preceq S_\psi\preceq(e_*+\zeta_n)\id,
\qquad \zeta_n=O(n^{-3}+m_n/D).
\label{eq:supp-auxiliary-bound}
\end{equation}
In particular, $B_\psi$ and $S_\psi$ have uniformly bounded norms on this
event.

It remains to control the full second moment. For the same Gaussian
vector $z$, set
$b_s(z)=z^\dagger H_sz-h_n$, $M_z=\int E(ds)b_s(z)^2$, and
$S_z=2\int E(ds)(z^\dagger H_s^2z-\nu_n)$.
Equation~\eqref{eq:supp-normal-order} with $A=B=H_s$ gives
\begin{equation}
M_z=\nu_n\id+S_z+Z_z.
\label{eq:supp-gaussian-second}
\end{equation}
Since $Z_z$ is Hermitian, $\|Z_z\|^4\le\Tr Z_z^4$.
Lemma~\ref{lem:supp-normal-remainder} and Markov's inequality give
\[
\Prob\{\|Z_z\|>n^{-3}\}
 \le n^{12}8!C_d^{8r}\Lambda_n^8\kappa_d^{n-8r}
 =e^{-\Omega(n)}.
\]
Write $\alpha=t^{-1}$. Direct substitution yields
$f_\psi=\alpha b_s+(\alpha-1)h_n$ and
$S_z=tS_\psi+2\nu_n(t-1)\id$, and hence
\begin{equation}
\begin{split}
M_\psi={}&\nu_n\id+\alpha S_\psi+\alpha^2Z_z
 +2(\alpha-1)h_nB_\psi
 -(\alpha-1)^2(\nu_n+h_n^2)\id.
\end{split}
\label{eq:supp-haar-second}
\end{equation}
Here $h_n,\nu_n=O(n^r)$ and $|\alpha-1|=O(n^{-r-4})$ on the radial event.
The last two terms are respectively $O(n^{-4})$ and $O(n^{-8})$ in norm
because $\|B_\psi\|=O(1)$. Also $\|(\alpha-1)S_\psi\|=O(n^{-r-4})$
by Eq.~\eqref{eq:supp-auxiliary-bound}. With $\|Z_z\|\le n^{-3}$ this
gives $\|M_\psi-\nu_n\id-S_\psi\|=O(n^{-3})$. Applying both sides of
Eq.~\eqref{eq:supp-auxiliary-bound} proves Eq.~\eqref{eq:supp-uniform-second} with
$\varepsilon_n=O(n^{-3}+m_n/D)$.
Since $m_n=\exp[O((\log n)^2)]$ and $D=d^n$, both error sequences
are $O(n^{-3})$.

By a union bound, the probability that at least one of the high-weight,
Gram, radial, or remainder estimates fails is at most
\begin{equation}
O(n^{-8})+\frac{n^6m_n(m_n-1)}{D+1}
 +e^{-(\log d)n^3/4}
 +n^{12}8!C_d^{8r}\Lambda_n^8\kappa_d^{n-8r}
 \le n^{-2}
\label{eq:supp-score-event-failure}
\end{equation}
for sufficiently large $n$. Define $\mathcal G_n$ by the two deterministic
operator inequalities in the theorem, with the error sequences just
obtained. These conditions are measurable because $B_\psi$ and $M_\psi$
depend continuously on $P$. Whenever all the estimates hold, the
direction $\psi=z/\|z\|$ belongs to $\mathcal G_n$. Since this direction
is Haar distributed, Eq.~\eqref{eq:supp-score-event-failure} gives
$\Prob_{\mathrm{Haar}}(\mathcal G_n)\ge1-n^{-2}$.
\end{proof}

\subsection{A fixed-budget test}

Before accounting for finite-size errors, the lower envelope bound on the
mean score at fidelity $a$ minus the upper envelope bound at fidelity $b$ is
\begin{equation}
\Delta=\ell+(1-\ell)a-u-(1-u)b
=(1-b)(1-\ell)\left[g-\frac{u-\ell}{1-\ell}\right].
\label{eq:supp-envelope-gap}
\end{equation}

\begin{proposition}
\label{prop:supp-achievable-budget}
Suppose $\Delta>0$. There exists an integer $n_1\ge n_0$, depending only
on $d,r,\lambda,a,b$, such that for every $n\ge n_1$,
every $\psi\in\mathcal G_n$ of Theorem~\ref{thm:supp-uniform-score}, and
every $0<\delta<1/2$, there is a valid protocol whose copy count on every
run is bounded by
\begin{equation}
1548\,V_n\Delta^{-2}\log(1/\delta),\qquad
V_n=\nu_n+e_*+1.
\label{eq:supp-achievable-budget}
\end{equation}
\end{proposition}
\begin{proof}
The bounds $\ell\le0\le u$ give
\[
0<\Delta=(a-b)+\ell(1-a)-u(1-b)\le a-b\le1.
\]
Also $\Delta=(1-u)(1-b)-(1-\ell)(1-a)\le(1-u)(1-b)$, so $u<1$.
Tracing Eq.~\eqref{eq:supp-uniform-mean} against an arbitrary density
matrix gives
\begin{equation}
\ell+(1-\ell)F(\rho)-\eta_n
 \le\Ebr{f_\psi\condsep\rho}
 \le u+(1-u)F(\rho)+\eta_n.
\label{eq:supp-state-mean}
\end{equation}
The test uses the two operator bounds at sizes where $\eta_n\le\Delta/4$
and $\varepsilon_n\le1$. For the given thresholds $a,b$, choose
$n_1\ge n_0$ so that both conditions hold for all $n\ge n_1$.
The block decision threshold
\[
t_*=\frac{\ell+(1-\ell)a+u+(1-u)b}{2}
\]
then lies at least $\Delta/4$ below the mean for every $\rho\in\mathcal S_0$
and at least $\Delta/4$ above the mean for every $\rho\in\mathcal S_1$.
Equation~\eqref{eq:supp-uniform-second}
also gives, for all input states on the same target event,
\begin{equation}
\operatorname{Var}(f_\psi\condsep\rho)
 \le\Ebr{f_\psi^2\condsep\rho}\le V_n.
\label{eq:supp-score-variance}
\end{equation}

Generate records from distinct iid copies with independent basis choices,
and partition them into disjoint blocks of
$m=\lceil128V_n/\Delta^2\rceil$ records. Accept a block if its average score
is at least $t_*$. Chebyshev's
inequality bounds each of the two error probabilities by
$16V_n/(m\Delta^2)\le1/8$. Use the smallest odd integer
$k\ge8\log(1/\delta)$ and accept if a majority of the $k$ block decisions
accept. Let $J$ denote the final majority decision, with $J=0$ for
acceptance as in Sec.~\ref{sec:supp-model}. For $i\in\{0,1\}$ and
$\rho\in\mathcal S_i$, Hoeffding's inequality~\cite{Hoeffding1963} gives
\[
\Prob(J\ne i\condsep\rho)
 \le\exp[-2k(1/2-1/8)^2]
 =\exp(-9k/32)\le\delta.
\]

Since $V_n\ge1$ and $0<\Delta\le1$, and since
$\log(1/\delta)\ge\log2>1/2$, the integer choices satisfy
\[
m\le129V_n\Delta^{-2},\qquad
k<8\log(1/\delta)+2\le12\log(1/\delta).
\]
Every run uses exactly $mk$ copies, which proves
Eq.~\eqref{eq:supp-achievable-budget}.
\end{proof}

Apply this construction to the degree-$r$ optimizer of
Proposition~\ref{prop:supp-critical-sequence}. Its envelopes have
$\ell\le0$ and $u<1$, and
\begin{equation}
\Delta=(1-b)(1-\ell)\bigl[g-\gamma_r(d)\bigr].
\label{eq:supp-optimal-score-gap}
\end{equation}
Thus $\Delta>0$ for every fixed threshold pair with $g>\gamma_r(d)$.
For this optimizer, the binomial expansion in
Eq.~\eqref{eq:supp-radial-coefficients} determines the score weights
from $d$ and $r$. Since $\nu_n=O(n^r)$ and $e_*<\infty$,
Proposition~\ref{prop:supp-achievable-budget} gives, for every
$n\ge n_1$ and every $\psi\in\mathcal G_n$,
\begin{equation}
\Nmax\le C_{d,a,b,r}\,n^r\log(1/\delta)
\qquad(0<\delta<1/2).
\label{eq:supp-polynomial-upper}
\end{equation}
Here $\Prob_{\mathrm{Haar}}(\mathcal G_n)\ge1-n^{-2}$, and both
$C_{d,a,b,r}$ and $n_1$ are independent of $\delta$.
At fixed $d$, all these tests use the same distribution of local measurement
bases. The score is computed from the target, the record, and the chosen
weights. The fidelity thresholds enter both the block decision threshold
$t_*$ and the prescribed budget.

\subsection{Finite readout from complete MUBs}

\begin{proposition}
\label{prop:supp-mub-implementation}
Fix a complete set of $d+1$ mutually unbiased orthonormal bases in
$\mathbb C^d$. If the basis at each site on each copy is chosen
independently and uniformly from this set, all the score bounds and
achievable budgets above hold for this readout.
\end{proposition}
\begin{proof}
For a basis $\mathcal B$, let
$\mathcal D_{\mathcal B}(A)=\sum_{b\in\mathcal B}P_b\Tr(P_bA)$.
This map is the Hilbert--Schmidt orthogonal projection onto the matrices
diagonal in $\mathcal B$. Its range has a traceless subspace of
dimension $d-1$.
For distinct unbiased bases,
\[
\Tr[(P_b-\id/d)(P_c-\id/d)]
 =|\langle b,c\rangle|^2-1/d=0.
\]
The $d+1$ traceless diagonal subspaces are thus pairwise orthogonal.
Their dimensions sum to $d^2-1$, so they span the traceless matrix space.
Writing $A=(\Tr A/d)\id+A_0$ gives
\begin{equation}
\frac1{d+1}\sum_{\mathcal B}\mathcal D_{\mathcal B}(A)
 =\frac{\Tr A}{d}\id+\frac{A_0}{d+1}
 =\frac{A+\Tr(A)\id}{d+1}.
\label{eq:supp-mub-twirl}
\end{equation}
The left side equals $d\Ebr[s]{P_s\Tr(P_sA)}$ for the finite reference
ensemble. This proves the first twirl in Eq.~\eqref{eq:supp-local-twirls};
the complement twirl follows by subtraction. This second-moment identity
also expresses the projective 2-design property of the basis
vectors~\cite{KlappeneckerRoetteler2005Designs}.

The two twirls give the same Weyl expansions for $B_\psi$ and $S_\psi$.
For the full second-moment operator $M_\psi$, the finite-ensemble bound of
Lemma~\ref{lem:supp-local-contraction} supplies the strict contraction in
Lemma~\ref{lem:supp-normal-remainder}. The proof of
Theorem~\ref{thm:supp-uniform-score} then gives a Haar-typical target
family for this choice of bases, on which
Proposition~\ref{prop:supp-achievable-budget} gives the stated budget.
\end{proof}

\section{Hard states and uniform local distinguishability}
\label{sec:supp-hard-states}

For the lower bound, we construct a pair of states for each target in a
set of high Haar probability. We bound the difference between their outcome
distributions under every product measurement on a single copy, then
calibrate their fidelities while preserving the order of this bound.
Fix an integer $d\ge2$, put
$D=d^n$, and retain the Weyl operators and matrix-norm
conventions of Sec.~\ref{sec:supp-scores}.

\subsection{Reference distributions and complement kernels}

For a local POVM $M_i$, we use the outcome distribution of the maximally
mixed local state as the reference distribution, $\nu_i(A)=\Tr M_i(A)/d$.
Finite-dimensional Radon--Nikodym differentiation
gives a measurable positive matrix density
$M_i(\mathrm ds_i)=K_i(s_i)\nu_i(\mathrm ds_i)$, with
$\Tr K_i=d$ almost everywhere. Choose a measurable spectral decomposition
$K_i=\sum_{h=1}^d\lambda_{i,h}P_{i,h}$ into rank-one projectors. Such a
choice is obtained by ordering the eigenvalues, projecting the standard
basis vectors onto each eigenspace, and applying Gram--Schmidt to the
projected vectors in a fixed order, skipping zero residuals. The refined effects
and reference distribution are
\[
\widetilde M_i(\mathrm ds_i,h)
 =dP_{i,h}(s_i)\widetilde\nu_i(\mathrm ds_i,h),\qquad
\widetilde\nu_i(\mathrm ds_i,h)
 =\frac{\lambda_{i,h}(s_i)}d\nu_i(\mathrm ds_i).
\]
Forgetting $h$ recovers both $M_i$ and $\nu_i$. Tensoring these refinements
therefore gives a refined product POVM $\widetilde M$ and a map $\pi$
that forgets the spectral indices, with
$\pi_*\widetilde\nu=\nu_M:=q_M^{\id/D}=\bigotimes_i\nu_i$.

For fixed $M$, the squared $L^2(\nu_M)$ distance between the outcome
densities of two states $\omega,\omega'$ is
\begin{equation}
\|q_M^\omega-q_M^{\omega'}\|_{\nu_M^{-1}}^2
 :=\int\left|
   \frac{\mathrm d(q_M^\omega-q_M^{\omega'})}{\mathrm d\nu_M}
   \right|^2\mathrm d\nu_M.
\label{eq:supp-reference-quadratic}
\end{equation}
If
$f=\mathrm d(q_{\widetilde M}^\omega-q_{\widetilde M}^{\omega'})/
\mathrm d\widetilde\nu$, then $\widetilde\nu$-almost everywhere
\[
\frac{\mathrm d(q_M^\omega-q_M^{\omega'})}{\mathrm d\nu_M}\circ\pi
 =\Ebr[\widetilde\nu]{f\condsep\pi}.
\]
Indeed, for every bounded measurable function $\phi$ of the coarse outcome,
$\int(\phi\circ\pi)f\,\mathrm d\widetilde\nu
=\int\phi\,\mathrm d(q_M^\omega-q_M^{\omega'})$.
Conditional Jensen applied to the square gives the first inequality below.
The same conditional expectation identity holds for each Born density
separately. Applying conditional Jensen to these two densities with the
convex function $(u,v)\mapsto u\log(u/v)$ gives the second inequality.
\begin{equation}
\begin{gathered}
\|q_M^\omega-q_M^{\omega'}\|_{\nu_M^{-1}}^2
 \le\|q_{\widetilde M}^\omega-q_{\widetilde M}^{\omega'}
        \|_{\widetilde\nu^{-1}}^2,\\
D_{\mathrm{KL}}(q_M^\omega\|q_M^{\omega'})
 \le D_{\mathrm{KL}}(q_{\widetilde M}^\omega
                       \|q_{\widetilde M}^{\omega'}).
\end{gathered}
\label{eq:supp-povm-refinement-contraction}
\end{equation}
Thus it suffices to establish both bounds for rank-one local effects.

For a rank-one product POVM, suppress the refinement notation, put
$P_s=\bigotimes_iP_{i,s_i}$, and write
\begin{equation}
M_i(\mathrm ds_i)=dP_{i,s_i}\nu_i(\mathrm ds_i),\qquad
\int P_{i,s_i}\,\mathrm d\nu_i=\frac{\id}{d}.
\label{eq:supp-rank-one-reference}
\end{equation}
The Born density of any state $\omega$ is
\begin{equation}
\frac{\mathrm dq_M^\omega}{\mathrm d\nu_M}(s)=D\Tr(\omega P_s).
\label{eq:supp-rank-one-density}
\end{equation}
It integrates to one, lies between zero and $D$, and equals one for
$\omega=\id/D$.

Define the self-adjoint map
\[
\mathcal C(A)=\frac{\Tr(A)\id-A}{d-1}.
\]
Its eigenvalues on the scalar and traceless subspaces are $1$ and
$-1/(d-1)$, so it is a Hilbert--Schmidt contraction. Let $\mathcal C_J$
apply $\mathcal C$ at the sites in $J$ and the identity map elsewhere,
and set $\mathcal C_{n,k}=\binom nk^{-1}\sum_{|J|=k}\mathcal C_J$;
in particular $\mathcal C_{n,0}(A)=A$.
For a fixed finite set $\mathcal K\subset\mathbb N_0$ and fixed real
coefficients with $\sum_{k\in\mathcal K}\theta_k=1$, take
$n\ge\max\mathcal K$ and put
\begin{equation}
\begin{gathered}
\mathcal R_n=\sum_{k\in\mathcal K}\theta_k\mathcal C_{n,k},\qquad
B_s=\mathcal R_n(P_s),\\
L_R=\sum_{k\in\mathcal K}|\theta_k|,\qquad
v_n=\sum_{k\in\mathcal K}\frac{\theta_k^2}{\binom nk(d-1)^k}.
\end{gathered}
\label{eq:supp-shell-variance-scale}
\end{equation}
Here $B_s$ is an auxiliary Hermitian operator. The map
$\mathcal R_n$ is self-adjoint, preserves the trace and identity,
and satisfies $\|\mathcal R_n\|_{\mathrm{HS}\to\mathrm{HS}}\le L_R$.
If $s_0=\min\{k:\theta_k\ne0\}$, then
$v_n=\Theta_{d,\mathcal K,\theta}(n^{-s_0})$.

On the real Hilbert space of Hermitian matrices, with
$(A|B)=\Tr(AB)$, define
\[
S_i=\int|P_{i,s_i})(P_{i,s_i}|\,\mathrm d\nu_i,\qquad
S_M=\bigotimes_i S_i,\qquad
T_M=\int|B_s)(B_s|\,\mathrm d\nu_M.
\]
\begin{lemma}
\label{lem:supp-kernel-frames}
For every rank-one product POVM,
\begin{equation}
\begin{gathered}
\Tr B_s=1,\qquad \Tr B_s^2=v_n,\\
\int B_s\,\mathrm d\nu_M=\frac{\id}{D},\qquad
\int B_s^2\,\mathrm d\nu_M=\frac{v_n\id}{D}.
\end{gathered}
\label{eq:supp-signed-kernel-moments}
\end{equation}
The frame operators satisfy
\begin{equation}
\begin{gathered}
0\preceq S_i\preceq\frac{\id_{\mathrm{HS}}}{d},\qquad
\|S_M\|\le D^{-1},\\
T_M=\mathcal R_n S_M\mathcal R_n,\qquad
\Tr T_M=v_n,\qquad \|T_M\|\le\frac{L_R^2}{D}.
\end{gathered}
\label{eq:supp-local-frame-bounds}
\end{equation}
\end{lemma}
\begin{proof}
For a product rank-one projector, $\mathcal C_J(P_s)$ replaces each
factor on $J$ by $(\id-P_{i,s_i})/(d-1)$. The operators
$\mathcal C_J(P_s)$ for distinct $J$ have orthogonal supports.
Each has trace one and squared
Hilbert--Schmidt norm $(d-1)^{-|J|}$. Expanding $B_s$ gives its trace
and squared Hilbert--Schmidt norm in Eq.~\eqref{eq:supp-signed-kernel-moments}.
The average of either an unchanged or a complementary factor is $\id/d$.
The square of a complementary factor has average $\id/[d(d-1)]$.
Only equal-subset terms survive in $B_s^2$, proving the two mean identities.
For Hermitian $A$, rank-one Cauchy--Schwarz gives
\[
(A|S_i A)=\int[\Tr(AP_{i,s_i})]^2\,\mathrm d\nu_i
 \le\int\Tr(A^2P_{i,s_i})\,\mathrm d\nu_i=\frac{\Tr A^2}{d}.
\]
Tensorization proves the bound on $S_M$. Self-adjointness of
$\mathcal R_n$ gives the formula for $T_M$, whose trace is
$\int\Tr B_s^2\,\mathrm d\nu_M=v_n$ and whose norm has the stated bound.
\end{proof}

\subsection{Physical states from the finite dual}

For $0\le k\le n$, define the Weyl shell states by
\begin{equation}
\tau_k=\frac1{\binom nk(d^2-1)^k}\sum_{|W|=k}WPW^\dagger,
\qquad \tau_0=P,\qquad \tau_\infty=\frac{\id}{D}.
\label{eq:supp-weyl-shell-state}
\end{equation}
Each is a normalized convex combination of density matrices.

\begin{lemma}
\label{lem:supp-shell-readout}
For every product rank-one projector $P_s$ and $0\le k\le n$,
\begin{equation}
\Tr(\tau_kP_s)=(d+1)^{-k}\sum_{h=0}^k\binom kh d^h
                  \Tr\bigl[P\mathcal C_{n,h}(P_s)\bigr].
\label{eq:supp-shell-readout}
\end{equation}
\end{lemma}
\begin{proof}
For a one-site rank-one projector $B$, Weyl completeness
\eqref{eq:supp-weyl-identities} gives
\[
\frac1{d^2-1}\sum_{W\ne\id}W^\dagger BW
 =\frac{d\id-B}{d^2-1}=\frac{B+d\mathcal C(B)}{d+1}.
\]
On a support $J$ of size $k$, tensor this identity and expand over
$S\subseteq J$. A fixed $S$ of size $h$ appears in
$\binom{n-h}{k-h}$ supports. Averaging over $J$ gives coefficient
$(d+1)^{-k}d^h\binom{n-h}{k-h}\binom nh/\binom nk
=(d+1)^{-k}d^h\binom kh$. Taking the trace against $P$ proves the claim.
\end{proof}

Fix an auxiliary degree $j\ge0$. Choose a finite dual representation from
Theorem~\ref{thm:supp-extremal-duality}, with nodes
$k_1<\cdots<k_{j+1}$ and weights
$\mu_1,\ldots,\mu_{j+1},\mu_\infty$, and hold this choice fixed as $n$
grows. Take $n\ge k_*:=k_{j+1}$. Equation~\eqref{eq:supp-dual-part-masses}
gives positive mass $Z=1/\gamma_j(d)$ and negative mass $Z-1$,
including the atom at infinity. With $t_+=\max\{t,0\}$, set
\begin{equation}
\begin{aligned}
\sigma_H&=\frac1Z\left(P+\sum_{\mu_i<0}|\mu_i|\tau_{k_i}
                     +(-\mu_\infty)_+\tau_\infty\right),\\
\sigma_L&=\frac1Z\left(\sum_{\mu_i>0}\mu_i\tau_{k_i}
                     +(\mu_\infty)_+\tau_\infty\right).
\end{aligned}
\label{eq:supp-dual-physical-states}
\end{equation}
Both mixtures have nonnegative coefficients and trace one, and
$F(\sigma_H)\ge1/Z=\gamma_j(d)$. Exactly one has an explicit maximally
mixed component of weight
\begin{equation}
\kappa=\frac{|\mu_\infty|}{Z}>0.
\label{eq:supp-dual-background}
\end{equation}
Since $\operatorname{sgn}\mu_\infty=(-1)^{j+1}$ by
Eq.~\eqref{eq:supp-dual-alternation}, this component is in $\sigma_H$
for even $j$ and in $\sigma_L$ for odd $j$.

For $j+1\le k\le k_*$, define
\begin{equation}
a_k=-d^k\sum_i\mu_i(d+1)^{-k_i}\binom{k_i}{k},\qquad
\theta_k=\frac{a_k}{\mu_\infty},\qquad
\mathcal R_n=\sum_{k=j+1}^{k_*}\theta_k\mathcal C_{n,k}.
\label{eq:supp-residual-coefficients}
\end{equation}
Fix a rank-one product POVM $M$. Take the trace of $\sigma_H-\sigma_L$
in Eq.~\eqref{eq:supp-dual-physical-states} against $P_s$ and expand
each finite shell using Eq.~\eqref{eq:supp-shell-readout}. After factoring
out $1/Z$, the coefficient at radius zero is $1-\sum_i\mu_i(d+1)^{-k_i}=0$.
The coefficients at radii $1,\ldots,j$ also vanish by
Eq.~\eqref{eq:supp-dual-moments}. The same identities and the binomial
theorem give
\[
\sum_{k=j+1}^{k_*}a_k
 =1-\sum_i\mu_i(d+1)^{-k_i}(d+1)^{k_i}
 =1-\sum_i\mu_i=\mu_\infty,
\]
where the final equality is Eq.~\eqref{eq:supp-dual-mass}.
The maximally mixed term contributes $-\mu_\infty/(ZD)$ to the trace.
With $B_s=\mathcal R_n(P_s)$, multiplying the trace by $D$ as in
Eq.~\eqref{eq:supp-rank-one-density} gives, for $\nu_M$-almost every $s$,
\begin{equation}
\frac{\mathrm d(q_M^{\sigma_H}-q_M^{\sigma_L})}{\mathrm d\nu_M}(s)
 =\frac{\mu_\infty}{Z}\bigl[D\Tr(PB_s)-1\bigr].
\label{eq:supp-dual-readout-difference}
\end{equation}
The constant $-1$ comes from the maximally mixed term. Since
$\sum_k\theta_k=1$, the bracket is a weighted sum of centered terms
$D\Tr[P\mathcal C_{n,k}(P_s)]-1$ with $j+1\le k\le k_*$.

The first residual coefficient is nonzero. Let $L_i(x)$ be the
Lagrange basis on $k_1,\ldots,k_{j+1}$. By the moment identities,
the weights $\mu_i(d+1)^{-k_i}$ represent evaluation at zero on polynomials
of degree at most $j$. The Lagrange weights represent the same functional
on these nodes, so $\mu_i(d+1)^{-k_i}=L_i(0)$. For $f(x)=\binom{x}{j+1}$,
\[
f(x)-\sum_iL_i(x)f(k_i)
 =\frac1{(j+1)!}\prod_{i=1}^{j+1}(x-k_i).
\]
Evaluation at zero gives
\begin{equation}
a_{j+1}=(-1)^{j+1}\frac{d^{j+1}}{(j+1)!}
             \prod_{i=1}^{j+1}k_i\ne0.
\label{eq:supp-first-residual}
\end{equation}
It follows that $\theta_{j+1}>0$ and
\begin{equation}
v_n=\sum_{k=j+1}^{k_*}\frac{\theta_k^2}{\binom nk(d-1)^k}
 =\Theta_{d,j}(n^{-(j+1)}).
\label{eq:supp-dual-kernel-scale}
\end{equation}

\subsection{Uniform quadratic concentration}

Fix any kernel in Eq.~\eqref{eq:supp-shell-variance-scale}. We now bound
the mean square of $D\Tr(PB_s)-1$ under $\nu_M$ uniformly over rank-one
product POVMs $M$. We will also use the special case $\mathcal K=\{0\}$.
For $z\sim\mathcal N_{\mathbb C}(0,\id_D)$, put
\[
X_s=z^\dagger B_sz-1,\qquad
Q_M(z)=\int X_s^2\,\mathrm d\nu_M.
\]
For Hermitian $A,B$, Wick's formula gives
\begin{equation}
\operatorname{Cov}\bigl((z^\dagger Az-\Tr A)^2,
                        (z^\dagger Bz-\Tr B)^2\bigr)
 =2(\Tr AB)^2+2\Tr(ABAB)+4\Tr(A^2B^2).
\label{eq:supp-quadratic-covariance}
\end{equation}
The pair partitions contribute
$\Tr(A^2)\Tr(B^2)+2(\Tr AB)^2$, and the six four-cycles contribute
$2\Tr(ABAB)+4\Tr(A^2B^2)$; subtracting the product of the second moments
leaves the displayed covariance. Consequently,
\begin{equation}
\Ebr{Q_M}=v_n,\qquad
\operatorname{Var}(Q_M)
 \le\frac{2L_R^2v_n+6v_n^2}{D}
 \le\frac{8L_R^2v_n}{D}.
\label{eq:supp-shell-gaussian-moments}
\end{equation}
Indeed, the integral of $(\Tr B_sB_t)^2$ is
$\Tr T_M^2\le L_R^2v_n/D$ by
Lemma~\ref{lem:supp-kernel-frames}. Hilbert--Schmidt Cauchy--Schwarz gives
$|\Tr(ABAB)|\le\Tr(A^2B^2)$, and the remaining terms are bounded using
\[
\iint\Tr(B_s^2B_t^2)\,\mathrm d\nu_M(s)\mathrm d\nu_M(t)
 =\Tr\!\left[\left(\int B_s^2\,\mathrm d\nu_M\right)^2\right]
 =\Tr\!\left[\left(\frac{v_n\id}{D}\right)^2\right]
 =\frac{v_n^2}{D}.
\]
Also $v_n\le L_R^2$. All these integrals are finite since
$\|B_s\|_{\mathrm{HS}}^2=v_n$ and the reference measure is a probability.

\begin{lemma}
\label{lem:supp-uniform-shell}
For fixed $d,\mathcal K,\theta$ as in
Eq.~\eqref{eq:supp-shell-variance-scale}, there are constants $c,C>0$,
an integer $n_0$, and measurable target sets $\mathcal C_n$, depending
only on these fixed parameters, such that
$\Prob_{\mathrm{Haar}}(\mathcal C_n)\ge1-e^{-cn^3}$ for $n\ge n_0$.
With the supremum taken over rank-one product POVMs, every
$\psi\in\mathcal C_n$ satisfies
\begin{equation}
\sup_M
 \int\bigl[D\Tr(PB_s)-1\bigr]^2\,\mathrm d\nu_M(s)
 \le(1+C/n)v_n.
\label{eq:supp-uniform-shell}
\end{equation}
\end{lemma}
\begin{proof}
The centered variable $Q_M-v_n$ is a real polynomial of degree at most
four in the real and imaginary parts of $z$. Gaussian polynomial
hypercontractivity~\cite[Proposition~3.2]{GamarnikEtAl2024Hypercontractivity}
and Eq.~\eqref{eq:supp-shell-gaussian-moments} give
$\|Q_M-v_n\|_{L^m}\le5(m-1)^2\|Q_M-v_n\|_{L^2}$ for $m\ge2$.
Markov's inequality yields
\begin{equation}
\Prob\{Q_M>v_n(1+1/n)\}
 \le\left[\frac{5\sqrt8\,L_Rn(m-1)^2}{\sqrt{Dv_n}}\right]^m.
\label{eq:supp-shell-fixed-frame-tail}
\end{equation}
With $m=n^2$, the logarithm of the right-hand side of
Eq.~\eqref{eq:supp-shell-fixed-frame-tail} is
$-\tfrac12(\log d)n^3+O_{d,\mathcal K,\theta}(n^2\log n)$,
because $v_n$ has fixed polynomial order.

Use the same Gaussian vector for all measurements. Set
$t=\|z\|_2^2/D$ and $P=zz^\dagger/\|z\|_2^2$.
Equation~\eqref{eq:supp-signed-kernel-moments} gives
$\int X_s\,\mathrm d\nu_M=t-1$, hence the exact radial identity
\begin{equation}
H_P(M):=\int\bigl[D\Tr(PB_s)-1\bigr]^2\,\mathrm d\nu_M
 =t^{-2}\bigl[Q_M-(t-1)^2\bigr]\le t^{-2}Q_M.
\label{eq:supp-shell-haar-transfer}
\end{equation}
Lemma~\ref{lem:supp-common-radius} with $r=0$ gives
$|t-1|\le n^{-4}$ outside an event of probability at most
$\exp[-(\log d)n^3/4]$, for all sufficiently large $n$.

It remains to make the bound simultaneous in $M$. Let $\mathcal V_d$
be the set of local frame operators $S_i$ arising from probability
measures on rank-one projectors with first moment $\id/d$. Each such
measure $\mu$ defines a POVM on rank-one projectors $Q$ by
$M_i(\mathrm dQ)=d\,Q\,\mu(\mathrm dQ)$. Probability
measures on this compact projector space are weakly compact; the
first-moment constraint is closed and the second-moment map is continuous.
Thus $\mathcal V_d$ is compact. It lies in a real vector space of
dimension at most $d^4$, and every element has operator norm at most $1/d$.
A volumetric covering in this finite-dimensional normed space gives an
operator-norm $\epsilon_n$-net inside $\mathcal V_d$ of size at most
$(C_d/\epsilon_n)^{d^4}$. Before drawing $z$, fix one representing
rank-one POVM for each net point and form the $n$-fold product net.

The functional to be covered depends only on these frames:
\[
H_P(M)=D^2(\mathcal R_n(P)|S_M\mathcal R_n(P))-1.
\]
If each local frame changes by at most $\epsilon_n$, telescoping gives
$\|S_M-S_{M'}\|\le n\epsilon_n d^{-(n-1)}$.
Since $\|\mathcal R_n(P)\|_{\mathrm{HS}}\le L_R$, it follows that
\begin{equation}
|H_P(M)-H_P(M')|\le dDL_R^2n\epsilon_n.
\label{eq:supp-shell-off-net}
\end{equation}
Choose
\begin{equation}
\epsilon_n=\frac{v_n}{dDL_R^2n^2}.
\label{eq:supp-product-net-precision}
\end{equation}
The error in Eq.~\eqref{eq:supp-shell-off-net} is then at most $v_n/n$,
and the product net has logarithmic cardinality $O(n^2)$.
On the common radial event and the fixed-frame events at all net points,
\[
\sup_M H_P(M)
 \le\left[\frac{1+1/n}{(1-n^{-4})^2}+\frac1n\right]v_n
 =(1+O(1/n))v_n.
\]
By the union bound, the probability that the radial estimate or at least
one estimate on the net fails is at most
\begin{equation}
\left(\frac{C_d}{\epsilon_n}\right)^{d^4n}
 \left[\frac{5\sqrt8\,L_Rn(n^2-1)^2}{\sqrt{Dv_n}}\right]^{n^2}
 +\exp[-(\log d)n^3/4]\le e^{-cn^3}
\label{eq:supp-shell-event-failure}
\end{equation}
for some $c>0$ and sufficiently large $n$. The $O(n^2)$ logarithmic
covering cost is dominated by the negative cubic term in the tail.
Choose $C$ to cover the preceding $O(1/n)$ error and define
$\mathcal C_n$ by Eq.~\eqref{eq:supp-uniform-shell} on pure targets.
The supremum of the continuous functional $H_P$ over the compact set
$\mathcal V_d^n$ is continuous in $P$, so $\mathcal C_n$ is measurable.
Whenever all these estimates hold for $z$, its direction
$\psi=z/\|z\|_2$ belongs to $\mathcal C_n$. Since $\psi$ has the Haar
distribution, Eq.~\eqref{eq:supp-shell-event-failure} gives the claimed
probability bound.
\end{proof}

For $\mathcal K=\{0\}$ and $\theta_0=1$, we have $B_s=P_s$ and $v_n=1$.
Equations~\eqref{eq:supp-rank-one-density} and
\eqref{eq:supp-povm-refinement-contraction} therefore give, on the
corresponding common target event,
\begin{equation}
\sup_{M\in\mathcal M_n}\|q_M^P-\nu_M\|_{\nu_M^{-1}}^2\le1+C/n.
\label{eq:supp-zero-radius-quadratic}
\end{equation}

\subsection{Exact calibration and an ordered KL bound}

For probability distributions $p_1,p_2$ with densities $f_1,f_2$ relative
to a probability measure $\nu$, we use the usual extended-value conventions
at zero densities and write
\[
D_{\mathrm{KL}}(p_1\|p_2)=\int f_1\log\frac{f_1}{f_2}\,\mathrm d\nu,
\qquad
\chi^2(p_1\|p_2)=\int\frac{(f_1-f_2)^2}{f_2}\,\mathrm d\nu.
\]
We order the calibrated states so that the second outcome distribution
in the KL divergence comes from the state with a maximally mixed
component. For degree $j$ this gives
\begin{equation}
(\rho_1,\rho_2)=
\begin{cases}
(\rho_L,\rho_H),&j\ \text{even},\\
(\rho_H,\rho_L),&j\ \text{odd}.
\end{cases}
\label{eq:supp-hard-state-order}
\end{equation}

\begin{proposition}
\label{prop:supp-hard-states}
Fix integers $d\ge2$, $j\ge0$ and $0<b<1$. There are constants
$c,C,\beta>0$, an integer $n_0$, and measurable target sets
$\mathcal H_n$, all depending only on $d,j,b$, such that
$\Prob_{\mathrm{Haar}}(\mathcal H_n)\ge1-Ce^{-cn}$ for $n\ge n_0$.
For every such $n$ and $\psi\in\mathcal H_n$, there exist density matrices
$\rho_H,\rho_L$ that, with $(\rho_1,\rho_2)$ ordered as in
Eq.~\eqref{eq:supp-hard-state-order}, satisfy
\begin{gather}
F(\rho_L)=b,\qquad F(\rho_H)\ge b+(1-b)\gamma_j(d),
\label{eq:supp-hard-promises}\\
\sup_{M\in\mathcal M_n}
 \|q_M^{\rho_H}-q_M^{\rho_L}\|_{\nu_M^{-1}}
 \le Cn^{-(j+1)/2},
\label{eq:supp-hard-quadratic}\\
\rho_2\succeq\beta\frac{\id}{D},
\label{eq:supp-hard-background}\\
\sup_{M\in\mathcal M_n}
 D_{\mathrm{KL}}\bigl(q_M^{\rho_1}\|q_M^{\rho_2}\bigr)
 \le Cn^{-(j+1)}.
\label{eq:supp-hard-one-copy-kl}
\end{gather}
\end{proposition}
\begin{proof}
The Haar second-moment identity used in the proof of
Lemma~\ref{lem:supp-low-orbit}, applied to the traceless unitary Weyl
strings and averaged over a shell, gives
\begin{equation}
\Ebr[\mathrm{Haar}]{F(\tau_k)}=\frac1{D+1},\qquad 1\le k\le n.
\label{eq:supp-shell-fidelity}
\end{equation}
Let $\mathcal O_n$ be the target event on which
$F(\tau_{k_i})\le D^{-3/4}$ for all $j+1$ finite nodes.
Markov's inequality and a union bound give
\begin{equation}
\Prob_{\mathrm{Haar}}(\mathcal O_n^c)
 \le\frac{(j+1)D^{3/4}}{D+1}\le(j+1)D^{-1/4}.
\label{eq:supp-exponential-overlaps}
\end{equation}
Fix a target in $\mathcal O_n$. Every finite shell state appearing in $\sigma_L$ has
fidelity at most $D^{-3/4}$, and $F(\tau_\infty)=D^{-1}\le D^{-3/4}$.
Hence $t_L:=F(\sigma_L)\le D^{-3/4}$. Increase $n_0$ so that
$D^{-3/4}<b$ and $D^{-3/4}\le1/2$ for $n\ge n_0$. Define
\begin{equation}
\alpha=\frac{b-t_L}{1-t_L},\qquad
\rho_L=\alpha P+(1-\alpha)\sigma_L,\qquad
\rho_H=bP+(1-b)\sigma_H.
\label{eq:supp-calibrated-states}
\end{equation}
Here $0\le\alpha\le b<1$, so both states are positive and normalized.
Their fidelities satisfy
$F(\rho_L)=\alpha+(1-\alpha)t_L=b$ and
$F(\rho_H)=b+(1-b)F(\sigma_H)\ge b+(1-b)\gamma_j(d)$.
Thus they lie in opposite promise classes whenever $g\le\gamma_j(d)$,
including equality.

Put $e=b-\alpha=(1-b)t_L/(1-t_L)\le2D^{-3/4}$.
For each fixed product measurement $M$, the outcome distributions of
the calibrated states satisfy
\begin{equation}
q_M^{\rho_H}-q_M^{\rho_L}
 =(1-b)(q_M^{\sigma_H}-q_M^{\sigma_L})
      +e(q_M^P-q_M^{\sigma_L}).
\label{eq:supp-calibrated-difference}
\end{equation}
For even $j$, $\rho_2=\rho_H$ has maximally mixed weight $(1-b)\kappa$;
for odd $j$, $\rho_2=\rho_L$ has weight
$(1-\alpha)\kappa\ge(1-b)\kappa$. Hence
Eq.~\eqref{eq:supp-hard-background} holds with
$\beta=(1-b)\kappa>0$. For each $M\in\mathcal M_n$ and each measurable
outcome set $A$, positivity of $M(A)$ gives
$q_M^{\rho_2}(A)\ge\beta\nu_M(A)$. Thus, $\nu_M$-almost everywhere,
\begin{equation}
\frac{\mathrm dq_M^{\rho_2}}{\mathrm d\nu_M}\ge\beta.
\label{eq:supp-hard-denominator}
\end{equation}

First take a rank-one product POVM. Its Born densities belong to $[0,D]$
and integrate to one, so
$\|q_M^\omega-q_M^{\omega'}\|_{\nu_M^{-1}}^2\le2D$ for any two states.
On $\mathcal H_n=\mathcal C_n\cap\mathcal O_n$, where
$\mathcal C_n$ is the event of Lemma~\ref{lem:supp-uniform-shell} for the
dual kernel, Eqs.~\eqref{eq:supp-dual-readout-difference} and
\eqref{eq:supp-calibrated-difference} give
\begin{equation}
\begin{gathered}
\|q_M^{\rho_H}-q_M^{\rho_L}\|_{\nu_M^{-1}}
 \le(1-b)\kappa\sqrt{(1+C/n)v_n}+e\sqrt{2D}
   =O_{d,j,b}(n^{-(j+1)/2}),\\
\sqrt{\chi^2(q_M^{\rho_1}\|q_M^{\rho_2})}
 \le\beta^{-1/2}
       \|q_M^{\rho_H}-q_M^{\rho_L}\|_{\nu_M^{-1}}.
\end{gathered}
\label{eq:supp-calibrated-chi-square}
\end{equation}
Here $v_n=\Theta(n^{-(j+1)})$ by
Eq.~\eqref{eq:supp-dual-kernel-scale}, whereas
$e\sqrt{2D}=O(D^{-1/4})$ is smaller than any fixed inverse power of $n$.
The inequality $\log t\le t-1$ gives
$D_{\mathrm{KL}}(p_1\|p_2)\le\chi^2(p_1\|p_2)$.
Equation~\eqref{eq:supp-povm-refinement-contraction} transfers both the
quadratic and KL bounds to every product POVM on the same event.
This proves Eqs.~\eqref{eq:supp-hard-quadratic} and
\eqref{eq:supp-hard-one-copy-kl}, in the fixed ordering above.
Finally, $\mathcal O_n$ is measurable by continuity of its finitely many
shell fidelities, and $\mathcal C_n$ is measurable by the lemma.
Their combined failure probability is at most
$e^{-c_1n^3}+(j+1)d^{-n/4}\le Ce^{-cn}$ after increasing $n_0$.
Taking $C$ large enough gives all three upper bounds in the proposition.
\end{proof}
\section{Sequential information and optimal certification}
\label{sec:supp-hierarchy}

The tests of Sec.~\ref{sec:supp-scores} give upper bounds on $\Nmax$.
For the matching lower bounds, we use the state pairs from
Sec.~\ref{sec:supp-hard-states} and bound the information a protocol
can gain before it stops.

\subsection{Information with adaptation and random stopping}
\label{subsec:supp-sequential-information}

Fix a protocol $\Pi$ and an ordered pair of states $(\rho_1,\rho_2)$.
Let $\mathbb P_i$ be the distribution of the complete execution record
under $\rho_i$, and let $\mathbb E_i$ denote expectation under this
distribution. We use the seed, padded records, stopping time $N$, and
decision $J$ of Sec.~\ref{sec:supp-model}. Define
\begin{equation}
L_\delta=(1-2\delta)\log\frac{1-\delta}{\delta},
\qquad 0<\delta<\frac12.
\label{eq:supp-binary-information}
\end{equation}

\begin{lemma}
\label{lem:supp-sequential-information}
Suppose that, for some finite $\Lambda>0$,
\begin{equation}
\sup_{M\in\mathcal M_n}
 D_{\mathrm{KL}}(q_M^{\rho_1}\|q_M^{\rho_2})\le\Lambda.
\label{eq:supp-sequential-one-copy}
\end{equation}
Let $\Pi$ be an allowed protocol that stops almost surely under both states.
If, for some $\iota\in\{0,1\}$,
$\mathbb P_1(J\ne\iota)\le\delta$ and
$\mathbb P_2(J=\iota)\le\delta$, then
\begin{equation}
\Lambda\,\Ebr{N\condsep\Pi,\rho_1}\ge L_\delta.
\label{eq:supp-stopped-information}
\end{equation}
\end{lemma}
\begin{proof}
The conclusion holds when $\mathbb E_1N=\infty$, so assume this expectation
is finite. For an integer $m\ge1$, retain the seed and the first $m$ record
slots, including the stopping flags and any decision already made. Slots
after stopping contain the common empty record. This includes stopping and
deciding at time zero and uses exactly $N\wedge m$ copies. Write
$\mathbb P_i^{(m)}$ for the distribution of the truncated record under
$\rho_i$.

Given the same history $h_{<k}$, the protocol makes the same continuation
decision under both inputs. If it takes copy $k$, it selects the same
product POVM $M_k$, with Born kernels
\[
K_{i,k}(h_{<k},\mathrm dy)=\Tr[\rho_i M_k(h_{<k},\mathrm dy)].
\]
After stopping, both kernels are the point mass at the empty record.

To compare the two record distributions, we use the variational formula
for relative entropy~\cite{PolyanskiyWu2025}. For probability measures
$\mu,\nu$ on a measurable space,
\[
D_{\mathrm{KL}}(\mu\|\nu)
 =\sup_f\left\{\int f\,\mathrm d\mu-\log\int e^f\,\mathrm d\nu\right\},
\]
where $f$ ranges over bounded real-valued measurable functions.
For $\mu\ll\nu$, nonnegativity of the divergence from $\mu$ to the measure
with density $e^f/\int e^f\,\mathrm d\nu$ relative to $\nu$ bounds the
supremum from above by $D_{\mathrm{KL}}(\mu\|\nu)$.
For the reverse inequality, put $R=\mathrm d\mu/\mathrm d\nu$ and
truncate $\log R$ to $[-T,T]$, obtaining $f_T$. As $T\to\infty$,
$e^{f_T}\to R$ and $e^{f_T}\le 1+R$, so $\int e^{f_T}\,\mathrm d\nu\to1$.
The positive part of $f_T$ increases to $\log^+R$, and the negative part
is dominated under $\mu$ by $\log^-R$, with
$\int\log^-R\,\mathrm d\mu=\int R\log^-R\,\mathrm d\nu\le 1/e$.
Thus $\int f_T\,\mathrm d\mu\to D_{\mathrm{KL}}(\mu\|\nu)$, including
when the limit is infinite. If $\mu\not\ll\nu$, a set $A$ with
$\nu(A)=0<\mu(A)$ gives an infinite supremum by taking $f=T\mathbf1_A$.
Pulling back test functions gives data processing under measurable maps.
In particular, adjoining the settings, flags, and available decision
preserves the divergence, since they are determined by the seed and
padded outcomes and the original coordinates can be recovered by projection.

Now let $f(h_{<k},y)$ be bounded and measurable and put
$\varphi(h_{<k})=\log\int e^{f(h_{<k},y)}K_{2,k}(h_{<k},\mathrm dy)$.
This is again bounded and measurable. Applying the variational formula
at each history gives
\[
\int f(h_{<k},y)K_{1,k}(h_{<k},\mathrm dy)
 \le\varphi(h_{<k})+\Lambda\mathbf 1_{\{N\ge k\}}(h_{<k}).
\]
On stopped histories the kernels coincide, so the information term is zero.
Under input $\rho_i$, the joint distribution of the history and the
next slot is $\mathbb P_i^{(k)}=\mathbb P_i^{(k-1)}K_{i,k}$.
Integrating the preceding bound therefore gives
\[
\begin{aligned}
\int f\,\mathrm d\mathbb P_1^{(k)}
 -\log\int e^f\,\mathrm d\mathbb P_2^{(k)}
&\le\int\varphi\,\mathrm d\mathbb P_1^{(k-1)}
 -\log\int e^\varphi\,\mathrm d\mathbb P_2^{(k-1)}
 +\Lambda\mathbb P_1(N\ge k)\\
&\le D_{\mathrm{KL}}(\mathbb P_1^{(k-1)}\|\mathbb P_2^{(k-1)})
 +\Lambda\mathbb P_1(N\ge k).
\end{aligned}
\]
Taking the supremum over bounded measurable $f$ proves the KL-divergence recursion.
Both inputs use the same seed distribution, so their initial record
distributions agree and contribute zero divergence. Iteration yields
\begin{equation}
D_{\mathrm{KL}}(\mathbb P_1^{(m)}\|\mathbb P_2^{(m)})
 \le\Lambda\sum_{k=1}^m\mathbb P_1(N\ge k)
 =\Lambda\,\Ebr{N\wedge m\condsep\Pi,\rho_1}.
\label{eq:supp-truncated-information}
\end{equation}
Let $A$ be the event that the protocol stops with $J=\iota$, and put
$A_m=A\cap\{N\le m\}$. Its indicator is measurable from the truncated
record and equals zero if the protocol has not stopped by $m$. Write $p_m=\Prob_1(A_m)$, $q_m=\Prob_2(A_m)$,
$p=\Prob_1(A)$, and $q'=\Prob_2(A)$.
Almost-sure stopping under both inputs gives
\[
p_m\longrightarrow p\ge1-\delta,\qquad
q_m\longrightarrow q'\le\delta.
\]
For Bernoulli distributions, write
$\operatorname{kl}(x\|y)=x\log(x/y)+(1-x)\log[(1-x)/(1-y)]$,
with $0\log(0/y)=0$ and a positive mass divided by zero giving $+\infty$.
Data processing for the indicator of $A_m$, lower semicontinuity of binary
relative entropy~\cite{PolyanskiyWu2025}, and monotone convergence yield
\begin{equation}
\operatorname{kl}(p\|q')
 \le\liminf_{m\to\infty}\operatorname{kl}(p_m\|q_m)
 \le\Lambda\,\Ebr{N\condsep\Pi,\rho_1}.
\label{eq:supp-terminal-information}
\end{equation}
For $0\le y<x\le1$, binary relative entropy is nondecreasing in $x$
and nonincreasing in $y$, with endpoint values understood by limits. Hence
$\operatorname{kl}(p\|q')\ge\operatorname{kl}(1-\delta\|\delta)=L_\delta$,
which proves the claim.
\end{proof}

\subsection{Matching the hierarchy and confidence ranges}
\label{subsec:supp-matching}

Fix an integer $d\ge2$ and thresholds $0<b<a\le1$, and put
$g=(a-b)/(1-b)$ and $r=r_d(g)$. Proposition~\ref{prop:supp-elementary-decay}
ensures that $r$ is finite.

\begin{proposition}
\label{prop:supp-testing-scale}
There exist $A,C>0$, an integer $n_0$, and measurable target sets
$\mathcal G_n$, depending only on $d,a,b$, such that
$\Prob_{\mathrm{Haar}}(\mathcal G_n)\to1$ and, for every $n\ge n_0$,
every $\psi\in\mathcal G_n$, and every $0<\delta<1/2$,
\begin{equation}
A n^r L_\delta\le\Nexp\le\Nmax
 \le Cn^r\log(1/\delta).
\label{eq:supp-raw-hierarchy}
\end{equation}
\end{proposition}
\begin{proof}
Choose the degree-$r$ optimizer from
Proposition~\ref{prop:supp-critical-sequence} and its unique weights in
Eq.~\eqref{eq:supp-radial-coefficients}. These weights depend only on $d$ and $r$.
Substituting the optimizer's envelope endpoints into Eq.~\eqref{eq:supp-envelope-gap}
gives the separation between the mean-score bounds:
\[
\Delta=(1-b)(1-\ell)\,[g-\gamma_r(d)]>0.
\]
Theorem~\ref{thm:supp-uniform-score} and
Proposition~\ref{prop:supp-achievable-budget} give the upper bound in
Eq.~\eqref{eq:supp-polynomial-upper} on a measurable target set
$\mathcal A_n$ of probability at least $1-n^{-2}$, for all sufficiently
large $n$. The bound is a deterministic cap for the entire range $0<\delta<1/2$.
Also $\Nexp\le\Nmax$ by Sec.~\ref{sec:supp-model}.

If $r\ge1$, set $j=r-1$. Minimality of $r$ gives $g\le\gamma_j(d)$.
Write $C_h$ for the constant $C$ in Proposition~\ref{prop:supp-hard-states}.
For every target in its set $\mathcal H_n$, the calibrated pair satisfies
$F(\rho_L)=b$ and $F(\rho_H)\ge b+(1-b)\gamma_j(d)\ge a$.
Use its ordering in Eq.~\eqref{eq:supp-hard-state-order} and set
$\Lambda=C_hn^{-r}$. A valid protocol satisfies the hypotheses of
Lemma~\ref{lem:supp-sequential-information}, with $\iota=1$ when $j$ is
even and $\iota=0$ when $j$ is odd. Thus every valid $\Pi$ satisfies
\begin{equation}
\sup_{\rho\in\mathcal S}\Ebr{N\condsep\Pi,\rho}
 \ge\Ebr{N\condsep\Pi,\rho_1}
 \ge C_h^{-1}n^rL_\delta.
\label{eq:supp-worst-state-lower}
\end{equation}
Taking the infimum over valid protocols gives the lower bound with $A=C_h^{-1}$.

For $r=0$, use instead, for every target,
\begin{equation}
\rho_H=P,\qquad
\rho_L=\frac b2P+\left(1-\frac b2\right)\frac{\id}{D}.
\label{eq:supp-constant-hard-pair}
\end{equation}
These are density matrices. When $D^{-1}\le b/2$,
$F(\rho_H)=1\ge a$ and
$F(\rho_L)=b/2+(1-b/2)/D\le b$.
For any $M\in\mathcal M_n$, use the reference distribution
$\nu_M=q_M^{\id/D}$ and the densities
$p=\mathrm dq_M^P/\mathrm d\nu_M$ and
$s=\mathrm dq_M^{\rho_L}/\mathrm d\nu_M$. Set $B_b=4/b+2$.
Since $s\ge1-b/2>0$ and $s\ge(b/2)p$ almost everywhere,
\begin{equation}
D_{\mathrm{KL}}(q_M^P\|q_M^{\rho_L})
 \le\chi^2(q_M^P\|q_M^{\rho_L})
 \le2\int\frac{p^2}{s}\,\mathrm d\nu_M+2\int s\,\mathrm d\nu_M
 \le B_b.
\label{eq:supp-constant-information}
\end{equation}
Here $(p-s)^2\le2p^2+2s^2$ gives the second inequality.
Lemma~\ref{lem:supp-sequential-information}, with
ordered pair $(P,\rho_L)$ and $\iota=0$, now gives
$\Nexp\ge L_\delta/B_b$.

For $r\ge1$ take $\mathcal G_n=\mathcal A_n\cap\mathcal H_n$;
its failure probability is at most $n^{-2}+C_he^{-cn}$.
For $r=0$ take $\mathcal G_n=\mathcal A_n$.
Choose $n_0$ so that the preceding bounds, together with $D^{-1}\le b/2$
when $r=0$, hold for every $n\ge n_0$. The constants $A,C$, the size
cutoff $n_0$, and the family $(\mathcal G_n)_n$ depend only on $d,a,b$. In particular, $a=1$ has $g=1>\gamma_0(d)$
and is covered by the zero-degree upper test and this same lower-bound pair.
\end{proof}

\begin{proof}[Proof of Theorem~\ref{thm:hierarchy}]
\phantomsection\label{proof:supp-hierarchy}
Choose the Haar-typical family and the size cutoff $n_0$ from
Proposition~\ref{prop:supp-testing-scale}. Fix any $0<\delta_\star<1/2$.
To compare the two bounds in Eq.~\eqref{eq:supp-raw-hierarchy}, observe
that the positive continuous function
\begin{equation}
h(\delta)=\frac{L_\delta}{\log(1/\delta)}
 =(1-2\delta)\left[1+\frac{\log(1-\delta)}{\log(1/\delta)}\right],
\qquad 0<\delta<\frac12,
\label{eq:supp-confidence-ratio}
\end{equation}
has limit one as $\delta\downarrow0$. Set $h(0)=1$. Continuity and
positivity on $[0,\delta_\star]$ give a minimum $m_{\delta_\star}>0$. Taking
$c_{\delta_\star}=A m_{\delta_\star}$ in
Eq.~\eqref{eq:supp-raw-hierarchy} gives lower and upper comparison
constants $c_{\delta_\star}$ and $C$ for both $\Nexp$ and $\Nmax$.
Thus both are $\Theta(n^r\log(1/\delta))$ uniformly over
$0<\delta\le\delta_\star$ and $\psi\in\mathcal G_n$ for $n\ge n_0$.
\end{proof}

At $g=\gamma_j(d)$, the degree-$j$ hard pair remains in opposite promise
classes, while strict decrease gives $\gamma_{j+1}(d)<g$ and hence a
matching degree-$(j+1)$ test. Thus $r_d(\gamma_j(d))=j+1$.
Every fixed threshold pair has a finite optimal exponent. These
exponents are unbounded as the threshold pair varies, since each
$\gamma_j(d)>0$ is an attainable normalized gap.

\subsection{Uniform transition near a step point}
\label{subsec:supp-near-critical}

To prove Corollary~\ref{cor:near-critical}, fix a threshold pair $(a_0,b_0)$
whose normalized gap is a step point. We will choose an interval containing
$b_0$ in its interior and bound
$\Nexp$ and $\Nmax$ uniformly as $b$ varies in this interval and $g$ varies near
that step point. The bounds below apply to any closed interval
$[b_-,b_+]\subset(0,1)$ with $b_-<b_+$. Fix integers $d\ge2$ and $j\ge0$.
Write $\gamma_i=\gamma_i(d)$ for $i\ge0$, set $\gamma_{-1}=1$, and choose
\[
0<\eta<\tfrac12\min\{\gamma_{j-1}-\gamma_j,\ \gamma_j-\gamma_{j+1}\}.
\]
There exist constants $A,C>0$, an integer $n_0$, and a measurable
Haar-typical family $(\mathcal G_n)_n$, depending only on
$d,j,b_-,b_+,\eta$, with the following property. For every $n\ge n_0$,
$\psi\in\mathcal G_n$, $b\in[b_-,b_+]$, $|\epsilon|\le\eta$, and
$0<\delta<1/2$, take $a=b+(1-b)(\gamma_j+\epsilon)$ and
$\epsilon_+=\max\{\epsilon,0\}$. Then
\begin{equation}
A\frac{n^{j+1}}{1+n\epsilon_+^2}L_\delta
\le\Nexp\le\Nmax
\le C\frac{n^{j+1}}{1+n\epsilon_+^2}\log(1/\delta),
\label{eq:supp-near-critical-budgets}
\end{equation}
where $L_\delta$ is defined in Eq.~\eqref{eq:supp-binary-information}.
The choice of $\eta$ ensures $0<b<a<1$ throughout this region.

\begin{proof}[Proof of Eq.~\eqref{eq:supp-near-critical-budgets}
and Corollary~\ref{cor:near-critical}]
Choose the optimizing polynomials of degrees $j$ and $j+1$ from
Proposition~\ref{prop:supp-critical-sequence}, with the score weights
of Eq.~\eqref{eq:supp-radial-coefficients}. For $k=j,j+1$, write
$\ell^{(k)}=\ell_d(p_k)$ for the lower envelope of the optimizer $p_k$. Let $\mathcal A_n$ be the
intersection of the two measurable target events of
Theorem~\ref{thm:supp-uniform-score}. Its Haar probability is at least
$1-2n^{-2}$ for all sufficiently large $n$. For both scores, the errors
in the mean and second-moment bounds satisfy
$\eta_n^{(k)},\varepsilon_n^{(k)}=O(n^{-3})$.
Once $\varepsilon_n^{(k)}\le1$, the full second moment on $\mathcal A_n$
is bounded by $V_n^{(k)}=\nu_n^{(k)}+e_*^{(k)}+1=O(n^k)$ uniformly
over input states. These events and bounds depend only on $d$ and $j$.

For $\epsilon\ge n^{-1/2}$, Eq.~\eqref{eq:supp-optimal-score-gap} gives
the separation between the degree-$j$ mean-score bounds:
\[
\Delta_j=(1-b)(1-\ell^{(j)})\epsilon
\ge(1-b_+)(1-\ell^{(j)})\epsilon.
\]
Thus $\eta_n^{(j)}/\Delta_j=O(n^{-5/2})$ uniformly in this range.
One cutoff ensures $\eta_n^{(j)}\le\Delta_j/4$ and
$\varepsilon_n^{(j)}\le1$, the two error conditions in the proof of
Proposition~\ref{prop:supp-achievable-budget}. Its test has deterministic
copy cap $O(n^j\epsilon^{-2}\log(1/\delta))$. For the degree-$(j+1)$
score, throughout the entire region we have
\[
\begin{aligned}
\Delta_{j+1}
 &=(1-b)(1-\ell^{(j+1)})(\gamma_j+\epsilon-\gamma_{j+1})\\
 &\ge(1-b_+)(1-\ell^{(j+1)})(\gamma_j-\gamma_{j+1}-\eta)>0.
\end{aligned}
\]
A common cutoff therefore ensures the same two error conditions for this
score, giving a cap $O(n^{j+1}\log(1/\delta))$. Use the degree-$j$
test for $\epsilon\ge n^{-1/2}$ and the degree-$(j+1)$ test otherwise.
For $\epsilon>0$, the smaller of $n^j/\epsilon^2$ and $n^{j+1}$ lies
between $n^{j+1}/(1+n\epsilon^2)$ and twice that value. For
$\epsilon\le0$ the factor in Eq.~\eqref{eq:supp-near-critical-budgets}
is exactly $n^{j+1}$. This proves the uniform upper bound.

For the lower bound we calibrate the pair from
Proposition~\ref{prop:supp-hard-states} at each given $b$.
The target events, constants, and size cutoff can be chosen uniformly
over $[b_-,b_+]$. To see this, fix a dual for each required degree $k$ and
write $\mathcal H_n^{(k)}=\mathcal C_n^{(k)}\cap\mathcal O_n^{(k)}$
for the event in that proof. The first event is the uniform shell
bound of Lemma~\ref{lem:supp-uniform-shell} for the fixed dual kernel.
The second bounds the fidelities of its $k+1$ finite shell states.
These events do not depend on $b$, and Eq.~\eqref{eq:supp-exponential-overlaps}
and the lemma give failure probability at most $e^{-c_kn^3}+(k+1)D^{-1/4}$.
On $\mathcal H_n^{(k)}$, $t_L=F(\sigma_L)\le D^{-3/4}$.
Choose the cutoff so that $D^{-3/4}<b_-$ and $D^{-3/4}\le1/2$.
Then $\alpha_b=(b-t_L)/(1-t_L)$ satisfies $0\le\alpha_b\le b$ and
$e_b=b-\alpha_b=(1-b)t_L/(1-t_L)\le2D^{-3/4}$ throughout the interval.
Write $H_k,L_k$ for the states in Eq.~\eqref{eq:supp-calibrated-states}
at the chosen $b$. They satisfy
\[
F(L_k)=b,\qquad F(H_k)\ge b+(1-b)\gamma_k.
\]

Let $\kappa_k$ be the degree-$k$ background weight from
Eq.~\eqref{eq:supp-dual-background}. In the ordering of
Eq.~\eqref{eq:supp-hard-state-order}, the second state retains
maximally mixed weight $(1-b)\kappa_k$ for even $k$, and
$(1-\alpha_b)\kappa_k$ for odd $k$. Both are at least
$\beta_k=(1-b_+)\kappa_k>0$. For rank-one product POVMs,
Eq.~\eqref{eq:supp-calibrated-chi-square} gives
\[
\|q_M^{H_k}-q_M^{L_k}\|_{\nu_M^{-1}}
\le \kappa_k\sqrt{(1+C_k/n)v_n}+2\sqrt{2}\,D^{-1/4}
\le C'_k n^{-(k+1)/2},
\]
since $v_n=\Theta(n^{-(k+1)})$ by Eq.~\eqref{eq:supp-dual-kernel-scale}.
The constants are uniform in $b$. Refinement and
Eq.~\eqref{eq:supp-povm-refinement-contraction} give the same norm
bound for every product POVM $M$. The outcome density of the second
state in Eq.~\eqref{eq:supp-hard-state-order}, relative to $\nu_M=q_M^{\id/D}$,
is at least $\beta_k$ almost everywhere. The ordered KL divergence is
therefore at most $\beta_k^{-1}$ times the squared norm. In particular,
if $\epsilon\le0$, the degree-$j$ pair has $F(H_j)\ge a$ and $F(L_j)=b$,
and its ordered KL divergence is $O(n^{-(j+1)})$.

For $j\ge1$ and $\epsilon>0$, take a target in the intersection of the
two hard-pair events for $j$ and $j-1$. Using both pairs calibrated at
the same $b$, set
\[
t=\frac{\epsilon}{\gamma_{j-1}-\gamma_j}<\frac12,\qquad
H=(1-t)H_j+tH_{j-1},\qquad L=(1-t)L_j+tL_{j-1}.
\]
Then $F(L)=b$ and
$F(H)\ge b+(1-b)[(1-t)\gamma_j+t\gamma_{j-1}]=a$.
Order the pair by degree $j$, with $H$ second for even $j$ and $L$
second for odd $j$. The second state retains maximally mixed weight at
least $(1-t)\beta_j\ge\beta_j/2$. For each $M$, the reference distribution
$\nu_M$ is common to both degrees. The triangle inequality gives
\[
\|q_M^H-q_M^L\|_{\nu_M^{-1}}\le
 (1-t)C'_j n^{-(j+1)/2}+tC'_{j-1}n^{-j/2}
 \le C\bigl[n^{-(j+1)/2}+\epsilon n^{-j/2}\bigr].
\]
Squaring this estimate and using $D_{\mathrm{KL}}\le\chi^2$ with the
uniform background bound gives ordered divergence
$O(n^{-(j+1)}+\epsilon^2 n^{-j})$.

For $j=0$ and $\epsilon>0$, the degree-zero dual has weights $d+1,-d$
at $1,\infty$, respectively. Equations~\eqref{eq:supp-dual-physical-states}
and~\eqref{eq:supp-calibrated-states} give
\[
H_0=hP+(1-h)\id/D,\qquad h=b+(1-b)\gamma_0,\qquad
\gamma_0=1/(d+1).
\]
Take $H=aP+(1-a)\id/D$ and $L=L_0$. Then $F(L)=b$ and
$F(H)=a+(1-a)/D\ge a$. In the order $(L,H)$ the background weight
$1-a=(1-b)(1-\gamma_0-\epsilon)$ is at least
$(1-b_+)(1-\gamma_0-\eta)>0$. Since $a-h=(1-b)\epsilon$,
\begin{equation}
H-L=(H_0-L_0)+(1-b)\epsilon(P-\id/D).
\label{eq:supp-near-critical-zero-difference}
\end{equation}
Let $\mathcal Z_n$ be the target event in
Eq.~\eqref{eq:supp-zero-radius-quadratic}. On this event the $P-\id/D$
term satisfies $\sup_{M\in\mathcal M_n}\|q_M^P-\nu_M\|_{\nu_M^{-1}}=O(1)$.
On $\mathcal Z_n\cap\mathcal H_n^{(0)}$ this gives
$\|q_M^H-q_M^L\|_{\nu_M^{-1}}=O(n^{-1/2}+\epsilon)$ and
$D_{\mathrm{KL}}(q_M^L\|q_M^H)=O(n^{-1}+\epsilon^2)$ uniformly in
$M,b,\epsilon$ in this case.

In every case the ordered pair lies in opposite promise classes and
satisfies
\begin{equation}
\sup_{M\in\mathcal M_n}
D_{\mathrm{KL}}(q_M^{\rho_1}\|q_M^{\rho_2})
\le C_0\bigl[n^{-(j+1)}+\epsilon_+^2n^{-j}\bigr].
\label{eq:supp-near-critical-information}
\end{equation}
For any valid protocol $\Pi$, including those with adaptation across
copies and random stopping, Lemma~\ref{lem:supp-sequential-information} gives
\[
\Ebr{N\condsep\Pi,\rho_1}\ge
\frac{L_\delta}{C_0[n^{-(j+1)}+\epsilon_+^2 n^{-j}]}.
\]
Taking the worst promised input and then the infimum over protocols gives
the lower bound in Eq.~\eqref{eq:supp-near-critical-budgets}, and the
definitions give $\Nexp\le\Nmax$.

For $j\ge1$ take $\mathcal G_n=\mathcal A_n\cap\mathcal H_n^{(j)}
\cap\mathcal H_n^{(j-1)}$. For $j=0$ take
$\mathcal G_n=\mathcal A_n\cap\mathcal H_n^{(0)}\cap\mathcal Z_n$.
These finite intersections of measurable events have Haar probability
tending to one. Taking the largest of the finitely many cutoffs gives
one $n_0$ for all $a,b,\epsilon,\delta$ in the asserted ranges. For
each $0<\delta_\star<1/2$, Eq.~\eqref{eq:supp-confidence-ratio} bounds
$L_\delta/\log(1/\delta)$ below by $m_{\delta_\star}>0$ uniformly on
$0<\delta\le\delta_\star$, with the same target family and cutoff.

To obtain the corollary, take any threshold pair $(a_0,b_0)$ with
$g(a_0,b_0)=\gamma_j$. Since $0<\gamma_j<1$, we have $0<b_0<a_0<1$. Choose
$0<b_-<b_0<b_+<1$ and $\eta$ as above. The set
\[
\mathcal U=\bigl\{(a,b): b_-<b<b_+,\;
 \lvert(a-b)/(1-b)-\gamma_j\rvert<\eta\bigr\}
\]
is an open neighborhood of $(a_0,b_0)$ by continuity. It lies in the
region of Eq.~\eqref{eq:supp-near-critical-budgets}, so the same bounds
hold even if both thresholds vary with $n$ in $\mathcal U$.
Combining Eq.~\eqref{eq:supp-near-critical-budgets} with the uniform comparison of
$L_\delta$ and $\log(1/\delta)$ above gives Eq.~\eqref{eq:near-critical-cost},
with the stated uniformity.
\end{proof}

\subsection{Exponentially typical targets}

\begin{proposition}
\label{prop:supp-exponential-targets}
For fixed $d,a,b$, the target sets in Theorem~\ref{thm:hierarchy} can be
chosen to satisfy
\begin{equation}
\Prob_{\mathrm{Haar}}(\mathcal G_n)\ge1-e^{-c_0n}
\qquad(n\ge n_0)
\label{eq:supp-exponential-targets}
\end{equation}
for some $c_0>0$. Equation~\eqref{eq:supp-raw-hierarchy} holds on these
sets, with its constants $A,C$, the family $(\mathcal G_n)_n$, $c_0$,
and $n_0$ depending only on $d,a,b$. The uniform confidence comparison
of Sec.~\ref{subsec:supp-matching} also holds.
\end{proposition}
\begin{proof}
Use the degree-$r$ optimizing score from Sec.~\ref{subsec:supp-matching},
where $r=r_d(g)$ for the given thresholds $a,b$. Let
$H(x)=-x\log x-(1-x)\log(1-x)$, with $H(0)=H(1)=0$, and put
$f_d(x)=H(x)+x\log(d^2-1)$. Choose $0<\epsilon<1/2$ so small that
$f_d(\epsilon)<\tfrac12\log d$. The derivative
$f_d'(x)=\log[(1-x)(d^2-1)/x]$ is positive on $(0,\epsilon]$.
Replace the weight cutoff in the proof of
Theorem~\ref{thm:supp-uniform-score} by $L=\lfloor\epsilon n\rfloor$.
Using $\binom nw\le e^{nH(w/n)}$, the number of low-weight Weyl strings obeys
\begin{equation}
m_n=\sum_{w=0}^L\binom nw(d^2-1)^w
 \le(n+1)e^{nf_d(\epsilon)}.
\label{eq:supp-linear-weight-count}
\end{equation}
Consequently $m_n^2/D$ and $m_n/D$ decay exponentially.

For each of the coefficient sequences $a=c,e$ in
Eq.~\eqref{eq:supp-radial-coefficients}, its global bound
$a_*:=\sup_{w\in\mathbb N_0}|a_w|$ is fixed. Equation~\eqref{eq:supp-weight-sums}
and the degree-$r$ polynomial factor give
\begin{equation}
\begin{aligned}
\max_{L<w\le n}|a_w|
 &=O\bigl(n^r(d+1)^{-L}\bigr),\\
D^{-1}\sum_W|a_{|W|}|
 &=O(n^r),\\
\left(D^{-1}\sum_{|W|>L}a_{|W|}^2\right)^{1/2}
 &=O\bigl(n^r(d+1)^{-L/2}\bigr).
\end{aligned}
\label{eq:supp-linear-high-coefficients}
\end{equation}
For the square-sum bound, multiply the high-weight maximum by the total
absolute mass divided by $D$, then take a square root.
Equation~\eqref{eq:supp-choi-tails} bounds the full partial-transpose norm
by $O\bigl(n^r[2/(d+1)]^{n-r}\bigr)$. Removing the low-weight Bell terms changes it by
at most $a_*m_n/D$. Thus all three parameters in
Lemma~\ref{lem:supp-weyl-chaos} decay exponentially; its factor
$\log(2D)=O(n)$ preserves exponential decay.

Let $E_{a,z}$ be the centered high-weight operator of
Eq.~\eqref{eq:supp-high-weight-expectation}, now with this linear cutoff.
For some $c_1>0$ and all sufficiently large $n$,
$\mathbb E\|E_{a,z}\|\le e^{-c_1n}$.
Markov's inequality at $n^{-3}$ and a union bound for $a=c,e$ give
failure probability at most $2n^3e^{-c_1n}$.
Lemma~\ref{lem:supp-low-orbit}, with tolerance $n^{-3}$,
gives a failure probability at most
\[
\frac{n^6m_n(m_n-1)}{D+1}
 =O\!\left(n^8e^{-n[\log d-2f_d(\epsilon)]}\right).
\]
The Wick remainder in Lemma~\ref{lem:supp-normal-remainder} satisfies
\[
\Prob\{\|Z_z\|>n^{-3}\}
 \le n^{12}8!C_d^{8r}\Lambda_n^8\kappa_d^{n-8r}
 =O(n^{8r+12}\kappa_d^n),
\]
where $\kappa_d<1$ and $\Lambda_n=O(n^r)$.
Finally, for the same $z$ and $t=\|z\|^2/D$,
Lemma~\ref{lem:supp-common-radius} gives
$\Prob\{|t-1|>n^{-r-4}\}\le e^{-(\log d)n^3/4}$.
The sum of these four failure bounds is at most $e^{-c_un}$ for some
$c_u>0$ and all sufficiently large $n$.

These events imply the same deterministic operator estimates as before.
Indeed, the high-weight transfer is exactly
\[
\sum_{|W|>L}a_{|W|}WPW^\dagger-\sigma_a\id
 =t^{-1}E_{a,z}+(t^{-1}-1)\sigma_a\id,
\qquad \sigma_a=D^{-1}\sum_{|W|>L}a_{|W|}.
\]
Since $\sigma_a=O(n^r)$, the radial correction is $O(n^{-4})$.
Orthonormalizing the low-weight columns changes the weighted operator by
$O(n^{-3})$ and its target column projector by $O(n^{-3})$.
The remaining scalar contribution is bounded by $a_*m_n/D$.
As in the spectral comparison leading to
Eqs.~\eqref{eq:supp-uniform-mean} and~\eqref{eq:supp-auxiliary-bound},
this gives the mean envelope and the bounds on $S_\psi$ with errors
$O(n^{-3}+m_n/D)$. For the full second moment, the exact identity
\eqref{eq:supp-haar-second} still applies. Its last two radial terms are
$O(n^{-4})$ and $O(n^{-8})$, while $\|Z_z\|\le n^{-3}$.
It therefore gives Eq.~\eqref{eq:supp-uniform-second} with an error
of the same order.

Define $\mathcal A_n^{\mathrm{exp}}$ by these deterministic mean and
second-moment inequalities, choosing fixed error bounds
$\eta_n,\varepsilon_n=O(n^{-3}+m_n/D)=o(1)$.
The operators depend continuously on the target, so these sets are
measurable. Whenever all these estimates hold for $z$, its direction
$\psi=z/\|z\|$ belongs to $\mathcal A_n^{\mathrm{exp}}$. Since $\psi$ is
Haar distributed, $\Prob_{\mathrm{Haar}}(\mathcal A_n^{\mathrm{exp}})
\ge1-e^{-c_un}$. For all sufficiently large $n$, the errors satisfy
$\eta_n\le\Delta/4$ and $\varepsilon_n\le1$.
The construction and constants of
Proposition~\ref{prop:supp-achievable-budget} then give the same cap bound
on this new family.

For $r\ge1$, Proposition~\ref{prop:supp-hard-states}, with $j=r-1$
and constant $C_h$, gives a measurable target set $\mathcal H_n$ with
failure probability at most $C_he^{-cn}$. On this set the calibrated
pair obeys the single-copy KL bound uniformly over $\mathcal M_n$.
Intersect $\mathcal H_n$ with
$\mathcal A_n^{\mathrm{exp}}$ and apply
Lemma~\ref{lem:supp-sequential-information} as above.
For $r=0$, the pair in Eq.~\eqref{eq:supp-constant-hard-pair}
works for every target, so use $\mathcal G_n=\mathcal A_n^{\mathrm{exp}}$.
The failure bounds sum to at most $e^{-c_0n}$ after decreasing $c_0>0$
and increasing $n_0$.
\end{proof}

\subsection{Certification on a fixed block partition}
\label{subsec:supp-fixed-blocks}

Fix an integer $k\ge1$. For each $m$, prescribe a partition of $Q=km$ qubits into
$m$ blocks of $k$ qubits, held fixed for all copies. Each copy may be
measured by an arbitrary POVM within each block, with a product of
these measurements across blocks. All block measurements are selected
before that copy is measured. Across copies, allow the
adaptation, random stopping, and classical processing of
Sec.~\ref{sec:supp-model}. A copy is the entire $Q$-qubit state.

\begin{proposition}
\label{prop:supp-fixed-blocks}
Fix $k$ and $0<b<a\le1$. Put $g=(a-b)/(1-b)$ and $s=r_{2^k}(g)$.
There exist $C,c_0>0$, an integer $m_0$, and measurable target sets
$\mathcal G_m$ of Haar probability at least $1-e^{-c_0m}$ for $m\ge m_0$,
depending only on the fixed parameters and prescribed partitions.
For every $0<\delta_\star<1/2$, there is $c_{\delta_\star}>0$ such that,
for every $m\ge m_0$, every $\psi\in\mathcal G_m$, and every
$0<\delta\le\delta_\star$,
\begin{equation}
c_{\delta_\star}m^s\log(1/\delta)
 \le\Nexp\le\Nmax\le Cm^s\log(1/\delta).
\label{eq:supp-fixed-block-budgets}
\end{equation}
The optimal exponent is zero exactly when
\begin{equation}
g>\frac1{2^k+1}.
\label{eq:supp-fixed-block-zero}
\end{equation}
At equality $s=1$; at or below this step point $\Nexp$ and $\Nmax$ both diverge with
$m$ for every fixed $0<\delta<1/2$.
\end{proposition}
\begin{proof}
Under the fixed identification
$(\mathbb C^2)^{\otimes km}\cong(\mathbb C^{2^k})^{\otimes m}$,
products of block POVMs are exactly the measurements in $\mathcal M_m$
with local dimension $d=2^k$. This identification preserves Haar measure,
density matrices, and fidelity. The same seed and relabeled complete records
give a bijection of protocols that preserves the outcome kernels and
the distributions of the complete execution records. Each run consistent
with the protocol rules uses the same number $N$ of whole-system copies
under this correspondence. Validity, deterministic caps, and
worst-case expected costs are therefore preserved, so the two optima
in Eqs.~\eqref{eq:supp-max-budget} and~\eqref{eq:supp-expected-budget}
agree with their qudit counterparts.
Pulling back the target sets of Proposition~\ref{prop:supp-exponential-targets}
gives Eq.~\eqref{eq:supp-fixed-block-budgets}.

By Proposition~\ref{prop:supp-envelope-equivalence},
$\gamma_0(2^k)=1/(2^k+1)$, which gives the strict condition
\eqref{eq:supp-fixed-block-zero}. At equality, strict decrease of the
sequence of step points makes $s=1$; below it $s\ge1$.
For any fixed $0<\delta<1/2$, take $\delta_\star=\delta$ in
Eq.~\eqref{eq:supp-fixed-block-budgets}. Its lower bound then diverges
with $m$ whenever $s\ge1$.
\end{proof}

For each fixed normalized gap $0<g\le1$, the choice
\begin{equation}
k=\max\{1,\lceil\log_2(1/g)\rceil\}
\label{eq:supp-sufficient-block-size}
\end{equation}
satisfies $2^k\ge1/g$ and hence $g>1/(2^k+1)$.
It has $k=O(1+\log(1/g))$ and is held fixed as $m$ grows.
Thus a sufficiently large fixed block size gives
$\Nexp,\Nmax=\Theta(\log(1/\delta))$ in each fixed confidence range above.
\section{Sharp decay and inversion of the sequence of step points}
\label{sec:supp-asymptotics}

Fix a real $d>0$ throughout this section and set $\alpha=(d+1)^{-1}\in(0,1)$.
We first derive the sharp decay of $\gamma_r(d)$ from
Wiener interpolation at a fixed repeated node. For an integer $r\ge0$, put $s=r+1$ and consider
power series $F(z)=\sum_{w\ge0}F_wz^w$ with absolutely summable coefficients.
Require $F_0=\alpha^s$ and $F^{(k)}(\alpha)=0$ for $0\le k<s$.
Over these series, define
\[
\phi_s(\alpha)=\inf_F\sum_{w\ge1}|F_w|.
\]
This is the extremum in Eq.~(3.1) of Ref.~\cite{SzehrZarouf2021}. Its Theorem~2 gives
$\phi_s(\alpha)\ge c_\alpha\sqrt{s}$; the classical Sch\"affer bound
recalled there gives $\phi_s(\alpha)\le\sqrt{es}$.

To relate this extremum to $B_r(d)$, choose a finite optimal dual
representation from Theorem~\ref{thm:supp-extremal-duality}. Write
$\mu_w$ for its weight at $w$, taking $\mu_w=0$ at all unselected
positive integers, and $\mu_\infty$ for its weight at infinity. Define
\[
F(z)=\alpha^s\left(1-\sum_{w\ge1}\mu_wz^w\right).
\]
This series has constant term $\alpha^s$, and the binomial moments in
Eq.~\eqref{eq:supp-dual-moments} give a zero of order at least $s$ at $\alpha$.
The mass identity \eqref{eq:supp-dual-mass} gives $F(1)=\alpha^s\mu_\infty$,
so the total variation of the chosen weights satisfies
\[
\alpha^{-s}\left(\sum_{w\ge1}|F_w|+|F(1)|\right)
=\sum_{w\ge1}|\mu_w|+|\mu_\infty|=B_r(d).
\]

Conversely, take any feasible series $F$ with real coefficients and set
$\mu_w=-\alpha^{-s}F_w$ for $w\ge1$ and $\mu_\infty=\alpha^{-s}F(1)$.
Absolute convergence at $1$ and $F_0=\alpha^s$ give total mass one.
Multiplying $F^{(k)}(\alpha)=0$ by $\alpha^k/k!$ gives the moments
in Eq.~\eqref{eq:supp-dual-moments}, now summed over all positive integers.
The polynomials $\binom{w}{k}$ for $0\le k\le r$ form a basis for
polynomials of degree at most $r$, so $\sum_{w\ge1}\mu_w\alpha^w R(w)=R(0)$
for every such $R$. Together with the mass identity this gives
\[
G(0)=\sum_{w\ge1}\mu_wG(w)+\mu_\infty G(\infty)
\qquad\text{for every }G\in V_r(d).
\]
Taking the supremum over $\|G\|_{\mathcal X}\le1$ gives
$B_r(d)\le\alpha^{-s}(\sum_{w\ge1}|F_w|+|F(1)|)$.
The series constructed from the finite optimal dual attains equality, so
\[
B_r(d)=\alpha^{-s}\inf_F
 \left(\sum_{w\ge1}|F_w|+|F(1)|\right).
\]
Here the constraints on $F$ are the same as for $\phi_s(\alpha)$.
Since $\alpha$ and $F_0$ are real, taking real parts of the coefficients
preserves feasibility. Neither $\sum_{w\ge1}|F_w|$ nor $|F(1)|$ increases,
so both infima are the same over real and complex coefficients.
The triangle inequality gives $|F(1)|\le\alpha^s+\sum_{w\ge1}|F_w|$, hence
\[
\alpha^{-s}\phi_s(\alpha)\le B_r(d)
 \le2\alpha^{-s}\phi_s(\alpha)+1.
\]
Thus Eq.~\eqref{eq:supp-envelope-equivalence} transfers the known
interpolation bounds to the sharp order of $\gamma_r(d)$.
We give a self-contained proof below, combining summable extrapolation
weights with a normalized polynomial construction, and then derive the
strict inversion used for the certification exponent.

\subsection{A summable extrapolation dual}

To construct the weights, define the Blaschke factor and its coefficients by
\begin{equation}
b_\alpha(z)=\frac{\alpha-z}{1-\alpha z},
\qquad
b_\alpha(z)^s=\sum_{w=0}^{\infty}c_wz^w,
\qquad s\ge1.
\label{eq:supp-blaschke-series}
\end{equation}
Here $s$ is an integer. The coefficients are real, $c_0=\alpha^s$,
and $|b_\alpha(e^{it})|=1$. The only pole is at $z=\alpha^{-1}>1$.
For any $1<R<\alpha^{-1}$, Cauchy's coefficient estimate on $|z|=R$ gives
$|c_w|\le C_{R,s}R^{-w}$. Thus the series and each of its fixed-order
derivatives converge absolutely and uniformly on the closed unit disk.

\begin{samepage}
\begin{lemma}
\label{lem:supp-blaschke-coefficients}
The coefficients in Eq.~\eqref{eq:supp-blaschke-series} satisfy
\begin{equation}
\sum_{w\ge0}|c_w|^2=1,
\qquad
\sum_{w\ge0}w^2|c_w|^2
=s^2\frac{1+\alpha^2}{1-\alpha^2}.
\label{eq:supp-blaschke-moments}
\end{equation}
Moreover,
\begin{equation}
\sum_{w\ge0}|c_w|\le D_d\sqrt s,
\qquad D_d=\sqrt{\frac{2+\pi}{1-\alpha^2}}.
\label{eq:supp-blaschke-l1}
\end{equation}
\end{lemma}
\end{samepage}

\begin{proof}
Parseval's identity with normalized measure $dt/(2\pi)$ gives the
first equality in Eq.~\eqref{eq:supp-blaschke-moments}. Differentiation gives
\[
b_\alpha'(z)=\frac{\alpha^2-1}{(1-\alpha z)^2},
\qquad
|b_\alpha'(e^{it})|
=\frac{1-\alpha^2}{|1-\alpha e^{it}|^2}=:P_\alpha(t).
\]
The boundary derivative of $b_\alpha^s$ therefore has modulus
$sP_\alpha(t)$. The absolutely convergent Poisson-kernel expansion
$P_\alpha(t)=1+2\sum_{k\ge1}\alpha^k\cos(kt)$ yields
\[
\frac1{2\pi}\int_0^{2\pi}P_\alpha(t)^2\,dt
=1+2\sum_{k\ge1}\alpha^{2k}
=\frac{1+\alpha^2}{1-\alpha^2}.
\]
Parseval applied to the differentiated series proves the second equality.

Apply weighted Cauchy--Schwarz to finite partial sums and pass to the
limit by monotone convergence. Using Eq.~\eqref{eq:supp-blaschke-moments},
we obtain
\[
\begin{aligned}
\left(\sum_{w\ge0}|c_w|\right)^2
&\le
\left(\sum_{w\ge0}\left(1+\frac{w^2}{s^2}\right)|c_w|^2\right)
\left(\sum_{w\ge0}\frac1{1+(w/s)^2}\right)\\
&\le \frac{2}{1-\alpha^2}
\left(1+\int_0^\infty\frac{dx}{1+(x/s)^2}\right)\\
&=\frac{2}{1-\alpha^2}\left(1+\frac{\pi s}{2}\right)
\le\frac{2+\pi}{1-\alpha^2}s.
\end{aligned}
\]
The integral comparison uses the decreasing integrand on $[0,\infty)$;
the last inequality uses $s\ge1$.
\end{proof}

\begin{proposition}
\label{prop:supp-sharp-lower}
For every integer $r\ge0$,
\begin{equation}
\gamma_r(d)\ge
\frac{2\alpha^{r+1}}{1+D_d\sqrt{r+1}}.
\label{eq:supp-sharp-lower}
\end{equation}
\end{proposition}

\begin{proof}
Set $s=r+1$ in Eq.~\eqref{eq:supp-blaschke-series}.
The function $b_\alpha^s$ has a zero of order $s$ at $\alpha$.
By induction, $(z\,d/dz)^k$ is a linear combination of ordinary
derivatives of orders at most $k$, with polynomial coefficients.
Consequently
\[
\left.(z\,d/dz)^k b_\alpha(z)^s\right|_{z=\alpha}=0,
\qquad 0\le k<s.
\]
The coefficient bound preceding Lemma~\ref{lem:supp-blaschke-coefficients}
justifies termwise differentiation and absolute convergence of the sums below.
Applying these identities to the monomials of a real polynomial $P$
of degree at most $r$ gives
\begin{equation}
\sum_{w\ge1}c_w\alpha^wP(w)=-\alpha^sP(0).
\label{eq:supp-infinite-dual-moments}
\end{equation}

Define weights on $\mathcal X$ by
\[
\mu_w=-\alpha^{-s}c_w\quad(w\ge1),
\qquad \mu_\infty=(-1)^s\alpha^{-s}.
\]
Since $b_\alpha(1)^s=(-1)^s$, summing the series at one gives
\[
\sum_{w\ge1}\mu_w+\mu_\infty
=-\alpha^{-s}\bigl((-1)^s-\alpha^s\bigr)
+(-1)^s\alpha^{-s}=1.
\]
For $G(x)=h+\alpha^xP(x)\in V_r(d)$, this mass identity and
Eq.~\eqref{eq:supp-infinite-dual-moments} imply
\begin{equation}
G(0)=\sum_{w\ge1}\mu_wG(w)+\mu_\infty G(\infty).
\label{eq:supp-infinite-extrapolation}
\end{equation}
These comparison weights are absolutely summable. Their total variation is
\begin{equation}
\begin{aligned}
\|\mu\|_{\mathrm{TV}}
&:=\sum_{w\ge1}|\mu_w|+|\mu_\infty|\\
&=\alpha^{-s}\left(1+\sum_{w\ge0}|c_w|\right)-1
\le \alpha^{-s}(1+D_d\sqrt s)-1,
\end{aligned}
\label{eq:supp-infinite-dual-variation}
\end{equation}
where the subtraction uses $c_0=\alpha^s>0$.
For $G\in V_r(d)$ with $\|G\|_{\mathcal X}\le1$,
Eq.~\eqref{eq:supp-infinite-extrapolation} gives
$G(0)\le\|\mu\|_{\mathrm{TV}}$. Hence
$1+B_r(d)\le \alpha^{-s}(1+D_d\sqrt s)$.
Substitution into Eq.~\eqref{eq:supp-envelope-equivalence} proves
Eq.~\eqref{eq:supp-sharp-lower}.
\end{proof}

\subsection{A polynomial with a matching bound}

The candidate polynomial will come from the positive-index coefficients
of a rational function. The following identity gives the polynomial's
value at zero, which we need for normalization.

\begin{lemma}
\label{lem:supp-rational-tail}
Let $r\ge0$ be an integer and let $N$ be a real polynomial of degree at
most $r+1$. Write
\[
F(z)=\frac{N(z)}{(1-\alpha z)^{r+1}}
=\sum_{w\ge0}f_wz^w.
\]
There is a real polynomial $Q$, of degree at most $r$, such that
\begin{equation}
f_w=\alpha^wQ(w)\quad(w\ge1),
\qquad Q(0)=F(0)-F(\infty),
\label{eq:supp-rational-tail}
\end{equation}
where $F(\infty)=\lim_{x\to+\infty}F(x)$ is the finite real limit
of this rational function.
\end{lemma}

\begin{proof}
Polynomial division and partial fractions give real constants
$L,A_1,\ldots,A_{r+1}$ such that
\[
F(z)=L+\sum_{a=1}^{r+1}\frac{A_a}{(1-\alpha z)^a},
\qquad L=F(\infty).
\]
For $w\ge1$, the constant term $L$ contributes no coefficient,
and expanding the remaining terms gives
\[
f_w=\alpha^w\sum_{a=1}^{r+1}A_a\binom{w+a-1}{a-1}.
\]
Define $Q(w)$ by the sum on the right. Each binomial polynomial has
degree $a-1$ and value one at zero. Thus $\deg Q\le r$ and
$Q(0)=\sum_a A_a=F(0)-L$.
\end{proof}

To bound the coefficients uniformly, we use an oscillatory integral
estimate whose amplitude vanishes with the curvature of the phase.

\begin{lemma}
\label{lem:supp-oscillatory-bound}
Let $A:[0,2\pi]\to\mathbb C$ be $C^1$ and
$\theta:[0,2\pi]\to\mathbb R$ be $C^2$.
Suppose $\theta'$ is monotone on each of $[0,\pi]$ and $[\pi,2\pi]$,
with either direction allowed on each interval, and
$|A(t)|\le C|\theta''(t)|$ for some $C\ge0$.
Set $S=\sup_{0\le t\le2\pi}|A(t)|$ and
$V=\int_0^{2\pi}|A'(t)|\,dt$.
For all real $m\ge1$ and $w\in\mathbb R$,
\begin{equation}
\left|\frac1{2\pi}\int_0^{2\pi}
A(t)e^{i(m\theta(t)-wt)}\,dt\right|
\le\frac{4C+12S+V}{2\pi\sqrt m}.
\label{eq:supp-oscillatory-bound}
\end{equation}
\end{lemma}

\begin{proof}
Put $f=m\theta'-w$ and $E(t)=e^{i(m\theta(t)-wt)}$, so that
$E'=ifE$ and $|A|\le(C/m)|f'|$.
For $\rho>0$, the regularized reciprocal
\[
R_\rho(x)=\frac{x}{i(x^2+\rho^2)}
\]
satisfies
\[
|R_\rho(x)|\le\frac1{2\rho},
\qquad
|R_\rho'(x)|=\frac{|\rho^2-x^2|}{(x^2+\rho^2)^2}
\le\frac1{x^2+\rho^2}.
\]
On either half-interval $[a,b]$, the function $f$ is monotone.
The fundamental theorem of calculus applied to $\arctan(f/\rho)$ gives
\[
\int_a^b\frac{|f'(t)|}{f(t)^2+\rho^2}\,dt
=\frac1\rho\left|\arctan\frac{f(b)}\rho
-\arctan\frac{f(a)}\rho\right|
\le\frac\pi\rho.
\]
Thus $J_\rho:=\int_0^{2\pi}|f'|/(f^2+\rho^2)\,dt\le2\pi/\rho$.

The identity $AE=A\rho^2E/(f^2+\rho^2)+A R_\rho(f)E'$
allows integration by parts over the whole interval:
\begin{equation}
\begin{aligned}
\int_0^{2\pi}AE\,dt
={}&\int_0^{2\pi}\frac{A\rho^2E}{f^2+\rho^2}\,dt
+[A R_\rho(f)E]_0^{2\pi}\\
&-\int_0^{2\pi}
\bigl(A'R_\rho(f)+A R_\rho'(f)f'\bigr)E\,dt.
\end{aligned}
\label{eq:supp-regularized-parts}
\end{equation}
All terms are continuous, since $\rho>0$. The first integral is bounded
in modulus by $(C\rho^2/m)J_\rho$, the two endpoint terms by $S/\rho$,
and the last integral by $V/(2\rho)+SJ_\rho$. Therefore
\begin{equation}
\begin{aligned}
\left|\int_0^{2\pi}AE\,dt\right|
&\le\left(\frac{C\rho^2}{m}+S\right)J_\rho
+\frac S\rho+\frac V{2\rho}\\
&\le\frac{2\pi C\rho}{m}
+\frac{(2\pi+1)S}{\rho}+\frac V{2\rho}.
\end{aligned}
\label{eq:supp-regularized-bound}
\end{equation}
Take $\rho=5\sqrt m/8$. Using $\pi<16/5$ and $C,S,V\ge0$ gives
\[
\left|\int_0^{2\pi}AE\,dt\right|
\le\frac{(5\pi/4)C+(8/5)(2\pi+1)S+(4/5)V}{\sqrt m}
\le\frac{4C+12S+V}{\sqrt m}.
\]
Dividing by $2\pi$ proves Eq.~\eqref{eq:supp-oscillatory-bound}.
The arctangent variation bound is independent of the shift $w$.
\end{proof}

\begin{proposition}
\label{prop:supp-sharp-upper}
There is a constant $K_d>0$, depending only on $d$, such that for every
integer $r\ge2$,
\begin{equation}
\gamma_r(d)\le
\frac{2K_d}{\sqrt{r-1}\,(\alpha^{-r-1}+\alpha^{r-1})}.
\label{eq:supp-sharp-upper}
\end{equation}
\end{proposition}

\begin{proof}
Consider
\begin{equation}
F_r(z)=\frac{1-z^2}{(1-\alpha z)^2}b_\alpha(z)^{r-1}
=\frac{(1-z^2)(\alpha-z)^{r-1}}{(1-\alpha z)^{r+1}}
=\sum_{w\ge0}f_wz^w.
\label{eq:supp-upper-generating-function}
\end{equation}
Lemma~\ref{lem:supp-rational-tail} gives a real polynomial $Q_r$ of
degree at most $r$ with $f_w=\alpha^wQ_r(w)$ for $w\ge1$.
The numerator and denominator in
Eq.~\eqref{eq:supp-upper-generating-function} have leading coefficients
$(-1)^r$ and $(-1)^{r+1}\alpha^{r+1}$, respectively. Hence
\begin{gather}
F_r(0)=\alpha^{r-1},\qquad F_r(\infty)=-\alpha^{-r-1},
\label{eq:supp-candidate-limits}\\
Q_r(0)=\alpha^{r-1}+\alpha^{-r-1}>0.
\label{eq:supp-candidate-normalization}
\end{gather}
It follows that $\widetilde p_r=Q_r/Q_r(0)$ belongs to $\mathcal P_r$.

For the coefficient estimate, put
$D_\alpha(t)=1-2\alpha\cos t+\alpha^2$ and define the smooth real phase
\begin{equation}
\theta(t)=\pi+\int_0^t\frac{1-\alpha^2}{D_\alpha(u)}\,du,
\qquad 0\le t\le2\pi.
\label{eq:supp-continuous-phase}
\end{equation}
Direct differentiation shows that
$(d/dt)b_\alpha(e^{it})=i\theta'(t)b_\alpha(e^{it})$.
Thus $b_\alpha(e^{it})e^{-i\theta(t)}$ has derivative zero and initial
value $b_\alpha(1)e^{-i\pi}=1$. This proves
$b_\alpha(e^{it})=e^{i\theta(t)}$ throughout the interval. Moreover,
\begin{equation}
\theta'(t)=\frac{1-\alpha^2}{D_\alpha(t)},
\qquad
\theta''(t)=-\frac{2\alpha(1-\alpha^2)\sin t}{D_\alpha(t)^2}.
\label{eq:supp-phase-derivatives}
\end{equation}
Consequently $\theta'$ decreases on $[0,\pi]$ and increases on
$[\pi,2\pi]$.

Set
\[
A(t)=\frac{1-e^{2it}}{(1-\alpha e^{it})^2},
\qquad C_\alpha^{\mathrm{amp}}
=\frac{(1+\alpha)^2}{\alpha(1-\alpha^2)}>0.
\]
Since $D_\alpha(t)\le(1+\alpha)^2$, we have for every $t$,
\[
|A(t)|=\frac{2|\sin t|}{D_\alpha(t)}
=\frac{D_\alpha(t)}{\alpha(1-\alpha^2)}|\theta''(t)|
\le C_\alpha^{\mathrm{amp}}|\theta''(t)|.
\]
Since $D_\alpha(t)=|1-\alpha e^{it}|^2\ge(1-\alpha)^2>0$, $A$ is smooth.
In particular, $S_\alpha=\sup_{[0,2\pi]}|A|$ and
$V_\alpha=\int_0^{2\pi}|A'|\,dt$ are finite. Define
\begin{equation}
K_d=\frac{4C_\alpha^{\mathrm{amp}}+12S_\alpha+V_\alpha}{2\pi}.
\label{eq:supp-coefficient-constant}
\end{equation}
The function $F_r$ is analytic on a neighborhood of the closed unit disk,
so its uniformly convergent boundary series gives its coefficients by
Fourier integration. Lemma~\ref{lem:supp-oscillatory-bound}, with
$m=r-1$, yields
\begin{equation}
|f_w|
=\left|\frac1{2\pi}\int_0^{2\pi}
A(t)e^{i((r-1)\theta(t)-wt)}\,dt\right|
\le\frac{K_d}{\sqrt{r-1}},\qquad w\ge0.
\label{eq:supp-upper-coefficients}
\end{equation}

For positive integers $w$, Eqs.~\eqref{eq:supp-candidate-normalization}
and \eqref{eq:supp-upper-coefficients} now give
\begin{equation}
|\alpha^w\widetilde p_r(w)|\le M_r,
\qquad M_r=\frac{K_d}{\sqrt{r-1}\,(\alpha^{-r-1}+\alpha^{r-1})}.
\label{eq:supp-candidate-envelope-bound}
\end{equation}
Because $\alpha^w\widetilde p_r(w)\to0$, its envelope satisfies
\[
-M_r\le\ell_d(\widetilde p_r)\le0
\le u_d(\widetilde p_r)\le M_r.
\]
In particular, $1-\ell_d(\widetilde p_r)\ge1$, and the definition of
$\gamma_r(d)$ gives
\[
\gamma_r(d)\le
\frac{u_d(\widetilde p_r)-\ell_d(\widetilde p_r)}
{1-\ell_d(\widetilde p_r)}\le2M_r.
\]
This is Eq.~\eqref{eq:supp-sharp-upper}.
\end{proof}

\subsection{Strict inversion}
\label{subsec:supp-inversion}

\begin{proposition}
\label{prop:supp-sharp-decay}
For every fixed real $d>0$, there are constants $c_d,C_d>0$ such that
\begin{equation}
c_d\frac{(d+1)^{-r}}{\sqrt r}
\le\gamma_r(d)\le C_d\frac{(d+1)^{-r}}{\sqrt r},
\qquad r\ge2.
\label{eq:supp-sharp-decay}
\end{equation}
Consequently,
\begin{equation}
\gamma_r(d)\longrightarrow0,
\qquad
\lim_{r\to\infty}\gamma_r(d)^{1/r}=(d+1)^{-1}.
\label{eq:supp-sharp-root-limit}
\end{equation}
\end{proposition}

\begin{proof}
For $r\ge2$, we have $\sqrt{r+1}\le\sqrt{3r/2}$ and
$\sqrt{r-1}\ge\sqrt{r/2}$. Propositions~\ref{prop:supp-sharp-lower}
and \ref{prop:supp-sharp-upper} therefore give
Eq.~\eqref{eq:supp-sharp-decay}, for example with
$c_d=2\alpha/(1+D_d\sqrt{3/2})$ and $C_d=2\sqrt2\,\alpha K_d$.
The upper bound tends to zero. Taking logarithms in the two-sided bound
and dividing by $r$ gives
$r^{-1}\log\gamma_r(d)=-\log(d+1)+O_d((\log r)/r)$,
which proves the root limit.
\end{proof}

\begin{proof}[Proof of Theorem~\ref{thm:sharp-hierarchy}]
\phantomsection\label{proof:supp-sharp-hierarchy}
For integer $d\ge2$, Eq.~\eqref{eq:supp-sharp-decay} gives Eq.~\eqref{eq:sharp-order}.
For the inversion, fix any real $d>0$. Since $\gamma_j(d)\to0$, the set
in Eq.~\eqref{eq:supp-critical-degree} is nonempty for every $g>0$.
Put $k=r_d(g)$ and $L=\log_{d+1}(1/g)$.
By the positivity and strict decrease established in
Proposition~\ref{prop:supp-critical-sequence}, $k\to\infty$ as
$g\downarrow0$: for each fixed $j$, $g\le\gamma_j(d)$ excludes
all indices at most $j$. For sufficiently small $g$, we thus have
$k\ge3$ and
\begin{equation}
\gamma_k(d)<g\le\gamma_{k-1}(d).
\label{eq:supp-strict-crossing}
\end{equation}
Use the lower bound in Eq.~\eqref{eq:supp-sharp-decay} for $\gamma_k(d)$
and the upper bound for $\gamma_{k-1}(d)$. Taking logarithms to base $d+1$ gives
\begin{equation}
\begin{aligned}
k+\frac12\log_{d+1} k&>L+\log_{d+1} c_d,\\
k-1+\frac12\log_{d+1}(k-1)&\le L+\log_{d+1} C_d.
\end{aligned}
\label{eq:supp-inversion-inequalities}
\end{equation}
The difference between the two left sides is
$1+\tfrac12\log_{d+1}(k/(k-1))$, which is bounded for $k\ge3$.
It follows that
\begin{equation}
k+\frac12\log_{d+1} k=L+O_d(1).
\label{eq:supp-implicit-inversion}
\end{equation}
First this yields $k\le L+O_d(1)$. Substituting that upper bound into
the logarithm gives $k\ge L-O_d(\log L)$, so
$k=L+O_d(\log L)$ and $k/L\to1$. Hence
$\log_{d+1} k=\log_{d+1} L+o(1)$, and Eq.~\eqref{eq:supp-implicit-inversion}
becomes
\begin{equation}
r_d(g)=L-\frac12\log_{d+1} L+O_d(1),
\qquad L=\log_{d+1}(1/g),\quad g\downarrow0.
\label{eq:supp-strict-inversion}
\end{equation}
At $g=\gamma_j(d)$, strict decrease excludes every index at most $j$
and includes $j+1$ in the defining set, so $r_d(g)=j+1$.
For integer $d\ge2$, Eq.~\eqref{eq:supp-strict-inversion} gives Eq.~\eqref{eq:degree-inversion}.
\end{proof}
\section{Exact computation and finite-degree certificates}
\label{sec:supp-exact-certificates}

The finite alternating structure of Sec.~\ref{sec:supp-duality} turns
optimality on the infinite integer domain into a finite certificate.
We use these certificates to obtain an exact algorithm and explicit
low-degree values.

\subsection{Finite rational certificates}

Fix a rational number $d>0$ and an integer $r\ge0$. Given positive integers
$j_1<\cdots<j_{r+1}$, set $h=(-1)^{r+1}$ and construct
$G(x)=h+(d+1)^{-x}R(x)\in V_r(d)$ by imposing
\begin{equation}
G(j_i)=(-1)^{i-1}\quad(1\le i\le r+1),
\qquad G(\infty)=h.
\label{eq:supp-certificate-contacts}
\end{equation}
In the basis $1,(d+1)^{-x},(d+1)^{-x}x,\ldots,(d+1)^{-x}x^r$, the evaluation
matrix at these nodes and $\infty$ has rational entries.
Lemma~\ref{lem:supp-interpolation} makes it invertible, so exact linear
algebra uniquely determines the rational coefficients of $R$.

The transposed evaluation system gives rational weights
$\mu_1,\ldots,\mu_{r+1},\mu_\infty$ such that
\begin{equation}
H(0)=\sum_{i=1}^{r+1}\mu_iH(j_i)+\mu_\infty H(\infty)
\qquad(H\in V_r(d)).
\label{eq:supp-certificate-representation}
\end{equation}
By Lemma~\ref{lem:supp-interpolation-signs}, their signs agree with
the contact values in Eq.~\eqref{eq:supp-certificate-contacts}.
Writing $\beta=G(0)$ and applying the representation to the constant
function and to $G$ gives
\begin{equation}
\sum_{i=1}^{r+1}\mu_i+\mu_\infty=1,
\qquad
\beta=\sum_{i=1}^{r+1}|\mu_i|+|\mu_\infty|>1.
\label{eq:supp-certificate-value}
\end{equation}
At least one weight is negative: it is the infinity weight for $r=0$
and the second finite weight for $r\ge1$. This proves the strict
inequality. Since $R(0)=\beta-h>0$, the normalized candidate
\begin{equation}
p(x)=\frac{R(x)}{\beta-h}
\label{eq:supp-certificate-normalization}
\end{equation}
satisfies $p(0)=1$ and $\deg p\le r$. To control its tail, define
$Q=(-1)^rp$ and $S(x)=(d+1)Q(x)-Q(x+1)$.

\begin{lemma}
\label{lem:supp-finite-certificate}
For a candidate constructed above, let $M\ge j_{r+1}$ be an integer.
Suppose $|G(w)|\le1$ for every integer $1\le w\le M$, and all
coefficients of $Q(M+t)$ and $S(M+t)$ are nonnegative. Then
\begin{equation}
B_r(d)=\beta,
\qquad \gamma_r(d)=\frac{2}{1+\beta},
\label{eq:supp-certified-critical-value}
\end{equation}
and $p$ is the unique optimal polynomial in $\mathcal P_r$.
\end{lemma}

\begin{proof}
The coefficient conditions give $Q(x),S(x)\ge0$ for all real $x\ge M$.
For integers $w\ge M$, put $a_w=(d+1)^{-w}Q(w)$. Then
\begin{equation}
a_w-a_{w+1}=(d+1)^{-(w+1)}S(w)\ge0,
\qquad 0\le a_w\le a_M,
\qquad a_w\longrightarrow0.
\label{eq:supp-certificate-tail}
\end{equation}
The limit holds because $Q$ is a polynomial and $d+1>1$. Since
$G(w)=h+(\beta-h)(-1)^ra_w$, its tail values lie between $G(M)$ and
$h$, both in $[-1,1]$. The checked prefix covers the remaining
positive integers, and $G(\infty)=h$, so $\|G\|_{\mathcal X}\le1$.

For every $H\in V_r(d)$ with $\|H\|_{\mathcal X}\le1$,
Eq.~\eqref{eq:supp-certificate-representation} bounds $H(0)$ by
the total variation in Eq.~\eqref{eq:supp-certificate-value}.
The feasible candidate attains this bound, proving $B_r(d)=\beta$.
Proposition~\ref{prop:supp-envelope-equivalence} gives the step
point in Eq.~\eqref{eq:supp-certified-critical-value} and identifies
Eq.~\eqref{eq:supp-certificate-normalization} as an optimal polynomial.
Its uniqueness follows from Proposition~\ref{prop:supp-critical-sequence}.
\end{proof}

A certificate records the integer nodes and $M$, the rational
coefficients of $R$, and the rational weights. The contact equations,
normalization, signs and total variation are finite checks. The
extrapolation identity is verified on a basis: in addition to the
total-mass identity in Eq.~\eqref{eq:supp-certificate-value}, check
\begin{equation}
\sum_{i=1}^{r+1}\mu_i(d+1)^{-j_i}j_i^k
=\mathbf 1_{\{k=0\}}
\qquad(0\le k\le r).
\label{eq:supp-certificate-moments}
\end{equation}
The decaying monomials vanish at $\infty$; the total mass includes
the weight at $\infty$. Linearity then gives the identity for every $H\in V_r(d)$.
Finally, verify that all coefficients of $Q(M+t)$ and $S(M+t)$ are nonnegative
and that $\lvert G(w)\rvert\le1$ for every integer $1\le w\le M$.
These operations use only finite rational arithmetic and integer powers.

\subsection{A terminating exact search}

\begin{lemma}
\label{lem:supp-certificate-existence}
For every rational $d>0$ and integer $r\ge0$, the optimal polynomial
has a certificate as in Lemma~\ref{lem:supp-finite-certificate}.
\end{lemma}

\begin{proof}
Theorem~\ref{thm:supp-extremal-duality} supplies an optimal function
with the contacts in Eq.~\eqref{eq:supp-certificate-contacts}.
Proposition~\ref{prop:supp-critical-sequence} and the rational
evaluation systems give its rational polynomial and weights.
Let $p$ be its normalization and put $Q=(-1)^rp$ as above.
This polynomial is nonzero because $p(0)=1$. Its leading coefficient
must be positive. Otherwise $Q(w)<0$ for all sufficiently large
integers, and
\[
G(w)=h\bigl[1-(\beta-h)(d+1)^{-w}Q(w)\bigr]
\]
would have absolute value greater than one.

Write $s=\deg Q$. The polynomial $S=(d+1)Q-Q(\,\cdot+1)$ also has
degree $s$, with leading coefficient $d$ times that of $Q$.
For $0\le k\le s$, both $Q^{(k)}$ and $S^{(k)}$ have positive
leading coefficients and are positive at every sufficiently large
real argument. The shifted coefficients are
\begin{equation}
[t^k]Q(M+t)=\frac{Q^{(k)}(M)}{k!},
\qquad
[t^k]S(M+t)=\frac{S^{(k)}(M)}{k!}.
\label{eq:supp-certificate-shifts}
\end{equation}
Choose an integer $M\ge j_{r+1}$ large enough for all these finitely
many derivatives to be positive. Coefficients above degree $s$ vanish.
For $s=0$, the two polynomials are positive constants, so the same
conclusion holds. The optimal function already satisfies every prefix
inequality, giving the required certificate.
\end{proof}

\begin{proposition}
\label{prop:supp-exact-computability}
For every rational $d>0$ and integer $r\ge0$, a terminating exact
algorithm returns $\gamma_r(d)$, its unique normalized optimal
polynomial, and a finite certificate.
\end{proposition}

\begin{proof}
Enumerate complete candidate records, including the tail cutoff $M$.
One explicit enumeration uses stages $N=1,2,\ldots$. At stage $N$,
list all increasing positive integer node tuples and integers $M$ with
$j_{r+1}\le M\le N$, together with all arrays of $r+1$ polynomial
coefficients and $r+2$ weights whose rational entries have numerator
of absolute value at most $N$ and positive denominator at most $N$.
Set $h=(-1)^{r+1}$ and perform the finite checks following
Lemma~\ref{lem:supp-finite-certificate}, rejecting records that fail
any check. Return the first passing record, with
$p=R/(\beta-h)$ and $\gamma_r(d)=2/(1+\beta)$.

Each stage contains finitely many records, and checking any record
terminates. Every fixed finite certificate appears at a finite stage:
choose $N$ larger than its nodes, $M$, and all numerator and denominator
bounds. Lemma~\ref{lem:supp-certificate-existence} provides such a
certificate, so a passing record is eventually reached.
Lemma~\ref{lem:supp-finite-certificate} proves the correctness of the
returned value and polynomial.
\end{proof}

\subsection{Low-degree values and the qubit specialization}

We first derive explicit low-degree formulas for real $d$ in the ranges
below, then illustrate the finite-certificate construction for qubits at $d=2$.

\begin{proposition}
\label{prop:supp-low-degree-values}
For real $d$ in the indicated ranges,
\begin{equation}
\begin{aligned}
\gamma_0(d)&=\frac{1}{d+1} &&(d>0),\\
\gamma_1(d)&=\frac{1}{(d+1)^2+1} &&(d\ge1),\\
\gamma_2(d)&=\frac{1}{(d+1)((d+1)^2+3)} &&(d\ge3).
\end{aligned}
\label{eq:supp-low-degree-values}
\end{equation}
For qubit targets,
\begin{equation}
\gamma_2(2)=\frac1{35}.
\label{eq:supp-qubit-quadratic-value}
\end{equation}
\end{proposition}

\begin{proof}
The degree-zero value is Eq.~\eqref{eq:supp-envelope-equivalence}.
For degree one, take $p(w)=1-w$. If $d\ge1$, the nonnegative
sequence $(w-1)(d+1)^{-w}$, $w\ge2$, is nonincreasing, since its
successive ratio is $w/[(d+1)(w-1)]\le2/(d+1)\le1$.
Thus $p$ has envelope $[-(d+1)^{-2},0]$ and objective $1/[(d+1)^2+1]$.
The extrapolation identities below follow from Lagrange interpolation
of the polynomial part and matching the constant term. For every
$G\in V_1(d)$,
\begin{equation}
G(0)=2(d+1)G(1)-(d+1)^2G(2)+d^2G(\infty).
\label{eq:supp-linear-extrapolation}
\end{equation}
The coefficient total variation is $2(d+1)^2+1$, so $B_1(d)\le2(d+1)^2+1$ and
Eq.~\eqref{eq:supp-envelope-equivalence} gives the matching lower bound
$\gamma_1(d)\ge1/[(d+1)^2+1]$.

For $d\ge3$, consider
\begin{equation}
p(w)=\frac{(w-2)(w-\xi)}{2\xi},\qquad
\xi=\frac{(d+1)^2+3}{(d+1)^2+1}.
\label{eq:supp-qudit-quadratic}
\end{equation}
It satisfies $p(0)=1$ and
\[
(d+1)^{-1}p(1)=(d+1)^{-3}p(3)=\frac{1}{(d+1)((d+1)^2+3)},
\qquad p(2)=0.
\]
The second root satisfies $1<\xi<2$.
For $w\ge3$, both factors in
\[
\frac{p(w+1)}{p(w)}
=\frac{w-1}{w-2}\frac{w+1-\xi}{w-\xi}
\]
are positive and decreasing. At $w=3$ their product is
$3+(d+1)^{-2}\le d+1$. The weighted positive tail is therefore
nonincreasing. Thus $\ell_d(p)=0$ and $u_d(p)=(d+1)^{-1}p(1)$.
For every $G\in V_2(d)$, the matching extrapolation identity is
\begin{equation}
G(0)=3(d+1)G(1)-3(d+1)^2G(2)+(d+1)^3G(3)-d^3G(\infty).
\label{eq:supp-quadratic-extrapolation}
\end{equation}
The coefficient total variation is $2(d+1)^3+6(d+1)-1$.
Substitution into Eq.~\eqref{eq:supp-envelope-equivalence} gives the
matching lower bound on $\gamma_2(d)$.

At $d=2$, take
\begin{equation}
p(w)=\frac{(w-2)(29w-35)}{70}.
\label{eq:supp-qubit-quadratic}
\end{equation}
Here $p(0)=1$, and the first four weighted values are
\[
\bigl(3^{-1}p(1),3^{-2}p(2),3^{-3}p(3),3^{-4}p(4)\bigr)
=\left(\frac1{35},0,\frac{26}{945},\frac1{35}\right).
\]
The roots are $2$ and $35/29$. The same two-factor ratio is decreasing
for $w\ge4$ and equals $55/27<3$ at $w=4$. Thus the weighted tail
decreases from its value at $4$, giving the envelope $[0,1/35]$.
Interpolation gives, for every $G\in V_2(2)$,
\begin{equation}
G(0)=8G(1)-18G(2)+27G(4)-16G(\infty).
\label{eq:supp-qubit-quadratic-extrapolation}
\end{equation}
The coefficient total variation is $69$, so
Eq.~\eqref{eq:supp-envelope-equivalence} gives $\gamma_2(2)\ge1/35$,
matching the candidate above.
\end{proof}

For a higher-degree example at $d=2$, the finite nodes $1,2,3,4,7$
give
\begin{equation}
p_4(w)=\frac{(w-2)(w-4)(143w^2-782w+654)}{5232}.
\label{eq:supp-quartic-polynomial}
\end{equation}
It has $p_4(0)=1$. Writing $c_w=3^{-w}p_4(w)$, its prefix is
\begin{equation}
(c_1,\ldots,c_7)
=\left(\frac5{1744},0,\frac5{1744},0,
\frac{319}{423792},\frac{185}{79461},\frac5{1744}\right).
\label{eq:supp-quartic-prefix}
\end{equation}
All these values lie in $[0,5/1744]$. At $M=7$ the two shifted
polynomials are
\begin{equation}
\begin{gathered}
5232p_4(7+t)=143t^4+2364t^3+14092t^2+35796t+32805,\\
5232\bigl[3p_4(7+t)-p_4(8+t)\bigr]
=286t^4+4156t^3+20234t^2+35744t+13215.
\end{gathered}
\label{eq:supp-quartic-tail}
\end{equation}
Their coefficients are positive, so Eq.~\eqref{eq:supp-certificate-tail}
shows that the weighted tail decreases to zero. The full integer
envelope consequently has endpoints $0$ and $5/1744$.

For the matching bound, now let $p$ be any polynomial in $\mathcal P_4$
and write $c_w=3^{-w}p(w)$. The coefficient of $x^5$ in the Lagrange
interpolation formula for $p$ at $0,1,2,3,4,7$ vanishes because $\deg p\le4$.
Multiplying the resulting identity by $840$
and using $p(0)=1$ gives
\begin{equation}
5+378c_2+945c_4=70c_1+945c_3+729c_7.
\label{eq:supp-quartic-interpolation}
\end{equation}
Using $\ell_2(p)\le c_w\le u_2(p)$ and $\ell_2(p)\le0$,
Eq.~\eqref{eq:supp-quartic-interpolation} gives
\begin{equation}
5+1739\ell_2(p)\le5+1323\ell_2(p)\le1744u_2(p),
\qquad
\frac{u_2(p)-\ell_2(p)}{1-\ell_2(p)}\ge\frac5{1744}.
\label{eq:supp-quartic-optimality}
\end{equation}
The polynomial in Eq.~\eqref{eq:supp-quartic-polynomial} attains this
bound, proving $\gamma_4(2)=5/1744$.

Table~\ref{tab:supp-critical-values} gives the exact qubit values
through degree ten. Each row is verified by the rational interpolation,
finite prefix and shifted-coefficient tests of
Lemma~\ref{lem:supp-finite-certificate}.

\begin{center}
\begin{minipage}{\textwidth}
\centering
\setlength{\tabcolsep}{8pt}
\renewcommand{\arraystretch}{2.2}
\begin{tabular}{@{}ccl@{}}
\toprule
$r$ & $\gamma_r(2)$ & Selected finite nodes\\
\midrule
0 & $\dfrac{1}{3}$ & $1$ \\
1 & $\dfrac{1}{10}$ & $1,2$ \\
2 & $\dfrac{1}{35}$ & $1,2,4$ \\
3 & $\dfrac{4}{427}$ & $1,2,3,5$ \\
4 & $\dfrac{5}{1744}$ & $1,2,3,4,7$ \\
5 & $\dfrac{35}{38672}$ & $1,2,3,4,6,9$ \\
6 & $\dfrac{8}{27667}$ & $1,2,3,4,5,7,11$ \\
7 & $\dfrac{1232}{13932665}$ & $1,2,3,4,5,6,9,13$ \\
8 & $\dfrac{286}{9700391}$ & $1,2,3,4,5,6,8,10,14$ \\
9 & $\dfrac{5005}{539669854}$ & $1,2,3,4,5,6,7,9,12,16$ \\
10 & $\dfrac{3536}{1153439109}$ & $1,2,3,4,5,6,7,9,11,14,18$ \\
\bottomrule
\end{tabular}

\par
\captionsetup{labelsep=period,hypcap=false,justification=centering}
\captionof{table}{Exact qubit step points $\gamma_r(2)$ and selected finite
alternating nodes for $0\le r\le10$. At $d=2$, the certificates in this table
use decay base~$3$ and tail cutoff $M=j_{r+1}$. Each full support also contains $\infty$.}
\label{tab:supp-critical-values}
\end{minipage}
\end{center}

\section*{Use of Artificial Intelligence}

Building on the authors' initial ideas, frontier large language models assisted in
developing complete proofs of the copy-complexity bounds and related
mathematical arguments. They were also used for literature searches and code
development. The main text was drafted by the authors and subsequently
polished with AI assistance. The Supplemental Material was largely
drafted and revised using these models under the authors' guidance. The
authors critically assessed and revised the AI-assisted material,
checked the arguments, computations, and cited sources, and accept full
responsibility for the final manuscript.

\fi

\end{document}